\documentclass{article}
\usepackage[utf8]{inputenc}
\usepackage[english]{babel}
\usepackage{graphicx}
\usepackage{amsmath, amssymb,xcolor,graphics,epsfig,amsbsy,latexsym, amsthm, amsfonts} 
\usepackage{hyperref}
\usepackage{apacite}
\usepackage{natbib}
\usepackage{tabularx,subfig,enumitem}
\usepackage{ragged2e}
\usepackage{color,soul, setspace}
\usepackage[letterpaper,top=2.54cm,bottom=2.54cm,left=2.54cm,right=2.54cm,marginparwidth=2.54cm]{geometry}
\usepackage[font=small, skip=0mm]{caption}
\usepackage{multirow}
\usepackage{footnote}
\usepackage{url}
\usepackage{placeins}
\title{Combining longitudinal cohort studies to examine cardiovascular risk factor trajectories across the adult lifespan}
\author{Zeynab Aghabazaz$^{1*}$,
Michael Joseph Daniels$^{2}$, Hongyan Ning$^{1}$, \\ 
Donald M. Lloyd-Jones$^{3}$,
Juned Siddique$^{1}$ \\
$^{1}$\footnotesize Department of Preventive Medicine, Northwestern University Feinberg School of Medicine, USA \\
$^{2}$\footnotesize Department of Statistics, University of Florida, USA \\  
$^{3}$\footnotesize \parbox{15cm}{Framingham Center for Population \& Prevention Research and Section of Preventive Medicine \& Epidemiology, 
Department of Medicine, Boston University Chobanian \& Avedisian School of Medicine, Boston, MA, USA} \\
\footnotesize *zeynab.aghabazaz@northwestern.edu}
\date{}
\begin{document}
\maketitle
\begin{abstract}
We introduce a statistical framework for combining data from multiple large longitudinal cardiovascular cohorts to enable the study of long-term cardiovascular health starting in early adulthood. Using data from seven cohorts belonging to the Lifetime Risk Pooling Project (LRPP), we present a Bayesian hierarchical multivariate approach that jointly models multiple longitudinal risk factors over time and across cohorts. 
Because few cohorts in our project cover the entire adult lifespan, our strategy uses information from all risk factors to increase precision for each risk factor trajectory and borrows information across cohorts to fill in unobserved risk factors. 
We develop novel diagnostic testing and model validation methods to ensure that our model robustly captures and maintains critical relationships over time and across risk factors. Our modeling reveals substantial age-related variation in risk factor trajectories, with patterns that differ across life stages, subgroups, and cohorts, thereby highlighting key periods for cardiovascular prevention and monitoring.

\noindent%
\end{abstract}
\maketitle
\section{Introduction}
\label{s:intro}
Cardiovascular disease (CVD) is the leading cause of death in the United States and is responsible for more than a third of all deaths each year \citep{ahmad2021leading}. The development of clinical CVD is a process that occurs across the lifespan, beginning early in life and spanning late into life as clinical event rates increase. Much of our understanding of the impact of CVD risk factors comes from studies examining the association between risk factor levels measured at a single point in time, often in middle age, with the incident disease over the short- to intermediate-term \citep{A2014}. However, risk factor levels in young adulthood are significantly associated with the development of CVD later in life \citep{yang2012trends}, and studies demonstrate that not only the levels at specific ages but also cumulative exposures and long-term trajectories in cardiovascular health are significantly related to the risk for subsequent CVD \citep{navar2015hyperlipidemia, pletcher2016young, pool2018use}. 
Therefore, a life course approach is critical in order to understand how CVD risk factors develop and impact an individual’s risk for CVD events later in life. 

Recent work has highlighted the need for integrative cross-cohort analyses to identify early-life determinants of disease and estimate risk trajectories across the lifespan \citep{zuber2023integrative}. Yet there is no single study that has collected detailed phenotypic data spanning young adulthood through old age on a broadly representative sample of the US population.

In this manuscript, we propose a statistical framework for combining longitudinal risk factor data from multiple large cohort studies to enable the study of long-term cardiovascular health starting in early adulthood. We use data from 7 contemporary cardiovascular cohort studies that are part of the Lifetime Risk Pooling Project (LRPP). The 7 cohorts contain $>$256k observations on repeated measures of CVD risk factors, detailed information about medication use, complete vital (death) status follow-up, with mortality ascertained via the National Death Index or cohort-specific procedures, and adjudicated CVD events \citep{wilkins2015data, bundy2020cardiovascular}.
Few cohorts in the LRPP cover the entire adult lifespan, our model allows us to consider risk factors at ages not included in each cohort study as missing data and to fill in unobserved measurements using multiple imputation. 

The traditional approach for combining information across multiple studies is meta-analysis, in which cohorts are analyzed separately, and inferences are averaged across cohorts. Using individual-level data as opposed to aggregate data has many advantages, including the ability to use common definitions/cutpoints, to adjust for variables at the individual level consistently across studies, to conduct time-to-event analysis, and the opportunity to examine heterogeneity at the individual or subgroup level.
However, the challenges involved with combining data from multiple studies are substantial. A key challenge is identifying and controlling for important sources of between-study heterogeneity. In CVD cohorts, this heterogeneity can be a result of differences in geography, historical period, and sample characteristics of the cohort, for example, all white or all African American cohorts \citep{curran2009integrative}.

There has been some work for handling these challenges in combining data. When sufficient overlap exists across ages, historical periods, and participant characteristics, multi-level models can be fit in order to capture between-study variability \citep{schafer2002computational, gelman2006data}.
Multiply imputed cohorts \citep{zeki2019use} and \citep{ siddique2019measurement} can help facilitate analyses by filling-in missing data at ages not captured by the individual cohorts. 
We extend these methodologies and develop a new approach to combine seven cohorts belonging to the LRPP and impute unobserved CVD risk factors.
Figure~\ref{suppfig:Hierarchical_structure} illustrates our proposed hierarchical risk factor model, with multiple risk factors measured repeatedly over time within the same individual and individuals clustered within cohort studies.
Features of our risk factor model include: i) multivariate; at a given age, the model captures correlations between risk factor slopes on the same individual, ii) longitudinal; for a given risk factor, the model captures correlation of risk factors trends
over time, iii) hierarchical; the model captures correlation between trends from different 
cohorts, and iv) error propagation; the model incorporate uncertainty due to incomplete data when borrowing information from cohorts to “fill-in” missing risk factor data.

The overall goal of this project is to identify and measure the characteristics of CVD risk factor trajectories across the adult lifespan. Measuring these characteristics can help identify critical periods for intervention, more precisely define thresholds for known risk factors, elucidate the role of lifestyle behaviors, explain differences in health among populations, and promote CVD prevention strategies at younger ages. 

The manuscript is organized as follows. Section \ref{sec2} provides a comprehensive description of the LRPP data. Section \ref{sec3} introduces our multivariate hierarchical Bayesian model. Section \ref{sec4} describes statistical inference for the longitudinal risk factors model. In Section~\ref{sec5}, novel model validation, posterior predictive checking, and simulation analyses are implemented to evaluate the model’s ability to impute missing risk factors and reproduce observed data patterns.
Section \ref{sec6} provides conclusions and future work.
\section{Application} \label{sec2}
\subsection{LRPP}
Our work is motivated by the LRPP, a well-established individual-level pooled data set from 20 community-based cardiovascular disease cohort studies conducted in the U.S. over the last 50 years. Cohorts were included in the LRPP if they met the following criteria: i) community- or population-based sampling or large volunteer cohort, not participants in a Randomized Control Trial (RCT), ii) availability of at least one baseline examination at which participants provided demographic, personal and medical history information and underwent direct measurement of physiologic and/or anthropometric variables (e.g., blood pressure, weight), iii) longitudinal follow-up of at least 10 years with complete or near-complete ascertainment of vital status. This inclusion criterion ensured that cohorts provided sufficient longitudinal coverage and reliable mortality data to support the analysis of within-subject risk factor trajectories. While 10 years of follow-up was a cohort-level requirement, individuals with shorter follow-up lengths were included; and iv) availability of cause-specific or cardiovascular mortality data with or without ascertainment of non-fatal CVD events.

\begin{figure}[!t]
		\centering
\includegraphics[width=12cm,height=4.5cm]{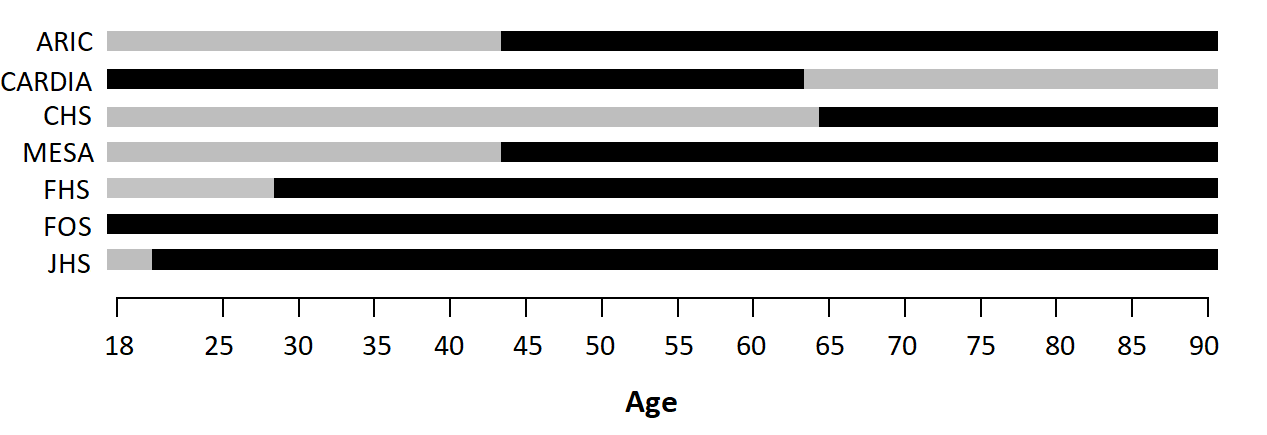}
		\caption{Age ranges in the LRPP (Black indicates ages that were included in each cohort)} 
		\label{figure:AgeRange}
	\end{figure}

For our analysis, we use data from 7 contemporary CVD cohorts, the Atherosclerosis Risk in Communities (ARIC) study,
Coronary Artery Risk Development in Young Adults (CARDIA),
Cardiovascular Health Study (CHS),
Multi-Ethnic Study of Atherosclerosis (MESA),
Framingham Heart Study (FHS), 
Framingham Offspring Study (FOS), and
the Jackson Heart Study (JHS). 
The dataset of each cohort is separately available on the BioLINCC data repository \citep{BioLINCC_Handbook_2023}. After obtaining the data, variables of interest from each data set were cleaned and renamed using a standardized protocol to allow for ease of use in pooling project analyses. Data have been aligned so that measurements are assigned to the age at which they were measured for each individual participant in each cohort. All data in the LRPP is de-identified.
\begin{table}[!t]
\caption{Demographic characteristics and mean follow-up duration for cohorts included in the LRPP. Race: White (W), Black (B), Other (Othr). Education: Less than high school ($<$HS), high school (HS), more than high school (HS+). Follow-up duration is reported as the mean (in years) at the individual level.}
 \label{table:DemographicDetails} 
	\centering          
 \resizebox{\textwidth}{!}{
    \begin{tabular}{llcccccc}
 \\	\hline
Cohort & Sex  & Number of & Age at  & Race/Ethnicity & Education & Length of & Total \\
 &   &  Individuals & Enrollment &  & Level & follow-up & Observations. \\
 \hline
 ARIC &  & & & & & &  56205\\
     & MEN & 5977 & 45-64 & 78\% W, 22\% B & 23\% -HS, 27\% HS, 50\% HS+ & 23.4 & 24349\\
    & WOMEN  & 7425 & 45-64 & 72\% W, 28\% B & 22\% -HS, 37\% HS, 41\% HS+ & 25.1 & 31856\\
 CARDIA  &  & & & & & &  35822\\
    & MEN & 2327 & 18-30 & 50\% W, 50\% B & 4\% -HS, 20\% HS, 76\% HS+ & 32.9 &  15874 \\
    & WOMEN & 2785 & 18-30 & 47\% W, 53\% B & 3\% -HS, 14\% HS, 83\% HS+ & 33.9 & 19948 \\
   CHS   &  & & & & & & 33003\\
  & MEN & 1666 & 65-90 & 85\% W, 14\% B, 1\% Othr & 30\% -HS, 23\% HS, 47\% HS+ & 12.7 & 12089  \\
    & WOMEN & 2625 &  65-90 & 84\% W, 15\% B, 1\% Othr & 27\% -HS, 31\% HS, 42\% HS+ & 14.9 & 20914 \\
  MESA  &  & & & & & &  28798\\
   & MEN & 3194 & 45–84 & 39\% W, 26\% B, 35\% Othr & 16\% -HS, 16\% HS, 68\% HS+ & 14.2 & 13510 \\
    & WOMEN & 3579 & 45–84 & 38\% W, 29\% B, 33\% Othr & 19\% -HS, 21\% HS, 60\% HS+ & 14.7 & 15288\\
FHS   &  & & & & & &  61649\\
   & MEN & 2157 & 28-62 & 100\% W & 45\% -HS, 27\% HS, 28\% HS+ & 31.1 & 25277 \\
  & WOMEN & 2652 & 28-62 & 100\% W & 41\% -HS, 31\% HS, 28\% HS+ & 36.5 & 36372 \\
 FOS  &  & & & & & & 26867 \\
   & MEN & 2005 & 5–70 &100\% W & 8\% -HS, 31\% HS, 61\% HS+ & 34.5 & 12372 \\
    & WOMEN & 2190 &  5–70 & 100\% W & 6\% -HS, 37\% HS, 57\% HS+ & 36.4 & 14495\\
   JHS &  & & & & & & 8203 \\
   & MEN & 1211 & 21-94 & 100\% B & 13\% -HS, 17\% HS, 70\% HS+ & 9.5 & 3043 \\
    & WOMEN & 2043 & 21-94 & 100\% B & 12\% -HS, 17\% HS, 71\% HS+ & 9.6 & 5160 \\
    \hline
\end{tabular} }
\end{table}
Table \ref{table:DemographicDetails} provides the number of individuals, age at enrollment, number of observations (the total number of exams), and demographic information of the 7 LRPP cohorts. All of the cohorts include detailed demographic data, including age, self-identified race/ethnicity, sex, and education levels. 
Figure \ref{figure:AgeRange} displays the age ranges in each of the cohorts included in the LRPP. The longest interval of follow-up, 59 years, comes from the FHS. Other cohorts, such as ARIC, CHS, and MESA, begin in middle age. Examination frequency is variable, with annual examinations in the CHS and longer intervals between examinations in many cohorts. With $>$256k observations, follow-up information across 40 to 50 years with overlapping age ranges, and high-quality, in-person phenotyping of risk factors during serial clinic visits over long follow-up periods, the LRPP provides us with an exceptional opportunity to introduce methods for combining multiple longitudinal cohort studies in order to examine patterns of CVD risk factor development from early adulthood through old age and the associations of these patterns with cardiovascular events later in life. More descriptive plots for the LRPP data are provided in Supplementary Materials~\ref{supp:Descriptive_Plots}.
Clinical risk factor information is available for all major cardiovascular risk factors. We include 7 risk factors: Systolic Blood Pressure (SBP), Diastolic Blood Pressure (DBP), Body Mass Index (BMI), Glucose (GLU), total cholesterol (TOTCHL), HDL cholesterol (HDL), and Triglycerides (TRIG). 

Birth year captures long-term secular trends in cardiovascular risk factors and varies substantially both within and across cohorts. We categorize Birth year into four groups based on quartiles ($<$1915, 1915–1929, 1929–1945, $>$1945) and include main effects and age interactions in the mean model to adjust for these generational differences. This allows the model to separate age-related change from period and cohort effects. Additional descriptive distributions of Birth year by cohort are provided in Supplementary Materials~\ref{supp:Birth_Year}.

In addition to risk factors at ages not covered by each cohort study, the LRPP data includes some risk factors that are unobserved at certain exams. Details on the proportion of missing values across risk factors, cohorts, and sex are provided in Supplementary Materials~\ref{supp:Missing_and}. Our model will leverage information across cohorts to address missing risk factor data; further discussion on this is provided in Section \ref{sec5}.
\section{Longitudinal Risk Factor Model} \label{sec3}
\sloppy
Let \( y_{\ell k(i)}(a_{ij}) \) represent the \( \ell \)th risk factor (\( \ell = 1, \ldots, L \)) for the \( i \)th participant (\( i = 1, \ldots, n_{k} \)) nested within the \( k \)th cohort (\( k = 1, \ldots, K \)) at age \( a_{ij} \) (\( j = 1, \ldots, J_{i} \)). We model \( y_{\ell k(i)} \) as 
\begin{equation}
y_{\ell k(i)}(a_{ij}) = \mu_{\ell k(i)}(a_{ij}) + \epsilon_{\ell k(i)}(a_{ij}) \label{RF:model},
\end{equation}
where \( \mu_{\ell k(i)}(a_{ij}) \) and \( \epsilon_{\ell k(i)}(a_{ij}) \) are the mean trajectory and error terms of risk factor \( \ell \) for
participant \( i \) at age \( a_{ij} \), respectively. 
To capture age-dependent changes in risk factors, we model \( \mu_{\ell k(i)}(a_{ij}) \) using a piecewise linear spline with \( P \) pre-selected breakpoints, 
dividing the age axis into partitions (or windows) defined by knots \( \{s_1, s_2, \dots, s_P\} \). 
We specify these breakpoints at 10-year intervals: 18, 28, 38, \ldots, 78. That is,
\begin{equation}
	\mu_{\ell k(i)}(a_{ij}~|\{s _{p}\}_{p=1}^{P}) =\beta_{\ell k(i)}^{(0)}+\beta_{\ell k(i)}^{(1)}a_{ij}+\sum_{p=1}^{P}\beta_{\ell k(i)}^{(p+1)}(a_{ij}- s_{p})_{+},
	\label{eq2} 
\end{equation}
\sloppy
where $(a - s_{p})_{+} = \max\{0,\, a - s_{p}\}$, which equals $a - s_{p}$ for ages beyond knot $s_{p}$ and 0 otherwise.
This model allows the rate of change (i.e., slope) of the risk factor to vary across age windows, providing flexibility to capture gradual shifts in risk factors over different life stages. The 10-year knots correspond to meaningful life stages, enabling interpretation of risk factor changes by decade. These intervals align with physiological and behavioral transitions, such as those related to midlife transitions or the onset of age-related health conditions. By allowing slopes to vary with covariates, the model captures cohort-specific and time-invariant influences on each risk factor, highlighting the impact of these factors over distinct periods. This specification supports interpretability and identifiability while remaining computationally efficient.

Let $\boldsymbol{A}(a_{ij})=(1,a_{ij},(a_{ij}- s_{1})_{+},\ldots
,(a_{ij}- s_{P})_{+})^{T}$ be a vector of basis functions in (\ref{eq2}), and let $\boldsymbol{\beta}_{\ell k(i)}=(\beta_{\ell k(i)}^{(0)},\beta_{\ell k(i)}^{(1)},
\beta_{\ell k(i)}^{(2)},\ldots,\beta_{\ell k(i)}^{(P+1)})^{T}$ 
be the corresponding vector of regression coefficients for risk factor $\ell$ and participant $i$ nested in cohort $k$. 
Equation (\ref{eq2}) can be re-written as 
\begin{eqnarray}
\mu_{\ell k(i)}(a_{ij}) = \boldsymbol{A}^{T}(a_{ij})\boldsymbol{\beta}_{\ell k(i)}.  \label{eq3}
\end{eqnarray}
We model the slopes, $\beta_{\ell k(i)}^{(p)}$ as 
\begin{equation} 
\beta_{\ell k(i)}^{(p)}=\left\{ 
\begin{array}{ll}
h_{\ell }^{(p)}(\boldsymbol{X}_{i}) + b_{i\ell }^{(p)} & \text{for } p=0,1 \\ 
h_{\ell }^{(p)}(\boldsymbol{X}_{i}) + b_{\ell k}^{(p)} & 
\text{for } p=2,\ldots,P+1,
\end{array} 
\right.
\label{slope}
\end{equation} 
where the first component $h_{\ell }^{(p)}(\boldsymbol{X}_{i})$ represents fixed effects and the second component the associated random effects. 
For $p = 0$, $h_{\ell }^{(0)}(\boldsymbol{X}_{i})$ includes fixed effects of the overall intercept, cohort indicators, and main effects of baseline covariates such as race, education, and Birth year. 
For $p = 1$, $h_{\ell }^{(1)}(\boldsymbol{X}_{i})$ includes interactions of age with these baseline covariates. 
This specification allows the mean level and overall rate of change in each risk factor to vary systematically with baseline characteristics. 
For example, we may write $h_{\ell }^{(p)}(\boldsymbol{X}_{i}) = \boldsymbol{X}_{i}^{T}\boldsymbol{\alpha}_{\ell }^{(p)}$, where $\boldsymbol{\alpha}_{\ell }^{(p)}$ denotes the corresponding fixed-effect coefficients.

For $p \ge 2$, $h_{\ell }^{(p)}(\boldsymbol{X}_{i})$ represents fixed effects for the spline-based slope increments that capture additional age-related changes beyond the global trend. 
These increments are modeled as constant across covariate groups,
\(
h_{\ell }^{(p)}(\boldsymbol{X}_{i}) = \alpha_{\ell }^{(p)}\), for \( p \ge 2
\),
and a first-order random-walk prior is imposed on 
$\alpha_{\ell }^{(p)}$, for \( p=2,\ldots,P+1\), to encourage smoothness across adjacent spline segments:
\(
\alpha_{\ell }^{(2)} \sim \mathrm{N}(0,\,10^2)\),
\(
\alpha_{\ell }^{(p)} \sim \mathrm{N}(\alpha_{\ell }^{(p-1)},\,\tau_{\ell}^{2})\), 
 \(p = 3,\ldots,P+1,
\)
where $\tau_{\ell}$ controls the smoothness of successive slope differences. 
This specification assigns a fixed effect to each spline segment and provides flexible slope changes with age while maintaining interpretability and parsimony; see Supplementary Materials~\ref{supp:Design_Matrix} for definitions of $h_{\ell }^{(p)}(\boldsymbol{X}_{i})$.

The second component in (\ref{slope}) includes the random effects associated with the $p$th coefficient of the $\ell$th risk factor for participant $i$ in cohort $k$. Specifically, $b_{i\ell }^{(0)}$ and $b_{i\ell }^{(1)}$ represent the \emph{subject-specific} deviation from the overall intercept or global age slope, respectively, and capture the correlation among different risk factor slopes within an individual. The random effect $b_{\ell k}^{(p)}$ represents the \emph{cohort-specific} deviation associated with $h_{\ell }^{(p)}(\boldsymbol{X}_{i})$ and captures the correlation in slope increments across cohorts. We introduce individual-specific random effects $b_{i\ell }^{(p)}$ only for the intercept and the overall age slope (i.e., \(p = 0, 1\)), while cohort-specific effects $b_{\ell k}^{(p)}$ are applied to the spline-based slope increments (i.e., \(p \ge 2\)). This specification enables us to capture individual baseline differences and general trends without introducing excessive complexity. Including cohort effects for the spline increments is important for modeling age-specific cohort influences and borrowing information across cohorts, while the individual effects focus on capturing participant-level deviations in the overall trajectory.

Adding individual-level random effects to each age window would lead to over-parameterization, as it would introduce additional, potentially redundant, sources of variation for each age interval. By limiting individual effects to the overall slope, we avoid this redundancy and preserve the model’s parsimony, maintaining a clear distinction between cohort-level variation across age windows and individual-level trends. Although cohort deviations are modeled across all age windows, this does not imply they are more important than individual differences; rather, individual heterogeneity is primarily captured through overall trends, which are most relevant for characterizing long-term trajectories.
\subsection{Specification of the Covariance Structures}
\label{3Spe} 
 We assume that a given risk factor's slopes $b_{i\ell }^{(p)}$, $p=0,1$, are correlated,  capturing the correlation of risk factor $\ell $ at age $a_{ij}$. 
Therefore, for $i$th individual and the $L$ risk factors, we have a random effects matrix $\boldsymbol{b}_{i}$ with dimension $ 2\times L$, i.e.
	\begin{eqnarray}
 \boldsymbol{b}_{i}=
	\left[ 
	\begin{array}{cccc}
	b_{i1}^{(0)} & b_{i2}^{(0)} &\ldots & b_{iL}^{(0)} \\ 
	b_{i1}^{(1)} & b_{i2}^{(1)} &\ldots & b_{iL}^{(1)} \\
	\end{array} \right], \label{eq5} \notag
	\end{eqnarray}
where the components of the first and second rows are the random intercepts and random slopes at the individual level. 
We assume $\mathrm{Vec}(\boldsymbol{b}_{i}) \sim \mathrm{N}(\boldsymbol{0}, \boldsymbol{\Sigma})$;
$\mathrm{Vec}(\cdot) $ is a vector with dimension $2L$. $\boldsymbol{\Sigma}$ is a $2L\times 2L$ covariance matrix of risk factor intercepts and slopes. 
The cohort-specific random effect is $ \boldsymbol{b}_{\ell }^{(p)} = (b_{\ell 1}^{(p)}\ldots, b_{\ell K}^{(p)})^{T}$ for $p \geq 2$. We assume that
$\boldsymbol{b}_{\ell }^{(p)} \sim \mathrm{N}(\boldsymbol{0},\boldsymbol{\Lambda}^{(\ell )}) \label{eq8}$, where $\boldsymbol{\Lambda}^{(\ell )}$ is a $K\times K$ covariance matrix with elements $\lambda_{kk^\prime}^{(\ell )}$, for ${k,k^\prime=1,\ldots, K}$, with $\lambda_{kk^\prime}^{(\ell)} = \lambda_{k^\prime k}^{(\ell)}$. Then, the covariance matrix $\boldsymbol{\Lambda}^{(\ell )}$ varies by risk factors but is constant over the $P$ age windows. 

Finally, the term $\epsilon_{\ell k(i)}(a_{ij})$ in (\ref{RF:model}) captures the residual variability within the $\ell$th risk factor of the $i$th individual nested in the $k$th cohort across different age windows. 
We assume $\epsilon_{\ell k(i)}(a_{ij}) \sim \mathrm{SN}\!\big(\xi_{\ell}^{(p)},\, \omega_{\ell}^{(p)},\, \psi_{\ell}\big)$, here SN denotes the skew-normal distribution with location parameter $\xi_{\ell}^{(p)} = -\,\omega_{\ell}^{(p)}\,\delta_{\ell}\sqrt{2/\pi}$, scale parameter $\omega_{\ell}^{(p)}$  for window $p$, and skewness parameter $\psi_{\ell}$, where $\delta_{\ell} = \psi_{\ell}/\sqrt{1+\psi_{\ell}^{2}}$. 
We use the mean-parameterized form of the skew-normal distribution, in which the location parameter $\xi_{\ell}^{(p)}$ is defined so that $E[\epsilon_{\ell k(i)}(a_{ij})]=0$. 
This parameterization centers the residual distribution while allowing asymmetry and age-specific residual variability across windows for each risk factor.

The individual-level slope variance is assumed constant across age (homoskedastic), which implies increasing variability in observed outcomes with age due to slope accumulation. This choice supports identifiability in our high-dimensional setting. However, age-specific residual variances help mitigate potential overdispersion at older ages.
In Supplementary Materials~\ref{supp:Properties_of}, we derive the covariances and variances that capture variability and dependencies across individuals, risk factors, age intervals, and cohorts.
\subsection{Observed Data Likelihood} \label{sec:3.4}
We use all available data and do not delete individuals with missing risk factors at some exams. We define \(\mathcal{B} = (\mathcal{B}_1, \mathcal{B}_2)\), where \(\mathcal{B}_1\) denotes the set of fixed (population-level) parameters, including regression coefficients, spline weights, and variance components, and \(\mathcal{B}_2\) represents subject- and cohort-specific random effects. A full list of the parameters included in \(\mathcal{B}\) is provided in Supplementary Materials~\ref{supp:Specification_of}. We define \(\mathcal{L} = \{1, \dots, L\}\) as the full set of common risk factors across all individuals. For each individual \(i\), \(\mathcal{L}_{\text{tot}}\) denotes the subset of risk factors that are 
measured at least once across all visits, and \(\mathcal{L}_{\text{obs}} \subseteq \mathcal{L}_{\text{tot}}\) 
denotes the subset measured at a given exam.
 The complete data likelihood for participant $i$ at the $j$th exam is
\begin{equation}
    \prod_{\ell \in \mathcal{L}_{\text{tot}}} \mathrm{Pr}\big(Y_{\ell k(i)}(a_{ij})|\mathcal{B}\big)\times \mathrm{Pr}(\mathcal{B}), \label{eq:likelihood}
\end{equation} 
where $\mathcal{L}_{\text{tot}}$ includes observed and unobserved risk factors.

We assume ignorable dropout and include all individuals, regardless of the number of visits, with the model borrowing strength across risk factors, age windows, and cohorts.
Therefore, conditional on $\mathcal{B}$, the observed data likelihood is
\begin{equation}
    \prod_{\ell \in \mathcal{L}_{\text{obs}}}{ \mathrm{Pr}(Y_{\ell k(i)}(a_{ij})|\mathcal{B})\times \mathrm{Pr}(\mathcal{B})}, \label{eq7.}
\end{equation}
where $\mathcal{L}_{\text{obs}} \subseteq \mathcal{L}_{\text{tot}}$, as defined above, and
includes all observed risk factors for $i$th individual at the $j$th exam. 
\subsection{Covariates and Fixed Effects}
We include covariates to capture demographic and temporal effects.
Education level (-HS, HS, and HS+) and race (Black vs.\ non-Black) enter the model as fixed main effects and also interact with age through the global age slope, allowing their associations with risk factor levels and age-related changes to vary across life stages. 
Birth year is included as a fixed effect and also interacts with age to capture secular and generational differences in both baseline levels and rates of change. Cohort membership is included as a fixed effect in the intercept term to account for systematic baseline differences across cohort studies.
\section{Bayesian Inference} \label{sec4}
\subsection{Prior Specification}
We use weakly informative priors for all model parameters to ensure stable estimation while avoiding strong assumptions. A complete description of all prior distributions, including those for skewness parameters, spline increments, covariance matrices, and residual variances, is provided in Supplementary Materials~\ref{supp:Full_Prior}.
\subsection{Posterior Estimation}
We employ Markov chain Monte Carlo (MCMC) algorithms to obtain samples from the
posterior distribution of the parameters, as implemented in Stan \citep{stan}.
To achieve robust convergence diagnostics in our Bayesian model, we employ the nested \( \hat{R} \) approach as described in \cite{margossian2024nested}, where subchains are nested within superchains. This is particularly advantageous in high-dimensional settings where the complexity of the parameter space and the dataset size can present challenges for efficient sampling. By using superchains, we reduce the computational demands of running long chains while maintaining reliable convergence checks. Implementing the nested \( \hat{R} \) approach requires access to high-performance parallel computing resources due to the intensive nature of managing and processing multiple superchains. Details of the nested 
$\hat{R}$ approach can be found in Supplementary Materials~\ref{supp:Nested_rhat}.

This diagnostic enables reliable convergence checks by examining consistency across superchains rather than individual chains, facilitating convergence to the stationary distribution with even shorter chain lengths. 
For our model, which includes 842 parameters, this setup yielded nearly all nested \( \hat{R} \) values below the standard threshold of 1.1, indicating satisfactory convergence. Posterior predictive checks to assess model fit are introduced in the next section.
\section{Model Validation} \label{sec5}
\subsection{Posterior Predictive Checking}
We evaluate our model to ensure that it accurately preserves key relationships in the observed data using posterior predictive checks \citep{gelman2013bayesian}. These checks involve comparing test statistics based on the observed data to the same statistics computed from data replicated from the posterior predictive distribution. 
Let $y^{\text{obs}}$ be the observed data, and recall that $\mathcal{B}$ denotes the full vector of model parameters (fixed and random effects), as defined in Section~\ref{sec:3.4}.
We define $y^{\text{rep}}$ as the replicated data that could have been observed with the same model and the same value of $\mathcal{B}$ that produced the observed data. We work with the distribution of $y^{\text{rep}}$ given the observed data, called the posterior predictive distribution
\begin{equation}
\mathrm{Pr}\left(y^{\text{rep}}|y^{\text{obs}}\right)=\int{\mathrm{Pr}\left(y^{\text{rep}}|\mathcal{B}\right)\mathrm{Pr}\left(\mathcal{B}|y^{\text{obs}}\right)d\mathcal{B}} . \notag
\end{equation}
We calculate a posterior predictive probability (PPP), which is defined as the probability 
that the replicated data is more extreme than the observed data,
\begin{eqnarray}  
\text{PPP} &=& \mathrm{Pr}\left(T(y^{\text{rep}}, \mathcal{B}) \geq T(y^{\text{obs}}, \mathcal{B})|y^{\text{obs}}\right) \notag \\
 &=& \int \int  I\left(T(y^{\text{rep}},\mathcal{B}) \geq T(y^{\text{obs}},\mathcal{B})\right) \mathrm{Pr}(y^{\text{rep}}|\mathcal{B}) \mathrm{Pr}(\mathcal{B}|y^{\text{obs}}) dy^{\text{rep}} d\mathcal{B},
\end{eqnarray}
where $I(\cdot)$ is the indicator function and $T(\cdot)$ is the test statistic. 
For $M$ draws from the posterior distribution of $\mathcal{B}$, the probability is estimated as the proportion of these $M$ draws for which the test quantity equals or exceeds its realized value, i.e., $T(y^{\text{rep}^{(m)}},\mathcal{B}^{(m)}) \geq T(y^{\text{obs}},\mathcal{B}^{(m)}), m = 1,\ldots,M$. For the longitudinal risk factors model described in Section \ref{sec3}, we consider numerous test statistics, which we describe below, to confirm that our model accurately captures the behavior of the risk factors.
\subsubsection{The variability of the longitudinal data around the mean risk trajectory} 
 We first examine whether the residual distribution provides a satisfactory fit. Recall the basis functions vector $\boldsymbol{A}(a_{ij})$ defined in (\ref{eq3}), and the regression coefficients vector $\boldsymbol{\beta}_{\ell k(i)}^{(m)}=(\beta_{\ell k(i)}^{(0)^{(m)}},\beta_{\ell k(i)}^{(1)^{(m)}},
\beta_{\ell k(i)}^{(2)^{(m)}},\ldots,\beta_{\ell k(i)}^{(P+1)^{(m)}})^{T}$, which represents the slopes over $P$ age intervals for risk factor $\ell $ and participant $i$ nested in cohort $k$ at iteration $m$. We define the \textit{mean risk trajectory} at age $a_{ij}$ as the latent, structured component of the model given by
\begin{eqnarray} 
\mu_{\ell k(i)}^{(m)}(a_{ij}) = \boldsymbol{A}^{T}(a_{ij})\boldsymbol{\beta}_{\ell k(i)}^{(m)},  \notag
\end{eqnarray}%
where 
\begin{equation}
\beta_{\ell k(i)}^{(p)^{(m)}}=\left\{ 
\begin{array}{ll}
h_{\ell }^{(p)^{(m)}}(\boldsymbol{X}_{i}) + b_{i\ell }^{(p)^{(m)}} & \text{for } p=0,1 \\ 
h_{\ell }^{(p)^{(m)}}(\boldsymbol{X}_{i}) + b_{\ell k}^{(p)^{(m)}} & 
\text{for } p=2,\ldots,P+1,
\end{array}%
\right. \notag
\end{equation} 
for $\ell =1,\ldots ,L$, $k=1,\ldots ,K$, $i=1,\ldots ,n_{k}$, $j=1,\ldots ,J_{i}$, and $m=1,\ldots,M$.
It captures the expected age-dependent pattern for risk factor \(\ell\) based on covariates and random effects, without adding the residual skew-normal error. 

Let $\sigma_{\ell }^{(m)}(p)$ be the standard deviation at age $a_{ij}$ that falls within the $p$th age window. 
We define the standardized residuals for replicated and observed risk factors as
\begin{eqnarray}
\mathcal{R}_{\ell k(i)}^{\text{rep}{(m)}}(a_{ij}) &=& \left(y_{\ell k(i)}^{\text{rep}(m)}(a_{ij}) - \mu_{\ell k(i)}^{(m)}(a_{ij})\right)/\sigma_{\ell }^{(m)}(p), \notag \\
\mathcal{R}_{\ell k(i)}^{\text{obs}(m)}(a_{ij}) &=& \left(y_{\ell k(i)}^{\text{obs}(m)}(a_{ij}) - \mu_{\ell k(i)}^{(m)}(a_{ij})\right)/\sigma_{\ell }^{(m)}(p) . \label{resrep}
\end{eqnarray} 

To assess model fit, we use QQ plots of these standardized residuals, comparing observed and replicated data quantiles for each risk factor; see Supplementary Materials~\ref{supp:Residual_QQ}. Points aligning closely with the 45-degree reference line indicate that the model adequately captures the variability around the mean risk trajectory. Deviations from this line, particularly in the tails, suggest areas where the model may over- or under-estimate variability.
The QQ plots serve as a diagnostic tool to validate the model’s distributional assumptions, confirming that the residuals of replicated data are consistent with observed data across age intervals and demographic factors. This assessment supports our model’s capacity to capture within-subject correlations and longitudinal variability effectively.
\subsubsection{Predicted mean trajectories of CVD risk factors}
\label{sec:mean_traj}
We obtained predicted mean trajectories for each CVD risk factor using posterior means from the longitudinal model. 
For each participant, predicted values were calculated across ages 18--88~years conditional on observed baseline covariates, and then averaged to yield population-level trajectories. 
The goal of this analysis was to assess whether the fitted mean structure reproduces realistic age patterns in the risk factors. 

The resulting trajectories (Figure~\ref{fig:mean_traj_men} and Figure~\ref{fig:mean_traj_women}) display smooth, biologically plausible trends for all seven risk factors, with 95\% credible intervals that reflect uncertainty and vary across age in accordance with the differing levels of information available across the age span.
These patterns align with established epidemiologic evidence from large cohort studies, supporting the adequacy and interpretability of the mean model. 
Additional details, and relevant references are provided in Supplementary Materials~\ref{supp:Predicted_Mean}.
\subsubsection{Mean of variance ratio} 
 To evaluate the model’s ability to accurately capture error variances across age windows, we calculate the mean variance ratio between observed and replicated residuals. This ratio is defined as 
\begin{equation} \overline{\text{Ratio}}_{\ell}^{(p)} = \frac{\sum_{m=1}^{M}\left( \mathrm{Var}(\mathcal{R}_{\ell}^{\text{obs}(m)}(p)) /
 \mathrm{Var}(\mathcal{R}_{\ell}^{\text{rep}(m)}(p))\right)}{M},
 \end{equation}
 where \( \mathcal{R}_{\ell}^{\text{obs}(m)}(p) \) and \( \mathcal{R}_{\ell}^{\text{rep}(m)}(p) \) represent the observed and replicated residuals for risk factor \( \ell \) in age window \( p \) at iteration \( m \) defined in (\ref{resrep}). 

 The resulting plots for men and women, displayed in Supplementary Materials~\ref{supp:Variance_Ratios}, illustrate the mean variance ratios across different risk factors and age windows. For most risk factors, the variance ratios cluster close to the reference line at 1 (indicated by the dashed horizontal line), suggesting that the model successfully captures the variability observed in the data. Deviations from this line, particularly for some age windows and specific risk factors (e.g., BMI and GLU), highlight areas where the model may slightly under- or overestimate variability. Overall, the consistency of variance ratios near 1 across age windows indicates that the model effectively captures the error variances, providing a reliable fit to the data across gender and risk factor categories.
 \subsubsection{Correlation between the Rate of Change Across Risk Factors}
We next examine the correlation between the same or different risk factors at the same or different ages to assess whether the covariance structure of the random effects is adequate. To achieve this, we refit the model without cohort effects, isolating the covariance structure of the random effects. This approach enables us to focus on the relationships among risk factors while excluding cohort-level variability, ensuring the random-effects structure accurately reflects these correlations across different ages and rates of change.

For $\ell =1,\ldots ,L$, $k=1,\ldots ,K$, $i=1,\ldots ,n_{k}$, $j=1,\ldots ,J_{i}$, and $m=1,\ldots,M$, we define the coefficients vector 
$\boldsymbol{\beta}_{\ell k(i)}^{\text{nc}(m)}=(\beta_{\ell k(i)}^{(0)^{\text{nc}(m)}},\beta_{\ell k(i)}^{(1)^{\text{nc}(m)}},
\beta_{\ell k(i)}^{(2)^{\text{nc}(m)}},\ldots,\beta_{\ell k(i)}^{(P+1)^{\text{nc}(m)}})^{T}$, with elements
\begin{equation}
\beta_{\ell k(i)}^{(p)^{\text{nc}(m)}}=\left\{ 
\begin{array}{ll}
h_{\ell }^{(p)^{(m)}}(\boldsymbol{X}_{i}) + b_{i\ell }^{(p)^{(m)}} & \text{for } p=0,1 \\ 
h_{\ell }^{(p)^{(m)}}(\boldsymbol{X}_{i}) & 
\text{for } p=2,\ldots,P+1,
\end{array}%
\right. \notag
\end{equation} 
and 
\begin{eqnarray}
\mu_{\ell k(i)}^{\text{nc}(m)}(a_{ij}) = \boldsymbol{A}^{T}(a_{ij})\boldsymbol{\beta}_{\ell k(i)}^{\text{nc}(m)}.  \label{2}
\end{eqnarray}%
Here, the superscript \( nc(m) \) denotes the \( m^{\text{th}} \) posterior draw from the model without cohort-level random effects.
The standardized residuals are
\begin{eqnarray}
\mathcal{R}_{\ell k(i)}^{\text{nc-rep}{(m)}}(a_{ij}) &=& \left(y_{\ell k(i)}^{\text{nc-rep}(m)}(a_{ij}) - \mu_{\ell k(i)}^{\text{nc}(m)}(a_{ij})\right)/\sigma_{\ell}^{(m)}(p) , \notag \\
\mathcal{R}_{\ell k(i)}^{\text{nc-obs}(m)}(a_{ij}) &=& \left(y_{\ell k(i)}^{\text{nc-obs}(m)}(a_{ij}) - \mu_{\ell k(i)}^{\text{nc}(m)}(a_{ij})\right)/\sigma_{\ell}^{(m)}(p) . 
\end{eqnarray}

To ensure that our model can accurately capture within individual correlations, we define the test statistic to be the correlation between residuals. That is, 
\begin{eqnarray}
 \rho^{\text{nc-rep}(m)} &=& \mathrm{Corr}\left(\mathcal{R}_{\ell k(i)}^{\text{nc-rep}{(m)}}(a_{ij}), \mathcal{R}_{\ell ^\prime k(i)}^{\text{nc-rep}{(m)}}(a_{ij^\prime})\right)_{L\times L} ,\notag \\
  \rho^{\text{nc-obs}(m)}  &=& \mathrm{Corr}\left(\mathcal{R}_{\ell k(i)}^{\text{nc-obs}{(m)}}(a_{ij}), \mathcal{R}_{\ell ^\prime k(i)}^{\text{nc-obs}{(m)}}(a_{ij^\prime})\right)_{L\times L} . \label{eq12}
\end{eqnarray}
Equation (\ref{eq12}) is the correlation between risk factors at the same age of an individual if $a_{ij}=a_{ij^\prime}$ and the correlation between risk factors at different ages of an individual if $a_{ij}\neq a_{ij^\prime}$.
We calculate the PPP as
\begin{equation}
 \mathrm{Pr}\left( \rho^{\text{nc-rep}(m)} \geq \rho^{\text{nc-obs}(m)} \mid y^{\text{obs}} \right) . \notag
\end{equation}
where $y^{\text{obs}}$ denotes the observed data. 
These checks evaluate the model's ability to capture complex interdependencies between risk factors across different ages within individuals.  
PPP are reported Table \ref{pp_summary}. With most values close to $0.5$, our results indicate that the model’s covariance structure accurately mirrors observed data patterns.
\begin{table}[!t]
\centering
\caption{Posterior predictive probability for Correlation of Risk Factors at Same and Different Ages by Sex}
\label{pp_summary}
\begin{tabular}{p{3.5cm} p{1cm} p{1cm} p{1cm} p{1cm}}
    \hline
    & \multicolumn{2}{c}{At the Same Age} & \multicolumn{2}{c}{At Different Ages}  \\ 
    \cline{2-3} \cline{4-5}
    Risk Factor Pair & Men & Women & Men & Women \\
    \hline
    SBP–DBP       & 0.18 & 0.13 & 0.16 & 0.13 \\
    SBP–BMI       & 0.39 & 0.41 & 0.40 & 0.41 \\
    SBP–TOTCHL    & 0.30 & 0.32 & 0.28 & 0.32 \\
    SBP–GLU       & 0.67 & 0.61 & 0.71 & 0.65 \\
    SBP–HDL      & 0.36 & 0.52 & 0.41 & 0.52 \\
    SBP–TRIG      & 0.42 & 0.34 & 0.43 & 0.34 \\
    DBP–BMI       & 0.27 & 0.48 & 0.30 & 0.47 \\
    DBP–TOTCHL    & 0.24 & 0.34 & 0.27 & 0.31 \\
    DBP–GLU       & 0.69 & 0.68 & 0.75 & 0.69 \\
    DBP–HDL      & 0.43 & 0.59 & 0.44 & 0.56 \\
    DBP–TRIG      & 0.39 & 0.44 & 0.39 & 0.44 \\
    BMI–TOTCHL    & 0.48 & 0.52 & 0.47 & 0.52 \\
    BMI–GLU       & 0.39 & 0.40 & 0.45 & 0.41 \\
    BMI–HDL      & 0.63 & 0.60 & 0.59 & 0.59 \\
    BMI–TRIG      & 0.37 & 0.41 & 0.38 & 0.41 \\
    TOTCHL–GLU    & 0.65 & 0.64 & 0.72 & 0.71 \\
    TOTCHL–HDL   & 0.41 & 0.34 & 0.38 & 0.34 \\
    TOTCHL–TRIG   & 0.33 & 0.21 & 0.33 & 0.22 \\
    GLU–HDL      & 0.53 & 0.70 & 0.52 & 0.69 \\
    GLU–TRIG      & 0.27 & 0.24 & 0.28 & 0.27 \\
    HDL–TRIG     & 0.60 & 0.76 & 0.60 & 0.74 \\
    \hline
\end{tabular}
\end{table}
\subsubsection{Correlation between the rate of change for the same 10-year windows across cohorts}
We now assess the covariance structure across cohorts. In this case, we refit the model excluding the individual random effects. By removing individual effects, we examine the covariance structure specifically across cohorts. This approach allows us to determine whether the covariance patterns are consistent across cohorts, free from the noise introduced by individual variability. It also ensures that the model adequately captures cohort-level trends, providing a clearer understanding of how risk factor trajectories differ across population subgroups.
For $\ell =1,\ldots ,L$, $k=1,\ldots ,K$, $i=1,\ldots ,n_{k}$, $j=1,\ldots ,J_{i}$, and $m=1,\ldots,M$, we assume that the vector of coefficients is 
$\boldsymbol{\beta}_{\ell k(i)}^{\text{np}(m)}=(\beta_{\ell k(i)}^{(0)^{\text{np}(m)}},\beta_{\ell k(i)}^{(1)^{\text{np}(m)}},
\beta_{\ell k(i)}^{(2)^{\text{np}(m)}},\ldots,\beta_{\ell k(i)}^{(P+1)^{\text{np}(m)}})^{T}$ with elements
\begin{equation}
\beta_{\ell k(i)}^{(p)^{\text{np}(m)}}=
h_{\ell }^{(p)^{(m)}}(\boldsymbol{X}_{i}) + b_{\ell k}^{(p)^{(m)}}  
~~ p=0,1,\ldots,P+1 ,\notag
\end{equation} 
and 
\begin{eqnarray}
\mu_{\ell k(i)}^{\text{np}(m)}(a_{ij}) = \boldsymbol{A}^{T}(a_{ij})\boldsymbol{\beta}_{\ell k(i)}^{\text{np}(m)}.  \notag
\end{eqnarray}
Here the superscript $np(m)$ indicates the \( m^{\text{th}} \) posterior draw from the model without participant random effects. The standardized residuals are then
\begin{eqnarray} 
\mathcal{R}_{\ell k(i)}^{\text{np-rep}{(m)}}(a_{ij}) &=& \left(y_{\ell k(i)}^{\text{np-rep}(m)}(a_{ij}) - \mu_{\ell k(i)}^{\text{np}(m)}(a_{ij})\right)/\sigma_{\ell}^{(m)}(p) , \notag \\
\mathcal{R}_{\ell k(i)}^{\text{np-obs}(m)}(a_{ij}) &=& \left(y_{\ell k(i)}^{\text{np-obs}(m)}(a_{ij}) - \mu_{\ell k(i)}^{\text{np}(m)}(a_{ij})\right)/\sigma_{\ell}^{(m)}(p) .
\end{eqnarray}
Next, we split residuals by the 10-year age intervals and define
\begin{eqnarray}
\overline{\mathcal{R}}_{\ell k}^{\text{np-rep}{(m)}}(p) &=&  \frac{\sum_{a_{ij}\in s_{p}}\mathcal{R}_{\ell k(i)}^{\text{np-rep}{(m)}}(a_{ij})}{\sum_{i\in k}I(a_{ij}\in s_{p})}, \notag \\
\overline{\mathcal{R}}_{\ell k}^{\text{np-obs}(m)}(p) &=& \frac{\sum_{a_{ij}\in s_{p}}\mathcal{R}_{\ell k(i)}^{\text{np-obs}{(m)}}(a_{ij})}{\sum_{i\in k}I(a_{ij}\in s_{p})} . \label{eq16}
\end{eqnarray}
We define the test statistic to be the correlation between residuals defined in (\ref{eq16}) 
\begin{eqnarray}
 \rho^{\text{np-rep}(m)}_{\ell} &=& \mathrm{Corr}\left(\overline{\mathcal{R}}_{\ell k}^{\text{np-rep}{(m)}}(p), \overline{\mathcal{R}}_{\ell k^\prime}^{\text{np-rep}{(m)}}(p)\right)_{K\times K} , \notag \\
  \rho^{\text{np-obs}(m)}_{\ell} &=& \mathrm{Corr}\left(\overline{\mathcal{R}}_{\ell k}^{\text{np-obs}(m)}(p), \overline{\mathcal{R}}_{\ell k^\prime}^{\text{np-obs}(m)}(p)\right)_{K\times K}.
\end{eqnarray}
Finally, we calculate PPP as
\begin{equation}
  \mathrm{Pr}\left( \rho^{\text{np-rep}(m)}_{\ell} \geq  \rho^{\text{np-obs}(m)}_{\ell} \mid y^{\text{obs}} \right) ,\notag
\end{equation}
where $y^{\text{obs}}$ is the observed data. 
\begin{table}[!t]
\centering
\caption{Posterior predictive probability for Cohorts Correlation Across 10-Year Windows by Risk Factor and Sex (M/W)}
\resizebox{\textwidth}{!}{%
\begin{tabular}{p{2.4cm} p{0.7cm} p{0.7cm} p{0.7cm} p{0.7cm} p{0.7cm} p{0.7cm} p{0.7cm} p{0.7cm} p{0.7cm} p{0.7cm} p{0.7cm} p{0.7cm} p{0.7cm} p{0.7cm}}
    \hline
    \multirow{2}{*}{Cohort Pair} & \multicolumn{2}{c}{SBP} & \multicolumn{2}{c}{DBP} & \multicolumn{2}{c}{BMI} & \multicolumn{2}{c}{TOTCHL} & \multicolumn{2}{c}{GLU} & \multicolumn{2}{c}{HDL} & \multicolumn{2}{c}{TRIG} \\ \cline{2-15}
    & M & W & M & W & M & W & M & W & M & W & M & W & M & W \\
    \hline
    ARIC–CA     & 0.23 & 0.25 & 0.22 & 0.30 & 0.08 & 0.27 & 0.35 & 0.25 & 0.09 & 0.10 & 0.18 & 0.23 & 0.31 & 0.23 \\
    ARIC–CHS    & 0.34 & 0.10 & 0.35 & 0.07 & 0.16 & 0.62 & 0.38 & 0.38 & 0.12 & 0.10 & 0.37 & 0.25 & 0.46 & 0.41 \\
    ARIC–MESA   & 0.40 & 0.22 & 0.32 & 0.14 & 0.69 & 0.70 & 0.62 & 0.54 & 0.34 & 0.09 & 0.59 & 0.62 & 0.45 & 0.47 \\
    ARIC–FHS    & 0.48 & 0.25 & 0.48 & 0.39 & 0.67 & 0.56 & 0.32 & 0.55 & 0.38 & 0.08 & 0.37 & 0.52 & 0.38 & 0.51 \\
    ARIC–FOS    & 0.79 & 0.14 & 0.76 & 0.20 & 0.74 & 0.62 & 0.64 & 0.87 & 0.55 & 0.39 & 0.68 & 0.64 & 0.78 & 0.76 \\
    ARIC–JHS    & 0.71 & 0.16 & 0.72 & 0.23 & 0.87 & 0.58 & 0.69 & 0.70 & 0.62 & 0.42 & 0.79 & 0.77 & 0.70 & 0.82 \\
    CA–CHS      & 0.08 & 0.20 & 0.23 & 0.21 & 0.14 & 0.11 & 0.11 & 0.19 & 0.10 & 0.05 & 0.23 & 0.16 & 0.27 & 0.25 \\
    CA–MESA     & 0.30 & 0.26 & 0.40 & 0.40 & 0.68 & 0.35 & 0.52 & 0.48 & 0.33 & 0.13 & 0.52 & 0.51 & 0.41 & 0.45 \\
    CA–FHS      & 0.46 & 0.31 & 0.66 & 0.55 & 0.74 & 0.23 & 0.49 & 0.53 & 0.34 & 0.05 & 0.38 & 0.47 & 0.45 & 0.51 \\
    CA–FOS      & 0.47 & 0.40 & 0.67 & 0.49 & 0.65 & 0.39 & 0.59 & 0.54 & 0.42 & 0.38 & 0.73 & 0.63 & 0.73 & 0.71 \\
    CA–JHS      & 0.45 & 0.33 & 0.62 & 0.47 & 0.77 & 0.65 & 0.59 & 0.50 & 0.37 & 0.39 & 0.79 & 0.75 & 0.73 & 0.68 \\
    CHS–MESA    & 0.05 & 0.06 & 0.09 & 0.07 & 0.30 & 0.05 & 0.13 & 0.05 & 0.20 & 0.11 & 0.13 & 0.16 & 0.04 & 0.10 \\
    CHS–FHS     & 0.44 & 0.20 & 0.38 & 0.40 & 0.72 & 0.12 & 0.55 & 0.43 & 0.36 & 0.13 & 0.23 & 0.23 & 0.10 & 0.11 \\
    CHS–FOS     & 0.60 & 0.24 & 0.61 & 0.33 & 0.39 & 0.45 & 0.33 & 0.22 & 0.21 & 0.41 & 0.40 & 0.24 & 0.23 & 0.20 \\
    CHS–JHS     & 0.52 & 0.38 & 0.44 & 0.37 & 0.78 & 0.63 & 0.41 & 0.38 & 0.30 & 0.30 & 0.45 & 0.34 & 0.57 & 0.38 \\
    MESA–FHS    & 0.38 & 0.15 & 0.22 & 0.43 & 0.17 & 0.07 & 0.50 & 0.37 & 0.03 & 0.02 & 0.15 & 0.10 & 0.02 & 0.03 \\
    MESA–FOS    & 0.57 & 0.24 & 0.41 & 0.26 & 0.30 & 0.36 & 0.34 & 0.21 & 0.14 & 0.27 & 0.23 & 0.24 & 0.26 & 0.20 \\
    MESA–JHS    & 0.52 & 0.27 & 0.49 & 0.27 & 0.53 & 0.65 & 0.63 & 0.39 & 0.39 & 0.29 & 0.29 & 0.28 & 0.55 & 0.31 \\
    FHS–FOS     & 0.10 & 0.10 & 0.23 & 0.13 & 0.44 & 0.41 & 0.23 & 0.38 & 0.06 & 0.13 & 0.11 & 0.11 & 0.27 & 0.05 \\
    FHS–JHS     & 0.15 & 0.27 & 0.52 & 0.56 & 0.33 & 0.63 & 0.28 & 0.33 & 0.36 & 0.38 & 0.32 & 0.24 & 0.48 & 0.28 \\
    FOS–JHS     & 0.10 & 0.10 & 0.31 & 0.27 & 0.28 & 0.20 & 0.09 & 0.23 & 0.33 & 0.24 & 0.21 & 0.15 & 0.38 & 0.15 \\
    \hline  
\end{tabular} 
}
\label{pp_cohort_pairs}
\end{table}
The PPP displayed in Table \ref{pp_cohort_pairs} are based on the correlation of cohort effects across 10-year age windows for each risk factor and sex. Values near 0.5 indicate that the model’s correlation structure for the cohort effects is well-calibrated with the observed data, meaning it accurately captures the temporal dependencies within each cohort. For most risk factors, including SBP, DBP, BMI, and TOTCHL, the PPP hover around $0.5$ across various cohort pairs and genders, suggesting strong model fit. In addition, some risk factors and cohort pairs have lower or higher PPP, highlighting specific areas where the model either slightly underestimates or overestimates correlation; however none of these probabilities are extreme. Overall, the PPP demonstrate good model fit in terms of cohort correlations across age windows, providing confidence in the model to represent the underlying longitudinal dynamics of these risk factors.
To further evaluate the model’s performance and its ability to recover cohort-level correlation structures, we conducted a simulation study based on the proposed model; additional details are provided in Supplementary Materials~\ref{supp:Simulation_Study}.
\subsection{Imputing Risk Factors at Younger Ages}
To evaluate the model’s predictive ability for risk factors outside a cohort’s age range, we excluded all ages below 40 from the FOS cohort and generated imputations from the posterior predictive distribution for these excluded ages. Because the FOS cohort covers the entire adult lifespan, it provides a strong reference point for validating imputed values in these unobserved age intervals. In Supplementary Materials~\ref{supp:Risk_Factor}, we include scatter plots of observed and imputed risk factors versus age for both men and women, allowing for a detailed examination of the imputed results across each risk factor by age. The trends confirm that the model can reasonably extrapolate risk factor values beyond observed age ranges, with some variability observed in younger ages. These results support the model’s robustness for extending predictions across unobserved age windows within cohort data. We have not included the QQ plots comparing observed and imputed residuals for ages under 40; however, these also indicate satisfactory model performance.
\subsection{Imputing Deleted Risk Factors}
To evaluate the model’s ability to retain associations between risk factors, we deleted all DBP values from the FOS cohort and generated samples from the posterior predictive distribution. This approach assesses how effectively the model can predict missing risk factors by imputing them based on observed data from \emph{other} risk factors.

We calculated posterior predictive probabilities to estimate the proportion of total iterations for which the posterior draw equals or exceeds the posterior mean calculated using the complete dataset. This probability reflects the model’s ability to preserve the relationships between DBP and other risk factors. Posterior probabilities close to 0.5 indicate that the model accurately maintains these associations, while significant deviations may suggest areas
where the model could be improved.

In Table~\ref{DBPdropped}, we summarize the posterior probabilities for the DBP fixed-effect 
coefficients at the intercept, linear age slope, and spline-increment levels, with results reported 
separately for men and women.  
Each entry represents the posterior probability that the 
corresponding coefficient differs from zero.
Posterior probabilities near 0.5 across most coefficients indicate that the model preserves the 
expected distribution of DBP fixed effects following imputation. The stability of the intercept 
and age-slope estimates across subgroups further suggests that the model maintains key 
relationships present in the observed data.
\begin{table}[!t]
	\caption{Comparing posterior probabilities of diastolic blood pressure fixed effects at the intercept, slope, and age window levels by sex. Birth year categories are defined as B1: $<$1915, B2: 1915–1929, B3: 1929–1945, and B4: $>$1945. The cohort reference is ARIC. } \label{DBPdropped}
	\centering  
\begin{tabular}{p{5cm}cc}
\hline
Coefficient  & Men & Women \\
\hline
\multicolumn{3}{l}{\textit{Intercept level} (\(p = 0\))} \\
\quad Intercept                           & 0.79 & 0.80 \\
\quad Race (Black)                        & 0.55 & 0.62 \\
\quad Edu: HS vs.\ -HS                     & 0.53 & 0.46 \\
\quad Edu: +HS vs.\ -HS                    & 0.34 & 0.54 \\
\quad Birth year: B2 vs.\ B1                       & 0.32 & 0.53 \\
\quad Birth year: B3 vs.\ B1                       & 0.06 & 0.11 \\
\quad Birth year: B4 vs.\ B1                       & 0.12 & 0.07 \\
\quad CARDIA                              & 0.51 & 0.72 \\
\quad CHS                                 & 0.45 & 0.46 \\
\quad MESA                                & 0.51 & 0.60 \\
\quad FHS                                 & 0.35 & 0.35 \\
\quad FOS                                 & 0.50 & 0.66 \\
\quad JHS                                 & 0.55 & 0.65 \\
[4pt]
\multicolumn{3}{l}{\textit{Slope level} (\(p = 1\))} \\
\quad Race (Black)                        & 0.51 & 0.51 \\
\quad Edu: HS vs.\ -HS                     & 0.57 & 0.58 \\
\quad Edu: +HS vs.\ -HS                    & 0.59 & 0.56 \\
\quad Birth year: B2 vs.\ B1                       & 0.44 & 0.49 \\
\quad Birth year: B3 vs.\ B1                       & 0.81 & 0.82 \\
\quad Birth year: B4 vs.\ B1                       & 0.80 & 0.90 \\
[4pt]
\multicolumn{3}{l}{\textit{Age window} (\(p \ge 2\))} \\
\quad \(s_1 = 18\)           & 0.35 & 0.43 \\
\quad \(s_2 = 28\)           & 0.55 & 0.59 \\
\quad \(s_3 = 38\)           & 0.55 & 0.57 \\
\quad \(s_4 = 48\)           & 0.51 & 0.44 \\
\quad \(s_5 = 58\)           & 0.49 & 0.53 \\
\quad \(s_6 = 68\)           & 0.49 & 0.55 \\
\quad \(s_7 = 78\)           & 0.48 & 0.50 \\
\hline
\end{tabular}
\end{table}
\subsection{Analysis of Imputed Risk Factors}
One of the motivations for developing our model was to use it to impute risk factors at ages not included in a cohort. To this end, we used data from the CARDIA cohort, which includes individuals aged 17 to 64. We deleted observations at ages below 40 and imputed TOTCHL values for these ages. A total of 128 imputed datasets were generated, each containing imputed TOTCHL values for observations under 40 while retaining the original values for those aged 40 and older.

To assess the impact of imputation on model performance, we jointly modeled TOTCHL as a longitudinal risk factor and time-to-CVD death using the \texttt{JMbayes2} package \citep{JMbayes2}. A key metric of interest was the coefficient of the area under the curve (AUC) feature for TOTCHL in the survival model.
We use the AUC of TOTCHL as a cumulative exposure measure, reflecting long-term risk. This choice is supported by prior studies showing that cumulative cholesterol burden is a stronger predictor of cardiovascular disease than single or proximal measures (\cite{domanski2020ldl}, \cite{domanski2023association}). Furthermore, a cumulative feature such as the AUC is generally more sensitive to misspecification of the imputation model compared with a current-value feature. While the AUC assumes equal weighting across ages, this simplification provides a consistent and interpretable summary. 

In applied settings, more flexible summaries, such as age-varying weights, could reflect life-stage–specific effects. 
The hazard model is given by
\begin{equation}
h_i(a) = h_0(a) \exp\left( \boldsymbol{X}_i^\top \boldsymbol{\gamma} + \eta \, \text{AUC}(\text{TOTCHL}_i(a)) \right),
 \notag
\end{equation}
where $h_0(a)$ is the baseline hazard, $\boldsymbol{\gamma}$ describe the impact of baseline covariates $\boldsymbol{X}_i$ on the risk of CVD death over time, and $\eta$ determines the association between the cumulative exposure to TOTCHL and the hazard of CVD death.

For the imputed datasets, we computed the mean and variance of the 128 AUC coefficients and combined them using Rubin’s rules and compared this value to that derived from the observed data. The post-imputation AUC coefficients were 0.41 (SE = 1.01) for men and 0.34 (SE = 1.77) for women, while the AUC coefficients from the real dataset were 0.42 (SD = 0.21) for men and 0.35 (SD = 0.38) for women. The close agreement between these values supports the robustness of the imputation process and the reliability of the estimated TOTCHL trajectories.
As expected, the standard deviation of the AUC coefficients in the imputed datasets was larger than in the observed data, reflecting the additional variability introduced through imputation. This is not surprising given the large amount of missing data. However, the stability of the mean AUC coefficient suggests that the imputed TOTCHL trajectories provided reasonable and precise predictions across the extended age range.

To further evaluate the robustness, we conducted the same analysis using the FOS cohort. Similarly, the AUC coefficients for the imputed and observed datasets in FOS were closely aligned, further supporting the reliability of the imputation process across different cohorts. A detailed summary of the FOS results is provided in Supplementary Materials~\ref{supp:Imputation_Results}.
\section{Discussion} \label{sec6}
We developed a complex hierarchical model to combine data from seven longitudinal cohort studies, to enhance our understanding of CVD risk factor trajectories across the life course and their association with CVD in a diverse sample of the United States population. The model leverages information from all risk factors to improve the precision of individual risk factor trajectories and borrows strength across cohorts in a data-driven manner. 

Preliminary analyses using single-cohort models often fail to detect important age-related patterns, such as the midlife rise in glucose or the late-life decline in HDL, due to limited age coverage and reduced sample sizes. In contrast, our integrated framework allows for more stable and complete estimation across the life course, emphasizing the advantages of pooling information across studies.
This approach is crucial since only a few cohorts in the study cover the entire adult lifespan.
We addressed computational challenges inherent in analyzing a multivariate longitudinal dataset, utilizing advanced methods to overcome these issues.

Death is not explicitly modeled in our framework, as the focus is on characterizing longitudinal risk factor trajectories. However, risk factor measurements are only available up to the time of the last visit for each participant, and the model uses all data collected prior to that point. Future work will extend this framework to jointly model longitudinal trajectories and time-to-event outcomes. Extensive model validation confirmed that the model fits the data well. We also evaluated the model's accuracy in predicting risk factors outside a cohort’s observed age range by deleting and imputing risk factors. Results demonstrated that the model accurately preserves critical relationships over time and across risk factors.

In broader applications, harmonization of risk factor definitions and measurement protocols may be necessary across cohorts. Here, we focused on seven risk factors that were consistently measured across participating cohorts and over time \citep{wilkins2015data}. In other applications, variable harmonization may be a critical step when integrating additional or more heterogeneous measures.

While the model offers flexibility and strong performance across diverse cohorts, it relies on several structural assumptions. These include the use of piecewise linear functions to model age-dependent trends, cohort-specific random effects for all age windows, and individual-specific random intercepts and slopes. We highlight that limiting individual deviations to an overall slope and intercept helps preserve identifiability in high-dimensional settings. Future extensions could consider more flexible modeling of individual-level trajectories and age-varying cohort covariances.

Selection into cohorts is not explicitly modeled. However, we mitigate differential selection by including baseline covariates and cohort-specific random effects that vary by age window, 
which absorb structural differences due to selection mechanisms such as geography, recruitment protocols, or demographic composition. While explicit modeling of selection remains an open area for development, our hierarchical framework helps reduce related biases.

Beyond its primary use for imputing trajectories, our model can support downstream analyses of clinical outcomes. Features derived from trajectories, such as cumulative exposure, age-specific summaries, or deviations from reference patterns, can be valuable inputs to assess life-course associations with health outcomes. The framework is flexible and can be applied to individual cohorts or pooled across multiple cohorts, depending on the research objective.

Although the present study focuses on CVD risk factors in the LRPP, the modeling framework is broadly applicable to longitudinal datasets with repeated measures and overlapping age ranges. Core components such as the piecewise linear age structure, hierarchical random effects, and multivariate sharing of information can be adapted to alternative contexts. Key aspects that may require modification include the number and placement of age windows, the choice of risk factors or outcomes, and the specification of prior distributions.

The structure of overlap across cohorts and the frequency of repeated measures per subject are critical for ensuring stable estimation and interpretability. At the cohort level, we recommend that at least three cohorts contribute data to each age interval to reduce reliance on extrapolation and enable reliable borrowing across studies. At the subject level, individuals should have at least three repeated measurements, ideally spanning different age windows; however, our model does not impose this as a requirement and remains well-defined even when subjects contribute fewer observations. To prevent over-reliance on between-subject variation, an average of 3--4 measurements per subject, distributed across age, supports robust estimation of within-subject change \citep{fitzmaurice2012applied}.

Future directions for this work include:
i) joint modeling of CVD risk factors, medication use, and time-to-events;
ii) informing treatment strategies by identifying clinically relevant features of longitudinal risk factor trajectories associated with CVD outcomes; and
iii) leveraging this work to facilitate the dissemination and use of synthetic LRPP data by the broader research community.
\section*{Acknowledgements}
This work was supported by NIH/NHLBI 1R01HL158963 (MPI: Siddique/Daniels).
\section*{Supplementary Materials}
Supplementary Materials S1--S7, Tables, and Figures referenced in Sections 2--5 are provided with this paper. These materials include additional details on cohort characteristics, design matrix structure and identifiability, covariance specification, model parameters and random effects, Bayesian estimation, model validation, and simulation studies. The Stan code used to fit the model is available from the authors upon request.
\section*{Data Availability}
The third-party Lifetime Risk Pooling Project (LRPP) data used in this study are available through the Biologic Specimen and Data Repository Information Coordinating Center (BioLINCC) (\url{https://biolincc.nhlbi.nih.gov}) and cannot be provided by the authors.
\bibliographystyle{chicago}
\bibliography{Reference}
\section*{Supplementary Materials} \label{supp:Supplementary_Materials}
Supplementary Materials S1--S7, Tables, and Figures referenced in Sections 2--5 of the main manuscript are provided below. These materials provide additional details on cohort characteristics, design matrix structure and identifiability, covariance specification, model parameters and random effects, Bayesian estimation, model validation, and simulation studies.
For reproducibility, the Stan code used to fit the proposed longitudinal risk factor model is available from the authors upon request.
\setcounter{section}{0}
\renewcommand{\thesection}{S.\arabic{section}}
\setcounter{figure}{0}
\renewcommand{\thefigure}{S\arabic{figure}}
\renewcommand{\thetable}{S\arabic{table}}
\setcounter{table}{0}

\section{Additional Descriptive Characteristics of LRPP Cohorts} \label{supp:Additional_Descriptive}
\subsection{Descriptive Plots} \label{supp:Descriptive_Plots}
\begin{figure}[!ht]
	\centering
	\includegraphics[width=15cm,height=7cm]{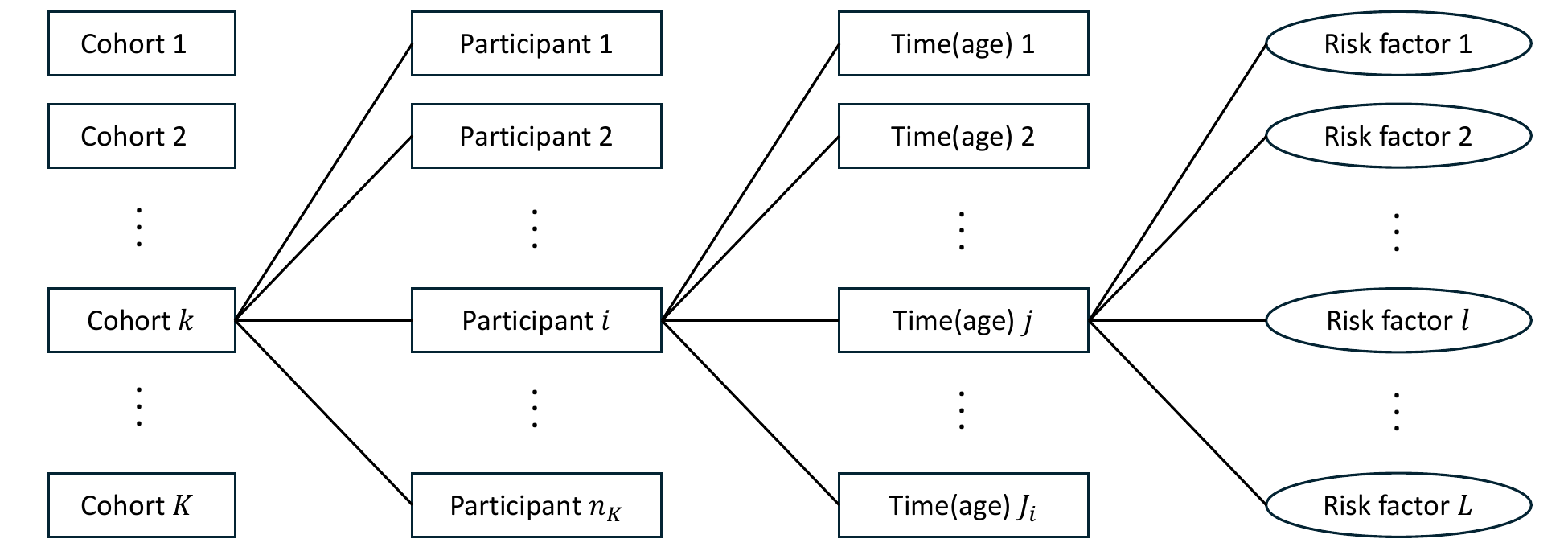}
	\caption{Hierarchical structure of the LRPP. Multiple risk factors at different age follow-ups are measured within participants who are nested within cohorts}
	\label{Schematic} \label{suppfig:Hierarchical_structure}
\end{figure}

  \begin{figure}[!ht]
		\centering
		\includegraphics[width=14cm,height=9cm]{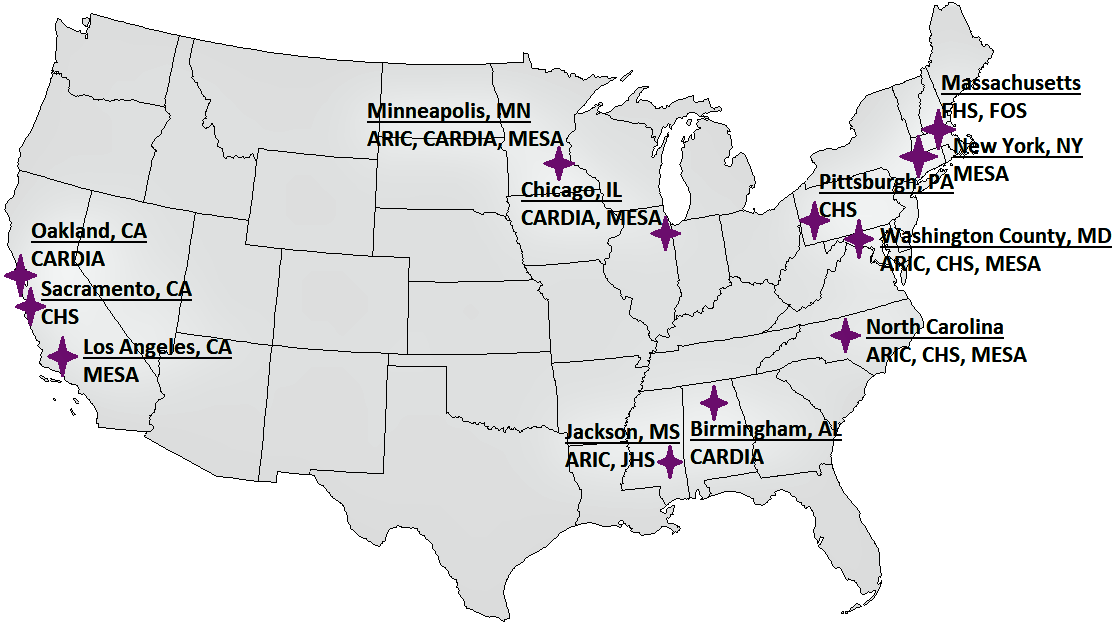}
		\caption{Geographical locations of cohorts included in the data } 
		\label{Map}
	\end{figure}

\begin{figure}[!t]
    \centering
\subfloat[ARIC]{{\includegraphics[width=0.49\textwidth,height=0.2\textheight]{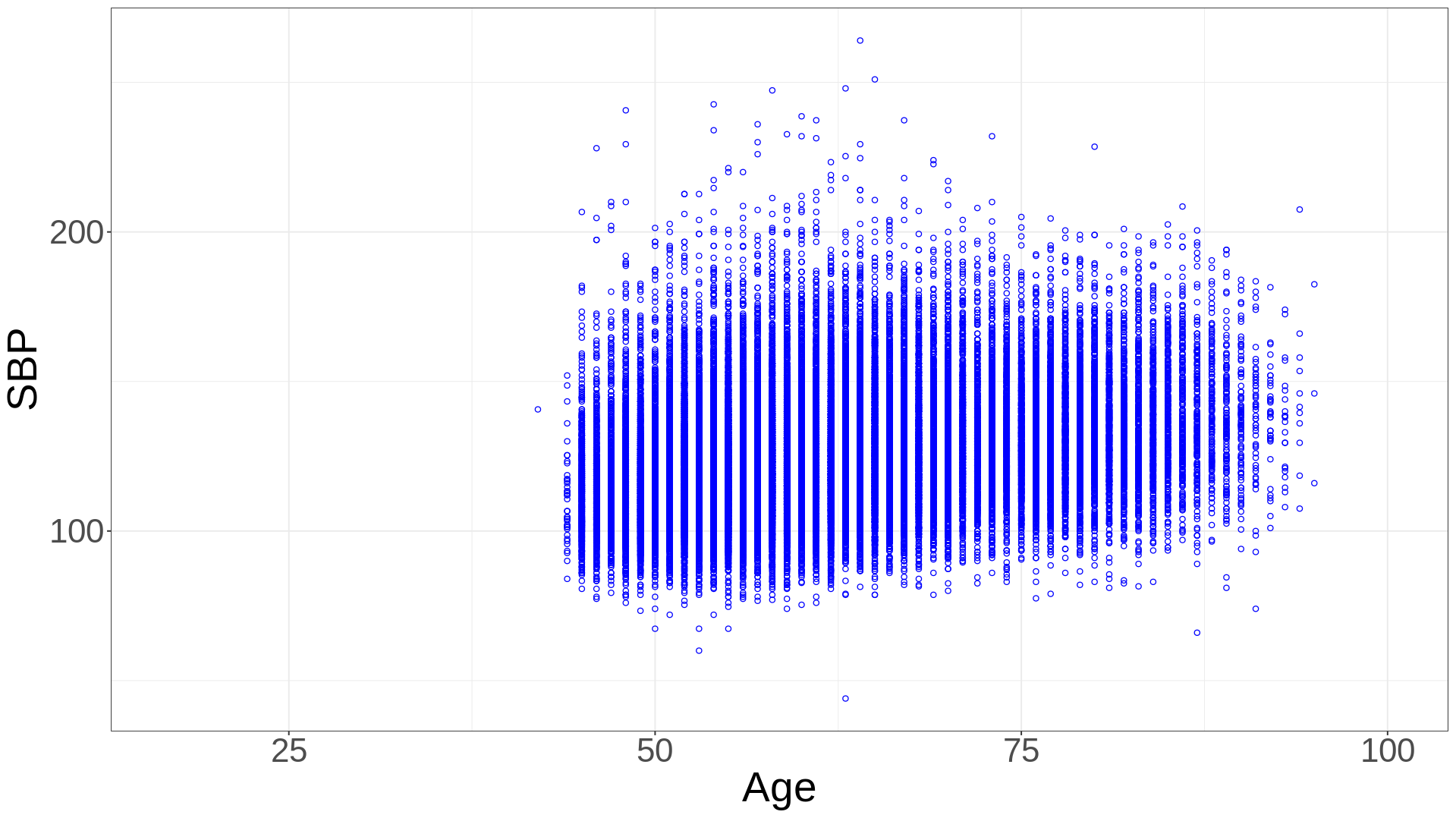}}}
\subfloat[CA]{{\includegraphics[width=0.49\textwidth,height=0.2\textheight]{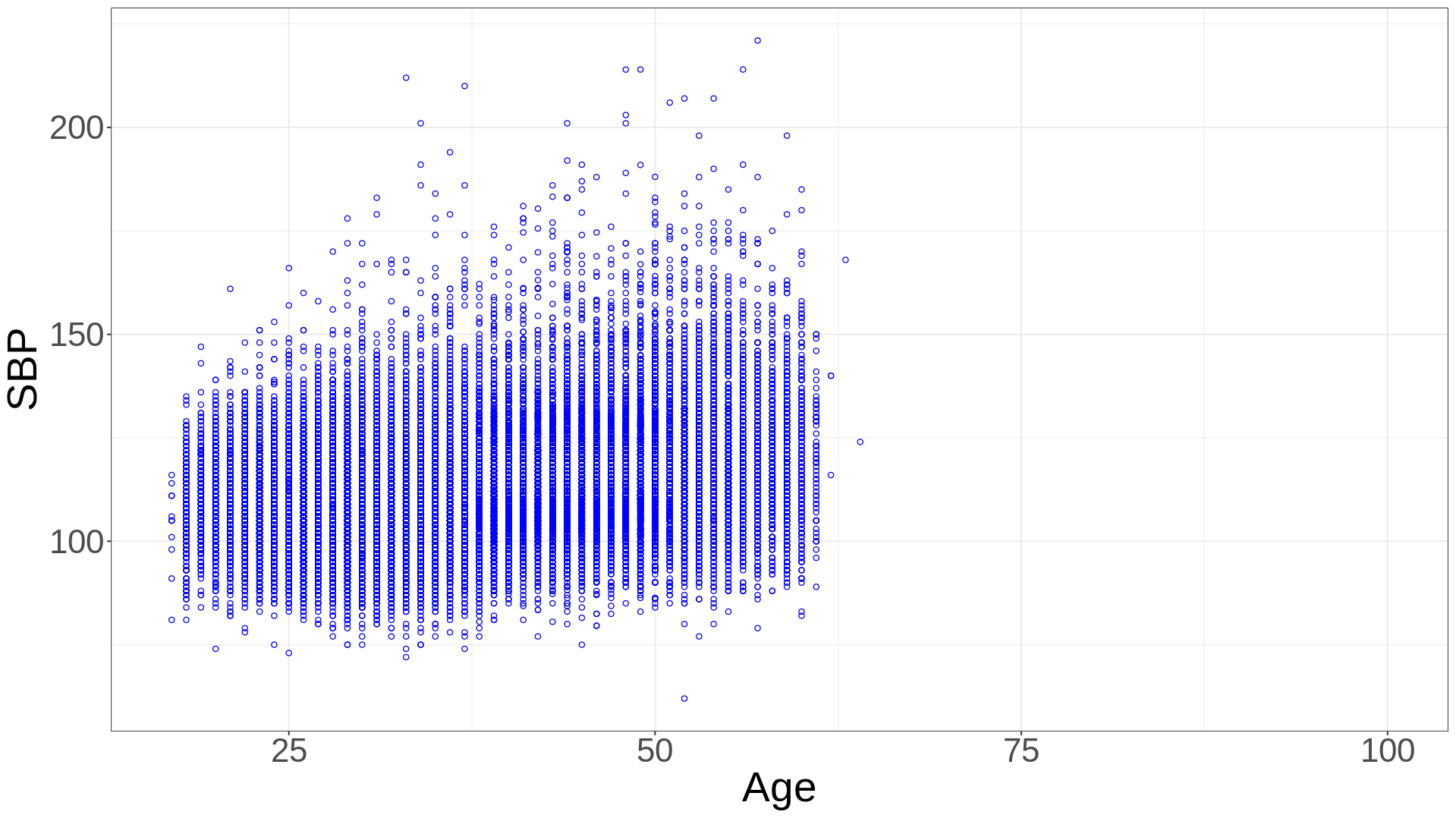}}} 

\subfloat[CHS]{{\includegraphics[width=0.49\textwidth,height=0.2\textheight]{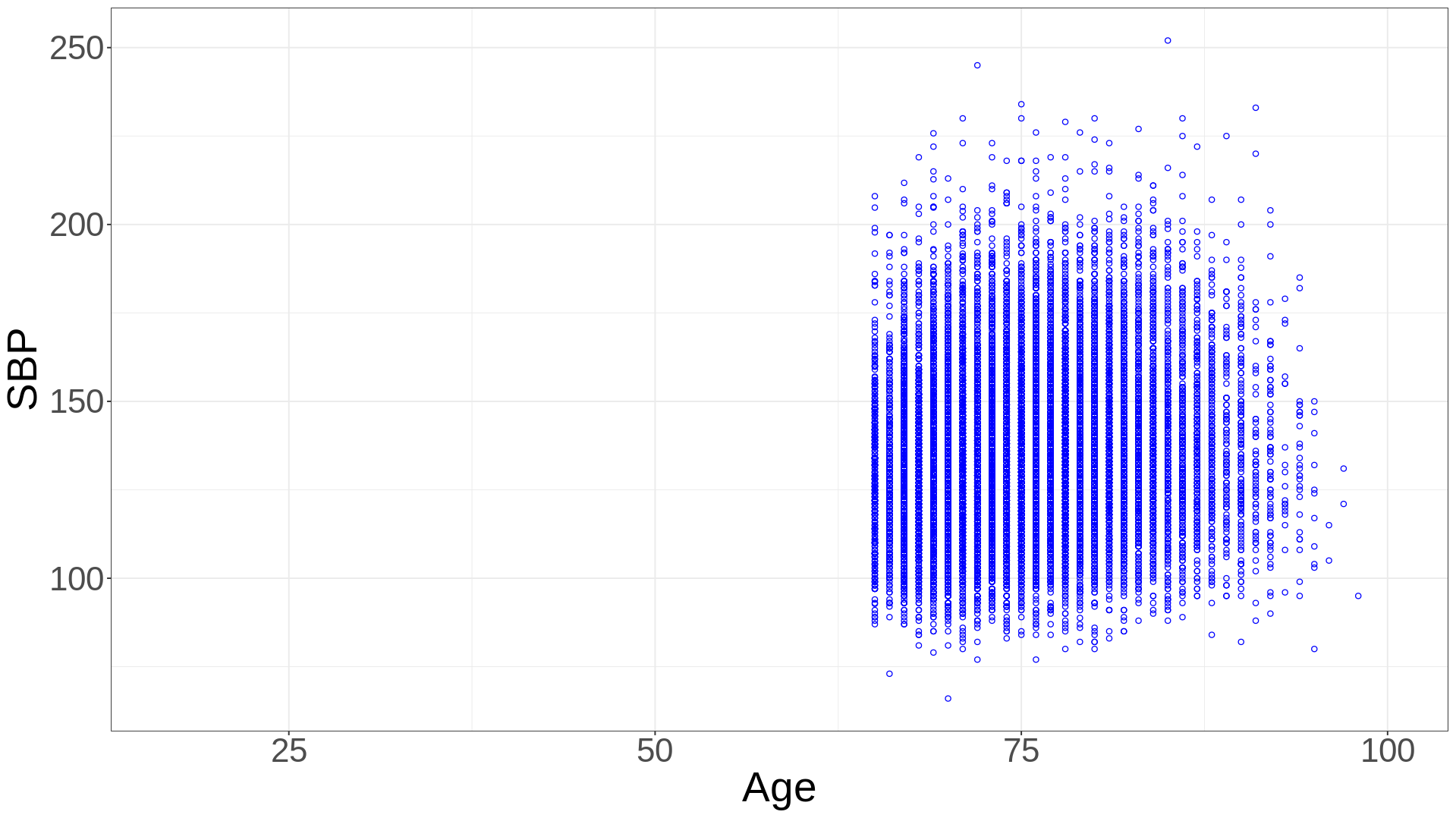}}}
\subfloat[MESA]{{\includegraphics[width=0.49\textwidth,height=0.2\textheight]{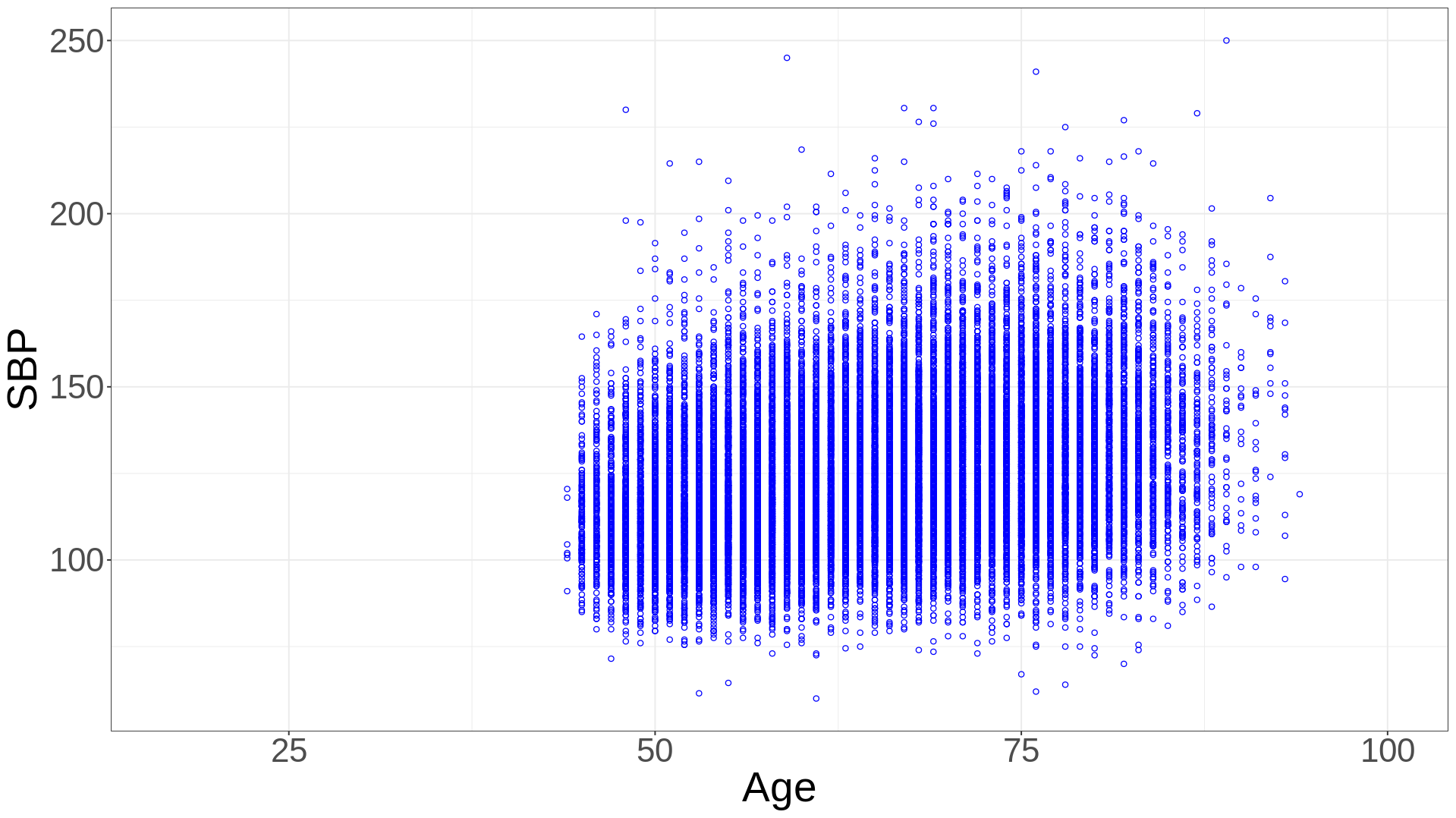}}} 

\subfloat[FHS]{{\includegraphics[width=0.49\textwidth,height=0.2\textheight]{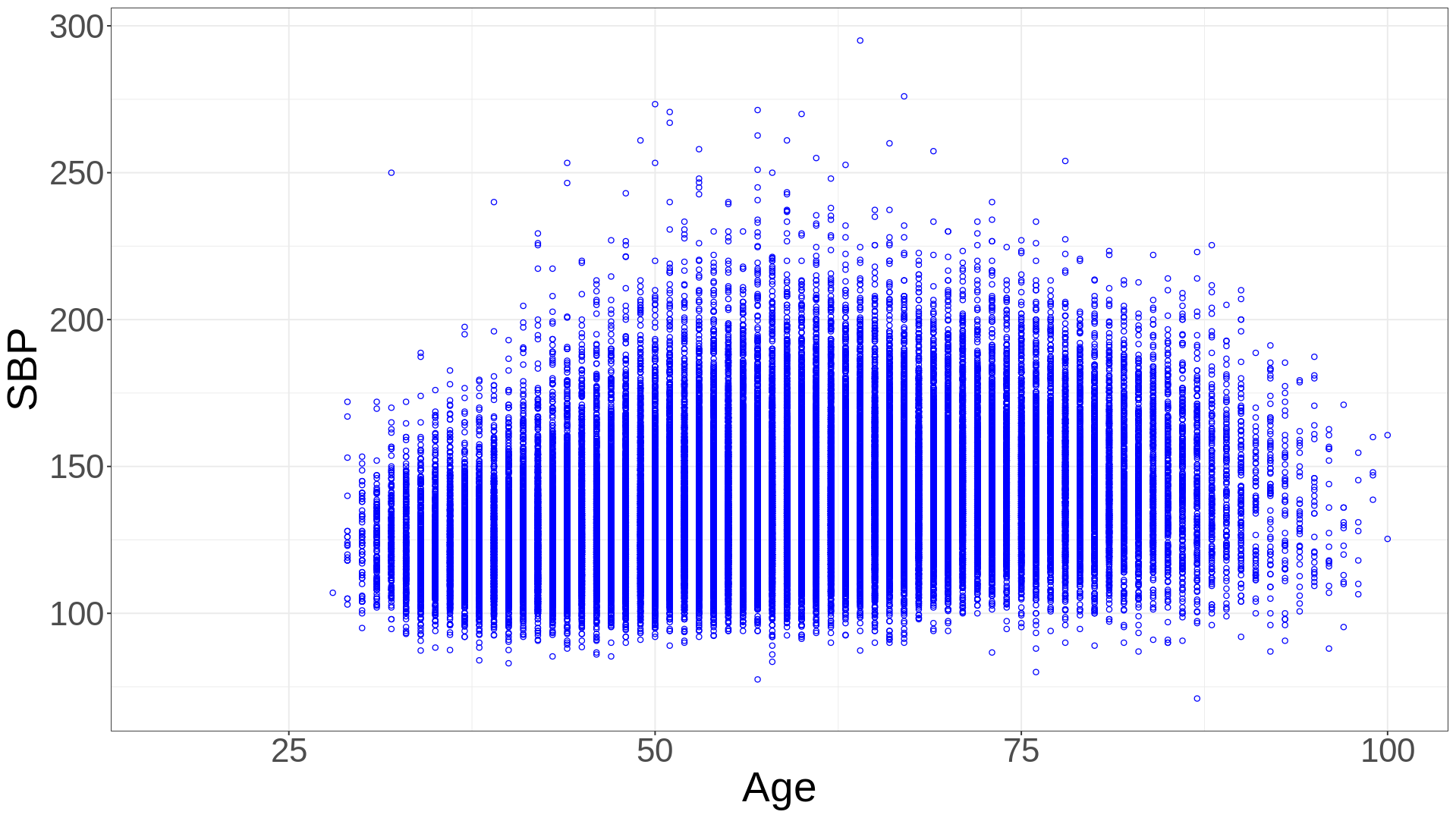}}}
\subfloat[FOS]{ {\includegraphics[width=0.49\textwidth,height=0.2\textheight]{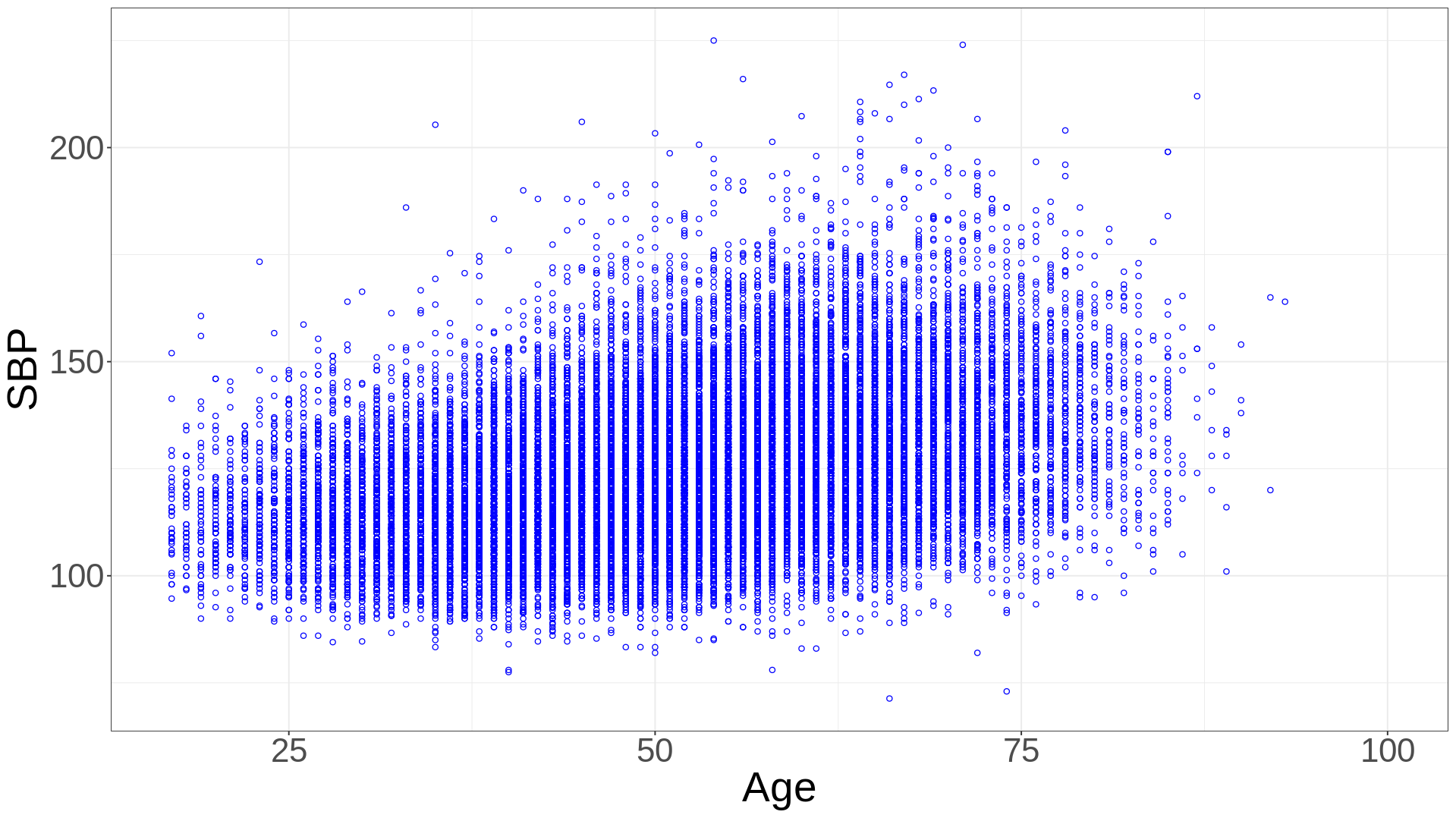}}} 

 \subfloat[JHS]{{\includegraphics[width=0.49\textwidth,height=0.2\textheight]{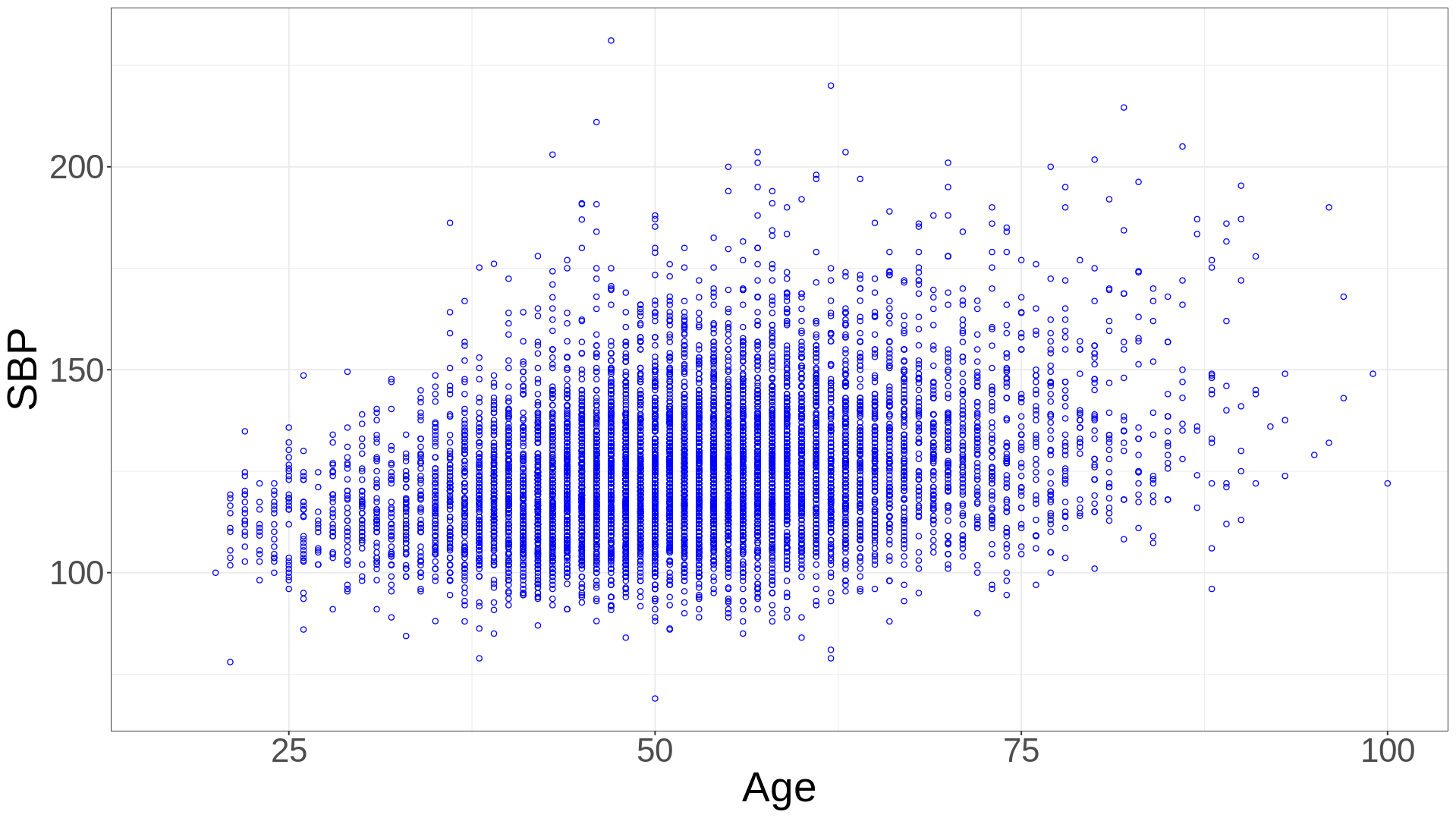}}}
    \caption{Scatter plots of systolic blood pressure against age across cohorts}
    \label{cohortsSBP}
\end{figure}
\subsection{Birth Year Effects in the LRPP} \label{supp:Birth_Year}
Over recent decades, there have been strong secular trends in the prevalence of CVD risk factors. For instance, rates of severe hypertension and high cholesterol have declined over time (notably over calendar years, rather than age). Until the 2000s, risk factor effects on events were largely time-constant, despite overall decreasing rates. 

Distinct cohort characteristics also shape these trends. For example, participants in the Framingham Study, which began enrollment in 1948, were exposed to substantially more cigarette smoke than those in later cohorts. Consequently, birth year effects and period adjustments are essential in modeling longitudinal risk factors accurately. Specifically, we account for two effects: i) age effects (e.g., two individuals born in the same year but measured at different ages), 
ii) birth year effects (e.g., two individuals measured at the same age but born in different years).

\begin{figure}[!t] 
\centering
\includegraphics[width=14cm,height=8cm]{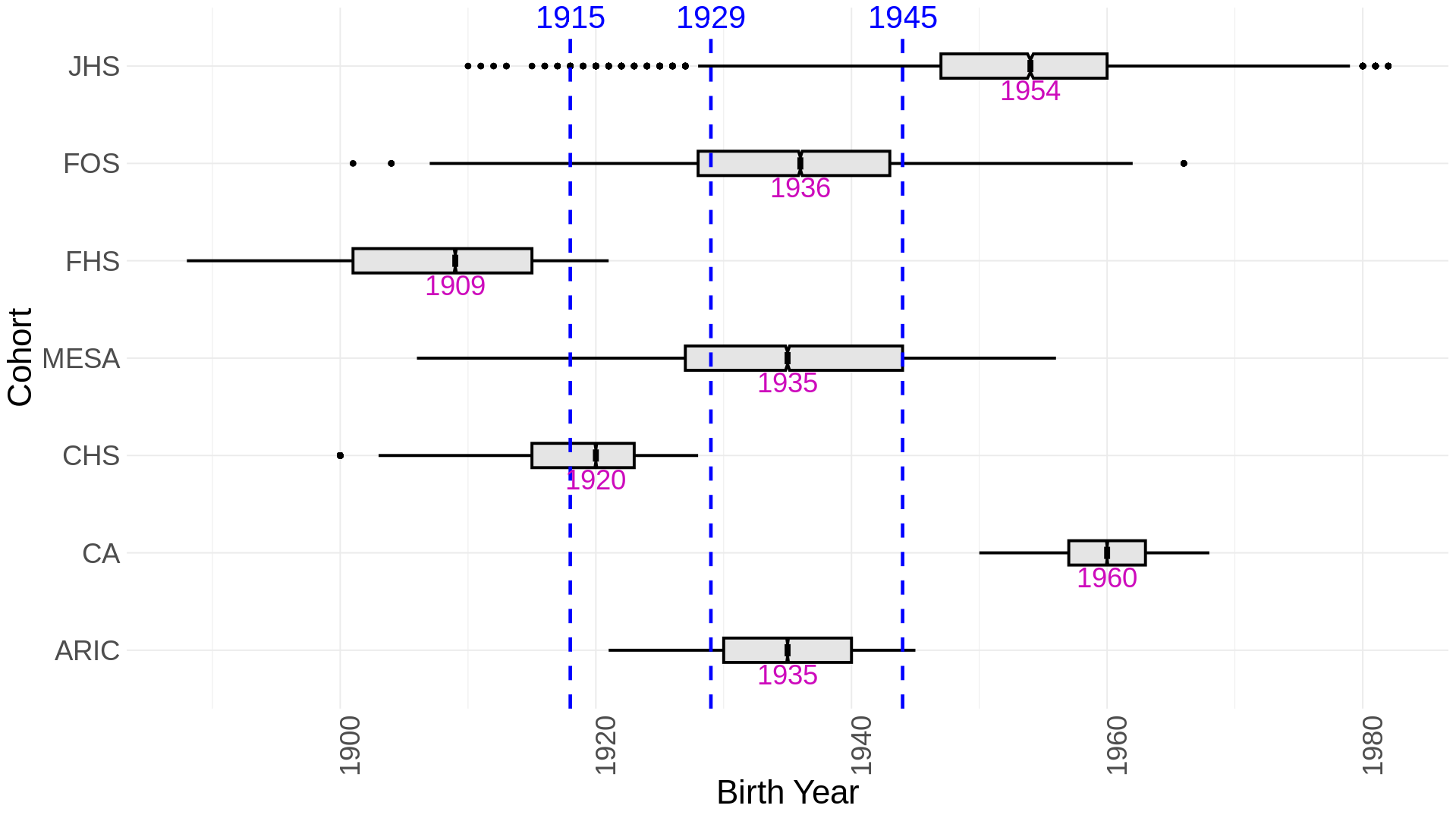}
\caption{Boxplots displaying the distribution of
birth year in each Lifetime Risk Pooling
Project cohort. Vertical dashed lines indicate
quantiles across all cohorts.} 
\label{BY}
\end{figure}

Our preliminary analyses confirm that adjustments should occur at the individual participant level rather than at the cohort level, as some cohorts (such as MESA, FOS, and JHS) cover a broad range of birth years. \cite{berry2012lifetime} stratified participants by birth year (e.g., before 1920) but did not include younger cohorts such as CARDIA and JHS, which are part of our study. Figure~\ref{BY} presents the birth year distributions for different cohorts included in the LRPP. We stratify participants into four categories based on quartiles of birth year: before 1915, 1915–1929, 1929–1945, and after 1945. These intervals align approximately with significant historical events, such as World War I (1918), the Great Depression (1929), the end of World War II, and the onset of the Baby Boomer generation (1946).
In our model, we incorporate the main effects of birth year categories and their interactions with age. This approach allows us to capture the cohort-level impact of birth year independently, allowing these effects vary across different age intervals.
\subsection{Missing and Intermittent Risk Factors Across Cohort Age Ranges} \label{supp:Missing_and}
In addition to risk factors at ages not covered by each cohort study, the LRPP data includes some risk factors that are unobserved at certain exams. Table~\ref{Missing} presents the percentage of unobserved values across risk factors, cohorts, and sex in the LRPP. Consequently, some risk factors for the $i$th individual at the $j$th exam were not recorded. The largest cohorts, ARIC and MESA, have minimal missing values (less than $2.6\%$). In contrast, CHS and FHS show the highest percentages of missingness, with approximately $77\%-86\%$ unobserved HDL and TRIG measurements, as these were not measured annually in these cohorts. Furthermore, lipid levels are not routinely measured in FHS and are missing at the baseline exam. TRIG testing only began at exam seven, with HDL/LDL cholesterol introduced even later (exam 9). Similarly, CHS has HDL/LDL cholesterol and triglyceride measurements in only 2–3 exams. Notably, FOS and JHS, which cover the full adult lifespan, show a low percentage of missing values. 
We also excluded a small number of observations where GLU levels were below 50 or above 300 (approximately 0.3\%) and where TRIG levels were below 50 or above 500 (approximately 0.2\%). Further details regarding outlying TRIG levels can be found at \cite{NHLBI_triglycerides}. 
\begin{table}[!t]
	\caption{Percentage of unobserved risk factors among participants who attended exams across seven cohorts} \label{Missing}
	\centering          
\begin{tabular}{ll|ccccccc}	\hline
Cohort & Sex & SBP & DBP & BMI & TOTCHL & GLU & HDL & TRIG \\
 \hline
 ARIC &  & & & & & & & \\
     & MEN & 0.11 & 0.11 & 0.48 & 0.90 & 5.10 & 0.94 & 8.40  \\
    & WOMEN & 0.12 & 0.12 & 0.53 & 1.46 & 5.58 & 1.53 & 7.40 \\
 CARDIA  &  & & & & & & & \\
    & MEN & 0.06 & 0.07 & 0.50 & 1.13 & 30.4 & 1.13 & 41.6  \\
    & WOMEN & 0.13 & 0.14 & 2.02 & 2.10 & 30.2 & 2.10 & 47.3  \\
   CHS   &  & & & & & & & \\
  & MEN  & 15.6 & 15.7 & 69.6 & 69.8 & 55.6 & 77.0 & 77.4 \\
    & WOMEN & 18.1 & 18.3 & 71.1 & 71.5 & 57.4 & 78.8 & 79.0   \\
  MESA   &  & & & & & & & \\
   & MEN  & 0.57 & 0.57 & 0.57 & 1.14 & 1.53 & 1.15 & 7.18 \\
    & WOMEN & 0.65 & 0.65 & 0.62 & 1.59 & 1.96 & 1.63 & 7.46  \\
FHS   &  & & & & & & & \\
   & MEN & 0.06 & 0.06 & 4.16 & 21.5 & 30.4 & 82.3 & 93.5 \\
  & WOMEN & 0.15 & 0.16 & 6.22 & 26.9 & 32.6 & 80.3 & 93.9  \\
 FOS  &  & & & & & & &  \\
   & MEN  & 0.02 & 0.03 & 1.25 & 1.15 & 13.01 & 1.36 & 21.5 \\
    & WOMEN & 0.04 & 0.05 & 1.81 & 3.23 & 15.2 & 3.50 & 28.7 \\
   JHS  &  & & & & & & &  \\
   & MEN  & 0.06 & 0.06 & 0.59 & 13.9 & 14.4 & 13.9 & 23.1 \\
    & WOMEN & 0.33 & 0.33 & 1.30 & 13.8 & 14.1 & 13.8 & 27.2  \\
    \hline
\end{tabular} 
\end{table}
\section{Design Matrix Structure and Identifiability} \label{supp:Design_Matrix}

This section describes the structure of the fixed‐effect design matrix in the piecewise linear spline model introduced in Section~3 and clarifies how identifiability is ensured through model construction. 
Recall that the mean trajectory for risk factor $\ell$ and participant $i$ nested within cohort $k$ is modeled as
\[
\mu_{\ell k(i)}(a_{ij}) 
= \beta_{\ell k(i)}^{(0)} 
+ \beta_{\ell k(i)}^{(1)} a_{ij} 
+ \sum_{p=1}^{P} \beta_{\ell k(i)}^{(p+1)} (a_{ij} - s_p)_{+},
\]
where $\{s_1, \ldots, s_P\}$ are pre‐specified knot locations along the age axis, and $(a - s_p)_{+} = \max\{0, a - s_p\}$ denotes the truncated linear spline basis.  
This formulation yields a continuous piecewise linear trajectory, where $\beta_{\ell k(i)}^{(0)}$ is the intercept, $\beta_{\ell k(i)}^{(1)}$ is the overall age slope, and $\beta_{\ell k(i)}^{(p)}$ for $p \ge 2$ represents incremental changes in slope beyond knot $s_p$.  

The truncated spline basis functions $(a - s_p)_{+}$ ensure identifiability of the fixed effects without requiring explicit constraints. Because each basis function is zero below its knot, the intercept and first slope are uniquely defined, and subsequent coefficients capture deviations in slope across adjacent segments. This construction yields a full-rank design matrix, and identifiability follows directly from the properties of truncated power spline bases \citep{ruppert2003semiparametric, hastie2009elements, wood2017generalized}. Additional regularization of the spline coefficients is introduced through the hierarchical prior specification described in Section~\ref{sec3} of the manuscript and Supplementary Materials~\ref{supp:Full_Prior}.
The fixed‐effect component of the model is expressed as
\[
h_{\ell }^{(p)}(\boldsymbol{X}_{i}) = \boldsymbol{X}_{i}^{T}\boldsymbol{\alpha}_{\ell }^{(p)}.
\]
Specifically, for $p=0$, $h_{\ell }^{(0)}(\boldsymbol{X}_{i})$ includes the overall intercept, and main effects of baseline covariates such as race/ethnicity, education, birth year, and cohort indicators,
\begin{eqnarray}
h^{(0)}_\ell(\boldsymbol{X}_{i}) &&= \alpha^{(0)}_{\ell 0}
+ \alpha^{(0)}_{\ell 1}\,\mathrm{RaceB}_i
+ \alpha^{(0)}_{\ell 2}\,\mathrm{Edu2vs1}_i
+ \alpha^{(0)}_{\ell 3}\,\mathrm{Edu3vs1}_i \notag \\
&& ~ + \alpha^{(0)}_{\ell 4}\,\mathrm{BY2vs1}_i
+ \alpha^{(0)}_{\ell 5}\,\mathrm{BY3vs1}_i
+ \alpha^{(0)}_{\ell 6}\,\mathrm{BY4vs1}_i
+ \sum_{c\neq \mathrm{ARIC}} \alpha^{(0)}_{\ell,c}\,\mathrm{Cohort}_c. \notag
\end{eqnarray}
For $p=1$, $h_{\ell }^{(1)}(\boldsymbol{X}_{i})$ includes interactions of the global age slope with these covariates, allowing the average rate of change in each risk factor to differ systematically across baseline characteristics,
\begin{eqnarray}
h^{(1)}_\ell(\boldsymbol{X}_{i}) &=&
\alpha^{(1)}_{\ell 1}\,\mathrm{RaceB}_i
+ \alpha^{(1)}_{\ell 2}\,\mathrm{Edu2vs1}_i
+ \alpha^{(1)}_{\ell 3}\,\mathrm{Edu3vs1}_i
\notag \\
&& ~
+ \alpha^{(1)}_{\ell 4}\,\mathrm{BY2vs1}_i
+ \alpha^{(1)}_{\ell 5}\,\mathrm{BY3vs1}_i
+ \alpha^{(1)}_{\ell 6}\,\mathrm{BY4vs1}_i. \notag
\end{eqnarray}
For $p \ge 2$, the fixed‐effect component represents the spline‐based slope increments that capture additional age‐specific deviations from the global trend, 
\[
h_{\ell }^{(p)}(\boldsymbol{X}_{i}) = \alpha^{(p)}_{\ell},
\qquad p \ge 2.
\]

This parameterization provides a clear interpretation of covariate effects across the life course.  
Cohort indicators and main effects determine baseline levels, covariate–age interactions govern global trends, and spline increments describe localized age‐specific deviations.  
Together, these components allow the model to capture both long‐term and age‐localized effects of cohort membership and individual characteristics on each risk factor trajectory, while maintaining parsimony and interpretability.

For modeling purposes, each continuous risk factor was scaled by its standard deviation to place coefficients on a comparable scale and to facilitate prior specification. Age enters the model as a continuous time-varying covariate; for computational stability and interpretability, age is centered at 58, the mean age across the seven cohorts, and scaled by 10, so that one unit represents a 10-year difference. The spline knots defining the piecewise linear age structure were likewise centered and scaled using this transformation to maintain consistency between the age variable and the spline basis. All posterior summaries and predicted trajectories are reported on the original measurement scales.
\section{Properties of the Covariance Structure} \label{supp:Properties_of}
In this section, we derive the variances and covariances of $Y_{\ell k(i)}$, capturing variability and dependencies across individuals, risk factors, age intervals, and cohorts while preserving model parsimony.

\subsection{Variances, Covariances, and Correlations} \label{supp:Variances_Covariances}
For individual \(i\) nested within cohort \(k\), and for \(L\) risk factors measured over \(P\) age intervals, let \(\boldsymbol{S}(a_{ij}) = \big((a_{ij}-s_{1})_{+}, \ldots, (a_{ij}-s_{P})_{+}\big)^{T}\) denote the spline subvector of \(\boldsymbol{A}(a_{ij})\), and let \(\boldsymbol{b}_i\) denote the \(2 \times L\) matrix of subject-specific random effects defined in Section 3.1 of the manuscript, with
\(
\mathrm{vec}(\boldsymbol{b}_i)
=
\big(b^{(0)}_{i1}, b^{(1)}_{i1}, \ldots, b^{(0)}_{iL}, b^{(1)}_{iL}\big)^T.
\)
We define
\(
\boldsymbol{\Sigma} = \mathrm{Cov}\big(\mathrm{Vec}(\boldsymbol{b}_i)\big),
\)
a \(2L \times 2L\) covariance matrix. This matrix can be written in \(L \times L\) block form as \(\boldsymbol{\Sigma} = (\boldsymbol{\Sigma}_{\ell\ell'})_{\ell,\ell'=1}^L\), where each \(2\times2\) block \(\boldsymbol{\Sigma}_{\ell\ell'}\) is given by
\[
\boldsymbol{\Sigma}_{\ell\ell'} =
\begin{pmatrix}
\sigma^{(00)}_{\ell\ell'} & \sigma^{(01)}_{\ell\ell'} \\
\sigma^{(10)}_{\ell\ell'} & \sigma^{(11)}_{\ell\ell'}
\end{pmatrix}
=
\begin{pmatrix}
\mathrm{Cov}\bigl(b^{(0)}_{i\ell}, b^{(0)}_{i\ell'}\bigr) &
\mathrm{Cov}\bigl(b^{(0)}_{i\ell}, b^{(1)}_{i\ell'}\bigr) \\
\mathrm{Cov}\bigl(b^{(1)}_{i\ell}, b^{(0)}_{i\ell'}\bigr) &
\mathrm{Cov}\bigl(b^{(1)}_{i\ell}, b^{(1)}_{i\ell'}\bigr)
\end{pmatrix}.
\]
Then, we have
\begin{itemize}[leftmargin=5mm]
\item[i)] For the $\ell$th risk factor of the $i$th individual nested within the $k$th cohort, the variance is
\begin{eqnarray}
\mathrm{Var}\big(Y_{\ell k(i)}\big)
&=&
\sigma^{(00)}_{\ell\ell}
+ 2a_{ij}\,\sigma^{(01)}_{\ell\ell}
+ a_{ij}^2\,\sigma^{(11)}_{\ell\ell}
+ \lambda_{kk}^{\ell}\boldsymbol{S}^{T}(a_{ij})\boldsymbol{S}(a_{ij})
+ \sigma_{\epsilon_{\ell}}^{2(p)} , \notag
\end{eqnarray}
where
\(
\sigma^{2(p)}_{\epsilon\ell} = \omega_{\ell}^{2(p)} \left(1 - 2\delta_\ell^2/{\pi} \right)
\)
denotes the marginal residual variance under the skew-normal specification with skewness parameter
\(
\delta_\ell = \psi_\ell/\sqrt{1 + \psi_\ell^2},
\)
for subjects whose age falls in window $p$. 
This variance captures the age-dependent variability of the $\ell$th risk factor, including individual-specific effects and cohort-level variability.

\item[ii)] For different risk factors $\ell$ and $\ell^\prime$, measured on the same individual $i$ in the $k$th cohort at ages $a_{ij}$ and $a_{ij^\prime}$, the covariance is
\begin{eqnarray} 
\mathrm{Cov}\big(Y_{\ell k(i)}, Y_{\ell^\prime k(i)}\big)
&=&
\sigma^{(00)}_{\ell\ell^\prime}
+ a_{ij}\,\sigma^{(01)}_{\ell\ell^\prime}
+ a_{ij^\prime}\,\sigma^{(10)}_{\ell\ell^\prime}
+ a_{ij}a_{ij^\prime}\,\sigma^{(11)}_{\ell\ell^\prime},
\notag
\end{eqnarray}
and the correlation is
\begin{eqnarray} 
\resizebox{.96\hsize}{!}{
$\mathrm{Corr}\big(Y_{\ell k(i)}(a_{ij}),Y_{\ell^\prime k(i)}(a_{ij^\prime})\big) =
\frac{
\sigma^{(00)}_{\ell\ell^\prime}
+ a_{ij}\,\sigma^{(01)}_{\ell\ell^\prime}
+ a_{ij^\prime}\,\sigma^{(10)}_{\ell\ell^\prime}
+ a_{ij}a_{ij^\prime}\,\sigma^{(11)}_{\ell\ell^\prime}
}{
\sqrt{
\sigma^{(00)}_{\ell\ell}
+ 2a_{ij}\,\sigma^{(01)}_{\ell\ell}
+ a_{ij}^2\,\sigma^{(11)}_{\ell\ell}
+ \lambda^{\ell}_{kk}\boldsymbol{S}^{T}(a_{ij})\boldsymbol{S}(a_{ij})
+ \sigma_{\epsilon_{\ell}}^{2(p)}
}
\sqrt{
\sigma^{(00)}_{\ell^\prime\ell^\prime}
+ 2a_{ij^\prime}\,\sigma^{(01)}_{\ell^\prime\ell^\prime}
+ a_{ij^\prime}^2\,\sigma^{(11)}_{\ell^\prime\ell^\prime}
+ \lambda^{\ell^\prime}_{kk}\boldsymbol{S}^{T}(a_{ij^\prime})\boldsymbol{S}(a_{ij^\prime})
+ \sigma_{\epsilon_{\ell^\prime}}^{2(p^\prime)}
}
}.
$
\notag
} 
\end{eqnarray}
They reflect shared biological or lifestyle influences between two different risk factors for the same individual and vary with the pair of ages considered.

\item[iii)] For the same risk factor $\ell$ across different individuals \(i\) and \(i^\prime\) in the same cohort \(k\), the covariance is
\begin{eqnarray}
\mathrm{Cov}\big(Y_{\ell k(i)}, Y_{\ell k(i^\prime)}\big)
&=&
\lambda_{kk}^{\ell} \boldsymbol{S}^{T}(a_{ij})\boldsymbol{S}(a_{i^\prime j^\prime}),
\notag
\end{eqnarray}
and the correlation is
\begin{equation} 
\resizebox{.96\hsize}{!}{
$\mathrm{Corr}\big(Y_{\ell k(i)},Y_{\ell k(i^\prime)}\big) =
\frac{\lambda^{\ell}_{kk}\boldsymbol{S}^{T}(a_{ij})\boldsymbol{S}(a_{i^\prime j^\prime})}
{\sqrt{
\sigma^{(00)}_{\ell\ell}
+ 2a_{ij}\,\sigma^{(01)}_{\ell\ell}
+ a_{ij}^2\,\sigma^{(11)}_{\ell\ell}
+ \lambda^{\ell}_{kk}\boldsymbol{S}^{T}(a_{ij})\boldsymbol{S}(a_{ij})
+ \sigma_{\epsilon_{\ell}}^{2(p)}
}
\sqrt{
\sigma^{(00)}_{\ell\ell}
+ 2a_{i^\prime j^\prime}\,\sigma^{(01)}_{\ell\ell}
+ a_{i^\prime j^\prime}^2\,\sigma^{(11)}_{\ell\ell}
+ \lambda^{\ell}_{kk}\boldsymbol{S}^{T}(a_{i^\prime j^\prime})\boldsymbol{S}(a_{i^\prime j^\prime})
+ \sigma_{\epsilon_{\ell}}^{2(p^\prime)}
}
}.
$
\notag
}
\end{equation} 
They quantify cohort-level shared variability in the same risk factor across different individuals and vary with the pair of ages considered.

\item[iv)] The covariance and correlation between different risk factors \(\ell\) and \(\ell^\prime\) for different individuals \(i\) and \(i^\prime\) in the same cohort \(k\), is
\begin{eqnarray}
\mathrm{Cov}\big(Y_{\ell k(i)}, Y_{\ell^\prime k(i^\prime)}\big) = \mathrm{Corr}(Y_{\ell k(i)},Y_{\ell ^{\prime}k(i^{\prime})}) =0.
\notag
\end{eqnarray}
This covariance is zero, indicating no direct relationship between different risk factors for different individuals in the same cohort. This assumption excludes biologically implausible dependencies, as any shared variability is assumed to be captured through cohort-level effects.

\item[v)] For the same risk factor (\(\ell = \ell^\prime\)) across individuals in different cohorts \(k\) and \(k^\prime\), the covariance is
\begin{eqnarray}
\mathrm{Cov}\big(Y_{\ell k(i)}, Y_{\ell k^\prime(i^\prime)}\big)
&=&
\lambda_{kk^\prime}^{\ell} \boldsymbol{S}^{T}(a_{ij}) \boldsymbol{S}(a_{i^\prime j^\prime}),
\notag 
\end{eqnarray}
and the correlation is
\begin{eqnarray} 
\resizebox{0.96\hsize}{!}{
$\mathrm{Corr}\big(Y_{\ell k(i)},Y_{\ell k^{\prime}(i^{\prime})}\big) =
\frac{\lambda^{\ell}_{kk^\prime}\boldsymbol{S}^{T}(a_{ij})\boldsymbol{S}(a_{i^\prime j^\prime})}
{\sqrt{
\sigma^{(00)}_{\ell\ell}
+ 2a_{ij}\,\sigma^{(01)}_{\ell\ell}
+ a_{ij}^2\,\sigma^{(11)}_{\ell\ell}
+ \lambda^{\ell}_{kk}\boldsymbol{S}^{T}(a_{ij})\boldsymbol{S}(a_{ij})
+ \sigma_{\epsilon_{\ell}}^{2(p)}
}
\sqrt{
\sigma^{(00)}_{\ell\ell}
+ 2a_{i^\prime j^\prime}\,\sigma^{(01)}_{\ell\ell}
+ a_{i^\prime j^\prime}^2\,\sigma^{(11)}_{\ell\ell}
+ \lambda^{\ell}_{k^\prime k^\prime}\boldsymbol{S}^{T}(a_{i^\prime j^\prime})\boldsymbol{S}(a_{i^\prime j^\prime})
+ \sigma_{\epsilon_{\ell}}^{2(p^\prime)}
}
}.
$
} \notag
\end{eqnarray}
They reflect cohort-specific trends in the same risk factor across individuals from different cohorts and vary with the pair of ages considered.

\item[vi)] The covariance between different risk factors \(\ell\) and \(\ell^\prime\) for different individuals \(i\) and \(i^\prime\) nested in different cohorts \(k\) and \(k^\prime\), is
\begin{eqnarray}
\mathrm{Cov}\big(Y_{\ell k(i)}, Y_{\ell^\prime k^\prime(i^\prime)}\big) = \mathrm{Corr}(Y_{\ell k(i)},Y_{\ell ^{\prime}k^{\prime}(i^{\prime})}) = 0.
\notag
\end{eqnarray}
This covariance is zero, reflecting the lack of direct dependence between different risk factors across individuals in different cohorts.
\end{itemize}

\subsection{Analytical Derivations} \label{supp:Analytical_Derivations}
For risk factors $\ell$ and $\ell^\prime$ and individuals $i$ and $i^\prime$ nested in cohorts $k$ and $k^\prime$, we have
\medmuskip=0.01mu
\thinmuskip=0.01mu
\thickmuskip=0.01mu
\nulldelimiterspace=0.01pt
\scriptspace=0.01pt
\begin{eqnarray}
\mathrm{Cov}\big(Y_{\ell k(i)},Y_{\ell^\prime k^\prime(i^\prime)}\big)
&=&
\mathrm{Cov}\big(\mu_{\ell k(i)}(a_{ij})+\epsilon_{\ell k(i)}(a_{ij}),\mu_{\ell^\prime k^\prime(i^\prime)}(a_{i^\prime j^\prime})+\epsilon_{\ell^\prime k^\prime(i^\prime)}(a_{i^\prime j^\prime})\big) \notag \\
&=&
\mathrm{Cov}\big(\mu_{\ell k(i)}(a_{ij}),\mu_{\ell^\prime k^\prime(i^\prime)}(a_{i^\prime j^\prime})\big) +
\mathrm{Cov}\big(\epsilon_{\ell k(i)}(a_{ij}),\epsilon_{\ell^\prime k^\prime(i^\prime)}(a_{i^\prime j^\prime})\big) \notag \\
&=&
\mathrm{Cov}\big(\boldsymbol{A}^{T}(a_{ij})\boldsymbol{\beta}_{\ell k(i)},\boldsymbol{A}^{T}(a_{i^\prime j^\prime})\boldsymbol{\beta}_{\ell^\prime k^\prime(i^\prime)}\big) +
\mathrm{Cov}\big(\epsilon_{\ell k(i)}(a_{ij}),\epsilon_{\ell^\prime k^\prime(i^\prime)}(a_{i^\prime j^\prime})\big) \notag \\
&=&
\boldsymbol{A}^{T}(a_{ij})\,
\mathrm{Cov}\big(\boldsymbol{\beta}_{\ell k(i)},\boldsymbol{\beta}_{\ell^\prime k^\prime(i^\prime)}\big)\,
\boldsymbol{A}(a_{i^\prime j^\prime})
+ \mathrm{Cov}\big(\epsilon_{\ell k(i)}(a_{ij}),\epsilon_{\ell^\prime k^\prime(i^\prime)}(a_{i^\prime j^\prime})\big). \notag
\end{eqnarray}
where $\mathrm{Cov}\big(\boldsymbol{\beta}_{\ell k(i)},\boldsymbol{\beta}_{\ell^\prime k^\prime(i^\prime)}\big)$ is a $(P+2)\times(P+2)$ matrix
\begin{eqnarray}
\mathrm{Cov}\big(\boldsymbol{\beta}_{\ell k(i)},\boldsymbol{\beta}_{\ell^\prime k^\prime(i^\prime)}\big)
=\left[ 
\begin{array}{cccc}
\mathrm{Cov}\big(\beta_{\ell k(i)}^{(0)},\beta_{\ell^\prime k^\prime(i^\prime)}^{(0)}\big) & \mathrm{Cov}\big(\beta_{\ell k(i)}^{(0)},\beta_{\ell^\prime k^\prime(i^\prime)}^{(1)}\big) &\ldots & \mathrm{Cov}\big(\beta_{\ell k(i)}^{(0)},\beta_{\ell^\prime k^\prime(i^\prime)}^{(P+1)}\big) \\ 
\mathrm{Cov}\big(\beta_{\ell k(i)}^{(1)},\beta_{\ell^\prime k^\prime(i^\prime)}^{(0)}\big) & \mathrm{Cov}\big(\beta_{\ell k(i)}^{(1)},\beta_{\ell^\prime k^\prime(i^\prime)}^{(1)}\big) &\ldots & \mathrm{Cov}\big(\beta_{\ell k(i)}^{(1)},\beta_{\ell^\prime k^\prime(i^\prime)}^{(P+1)}\big) \\
\vdots & \vdots & \ddots & \vdots \\ 
\mathrm{Cov}\big(\beta_{\ell k(i)}^{(P+1)},\beta_{\ell^\prime k^\prime(i^\prime)}^{(0)}\big) & \mathrm{Cov}\big(\beta_{\ell k(i)}^{(P+1)},\beta_{\ell^\prime k^\prime(i^\prime)}^{(1)}\big) & \ldots & \mathrm{Cov}\big(\beta_{\ell k(i)}^{(P+1)},\beta_{\ell^\prime k^\prime(i^\prime)}^{(P+1)}\big)
\end{array} \right]. \notag
\end{eqnarray}

Using equations (2) and (3) of the manuscript, the $(p,p^\prime)$th element of this covariance matrix can be written as
\begin{eqnarray}
\mathrm{Cov}\big(\beta_{\ell k(i)}^{(p)}, \beta_{\ell' k'(i')}^{(p')}\big)
=
\left\{
\begin{array}{ll}
\mathrm{Cov}\big(b_{i\ell}^{(p)}, b_{i'\ell'}^{(p')}\big), 
& \text{if } p,p' \in \{0,1\}, \\[0.5em]
\mathrm{Cov}\big(b_{\ell k}^{(p)}, b_{\ell' k'}^{(p')}\big), 
& \text{if } p,p' \geq 2, \\[0.5em]
0, 
& \text{if one of } p,p' \in \{0,1\} \text{ and the other } \geq 2.
\end{array}
\right.
\end{eqnarray}
Then, for different risk factors $\ell$ and $\ell^\prime$ for the $i$th individual in the $k$th cohort, we have 
\begin{eqnarray}
\boldsymbol{A}^{T}(a_{ij})\mathrm{Cov}\big(\boldsymbol{\beta}_{\ell k(i)},\boldsymbol{\beta}_{\ell ^\prime k(i)}\big)\boldsymbol{A}(a_{i j^\prime})
&=&
\sigma^{(00)}_{\ell\ell^\prime}
+ a_{ij}\,\sigma^{(01)}_{\ell\ell^\prime}
+ a_{ij^\prime}\,\sigma^{(10)}_{\ell\ell^\prime}
+ a_{ij}a_{ij^\prime}\,\sigma^{(11)}_{\ell\ell^\prime}, \notag \\
\boldsymbol{A}^{T}(a_{ij})\mathrm{Cov}\big(\boldsymbol{\beta}_{\ell k(i)},\boldsymbol{\beta}_{\ell k(i)}\big)\boldsymbol{A}(a_{i j})
&=&
\sigma^{(00)}_{\ell\ell}
+ 2a_{ij}\,\sigma^{(01)}_{\ell\ell}
+ a_{ij}^2\,\sigma^{(11)}_{\ell\ell}
+ \lambda^{\ell}_{kk}\boldsymbol{S}^{T}(a_{ij})\boldsymbol{S}(a_{ij})
+ \sigma_{\epsilon_{\ell}}^{2(p)}, \notag \\
\boldsymbol{A}^{T}(a_{ij^\prime})\mathrm{Cov}\big(\boldsymbol{\beta}_{\ell^\prime k(i)},\boldsymbol{\beta}_{\ell^\prime k(i)}\big)\boldsymbol{A}(a_{i j^\prime})
&=&
\sigma^{(00)}_{\ell^\prime\ell^\prime}
+ 2a_{ij^\prime}\,\sigma^{(01)}_{\ell^\prime\ell^\prime}
+ a_{ij^\prime}^2\,\sigma^{(11)}_{\ell^\prime\ell^\prime}
+ \lambda^{\ell^\prime}_{kk}\boldsymbol{S}^{T}(a_{ij^\prime})\boldsymbol{S}(a_{ij^\prime})
+ \sigma_{\epsilon_{\ell^\prime}}^{2(p^\prime)}, \notag
\end{eqnarray}
and
\begin{eqnarray} 
\resizebox{\hsize}{!}{
$\mathrm{Corr}\big(Y_{\ell k(i)},Y_{\ell^\prime k(i)}\big) =
\frac{
\sigma^{(00)}_{\ell\ell^\prime}
+ a_{ij}\,\sigma^{(01)}_{\ell\ell^\prime}
+ a_{ij^\prime}\,\sigma^{(10)}_{\ell\ell^\prime}
+ a_{ij}a_{ij^\prime}\,\sigma^{(11)}_{\ell\ell^\prime}
}{
\sqrt{
\sigma^{(00)}_{\ell\ell}
+ 2a_{ij}\,\sigma^{(01)}_{\ell\ell}
+ a_{ij}^2\,\sigma^{(11)}_{\ell\ell}
+ \lambda^{\ell}_{kk}\boldsymbol{S}^{T}(a_{ij})\boldsymbol{S}(a_{ij})
+ \sigma_{\epsilon_{\ell}}^{2(p)}
}
\sqrt{
\sigma^{(00)}_{\ell^\prime\ell^\prime}
+ 2a_{ij^\prime}\,\sigma^{(01)}_{\ell^\prime\ell^\prime}
+ a_{ij^\prime}^2\,\sigma^{(11)}_{\ell^\prime\ell^\prime}
+ \lambda^{\ell^\prime}_{kk}\boldsymbol{S}^{T}(a_{ij^\prime})\boldsymbol{S}(a_{ij^\prime})
+ \sigma_{\epsilon_{\ell^\prime}}^{2(p^\prime)}
}
}.
$
} \notag 
\end{eqnarray}
Similarly, for other combinations of risk factors and individuals, the corresponding expressions follow.

\section{Specification of Model Parameters and Random Effects} \label{supp:Specification_of}

This section summarizes the full set of unknown quantities in the proposed model, denoted by 
\(\mathcal{B} = (\mathcal{B}_1, \mathcal{B}_2)\), where \(\mathcal{B}_1\) includes fixed effects and hyperparameters, 
and \(\mathcal{B}_2\) comprises subject-specific and cohort-level random effects. 
All components of \(\mathcal{B}\) are treated as unknown in posterior inference.

\subsection{Fixed Parameters} \label{supp:Fixed_Parameters}
The fixed parameters in \(\mathcal{B}_1\) are specified as follows. 
For each risk factor \(\ell = 1,\ldots,L\), the vectors \(\boldsymbol{\alpha}^{(0)}_{\ell}\) and \(\boldsymbol{\alpha}^{(1)}_{\ell}\) represent the regression coefficients associated with the overall intercept and the global linear age component, respectively. 
These parameters define the baseline mean level and overall rate of change.

For the spline-based components, the age knots \(s_1,\ldots,s_P\) enter the mean model through the truncated linear terms \((a - s_p)_+\). 
For \(p \ge 2\), the corresponding coefficients are specified as intercept-only fixed effects \(\alpha^{(p)}_{\ell}\) (scalars), which govern localized deviations from the global age trend across the pre-specified age intervals. 
Thus, unlike the \(p=0,1\) components, no baseline covariate interactions are included in the spline increments.

For the residual distribution, we introduce age-window-specific scale and skewness parameters. 
Specifically, \(\omega_{\ell}^{(p)} > 0\), for \(p = 1,\ldots,P\), and \(\psi_{\ell} \in \mathbb{R}\) are the scale and skewness parameters 
of the skew-normal distribution for residual variation in risk factor \(\ell\). 
This specification allows both heteroscedasticity and asymmetry in the residuals to vary across age intervals.

The subject-level covariance matrix \(\boldsymbol{\Sigma} \in \mathbb{R}^{2L \times 2L}\) 
is symmetric and positive definite and models the joint variability in individual-specific 
intercepts and slopes across all \(L\) risk factors. 
This structure captures both within- and between–risk-factor dependencies in the latent age trajectories 
while maintaining interpretability and computational stability.

For each risk factor \(\ell\), the cohort-level covariance matrix \(\boldsymbol{\Lambda}^{(\ell)} \in \mathbb{R}^{K \times K}\) 
governs the variation in cohort-specific deviations from the fixed intercept and slope, 
and is also assumed to be symmetric and positive definite. 
This component captures persistent between-cohort heterogeneity in baseline levels and age-related changes 
in risk factor trajectories.

\subsection{Random Effects} \label{supp:Random_Effects}

The random effects in \( \mathcal{B}_2 \) consist of subject-specific and cohort-specific components. 
For each subject \( i = 1,\ldots,N \), the random intercepts and slopes for all risk factors are collected in the matrix

\[
\boldsymbol{b}_i = 
\begin{bmatrix}
    b_{i1}^{(0)} & \cdots & b_{iL}^{(0)} \\
    b_{i1}^{(1)} & \cdots & b_{iL}^{(1)}
\end{bmatrix} \in \mathbb{R}^{2 \times L},
\]
which is vectorized as \( \mathrm{Vec}(\boldsymbol{b}_i) \sim \mathrm{N}(\mathbf{0}, \boldsymbol{\Sigma}) \). 
These subject-specific random effects represent deviations from the fixed intercept and global linear age term 
and induce correlation among risk factor trajectories within individuals.
For each risk factor \( \ell = 1,\ldots,L \) and spline component \( p = 2,\ldots,P+1 \), 
the cohort-level random effect vector is defined as
\[
\boldsymbol{b}_\ell^{(p)} = \left( b_{\ell 1}^{(p)}, \ldots, b_{\ell K}^{(p)} \right)^\top \in \mathbb{R}^{K}.
\]
These effects capture deviations from the fixed spline coefficients \( \alpha_{\ell}^{(p)} \) at the cohort level and are modeled as
\(
\boldsymbol{b}_\ell^{(p)} \sim \mathrm{N}(\mathbf{0}, \boldsymbol{\Lambda}^{(\ell)}),
\)
where \( \boldsymbol{\Lambda}^{(\ell)} \in \mathbb{R}^{K \times K} \) is the risk factor-specific cohort covariance matrix. 
A common covariance structure \( \boldsymbol{\Lambda}^{(\ell)} \) is assumed across all spline components \( p \ge 2 \), 
reflecting the assumption that cohort-level deviations are persistent across age windows.

All random effects are treated as latent variables and are sampled jointly with the fixed parameters using MCMC. 
Posterior predictive checks and model-based replications are computed using full posterior draws of \( \mathcal{B} \), 
ensuring accurate representation of both subject-level and cohort-level variation.
\vspace{-3mm}
\section{Bayesian Estimation and Computational Details} \label{supp:Bayesian_Estimation}
\subsection{Full Prior Specification} \label{supp:Full_Prior}
We employ a Bayesian approach with weakly informative priors. Specifically, for the skewness parameters \( \psi_{\ell}, ~ \ell = 1,\ldots,L \), we use a normal prior with mean zero and standard deviation 10, i.e., \( \psi_{\ell} ~ \sim \mathrm{N}(0,10) \), allowing a broad range for skewness without directional constraints. For $p = 0,1$, the coefficient vectors 
$\boldsymbol{\alpha}^{(p)}_{\ell}$ are assigned weakly informative multivariate normal priors, 
\(
\boldsymbol{\alpha}^{(p)}_{\ell} \sim \mathrm{N}(\mathbf{0}, 10^2 \mathbf{I}),
\)
reflecting minimal assumptions on the baseline and age–covariate interaction effects.  
For $p \ge 2$, the spline-based slope increments are modeled as intercept-only fixed effects 
$h_{\ell }^{(p)}(\boldsymbol{X}_{i}) = \alpha_{\ell }^{(p)}$,  
and a first-order random-walk prior is imposed on 
$\alpha_{\ell }^{(p)}$, for \( p=2,\ldots,P+1\), to encourage smoothness across adjacent spline segments:
\(
\alpha_{\ell }^{(2)} \sim \mathrm{N}(0,\,10^2)\),
\(
\alpha_{\ell }^{(p)} \sim \mathrm{N}(\alpha_{\ell }^{(p-1)},\,\tau_{\ell}^{2})\), 
 \(p = 3,\ldots,P+1,
\)
where $\tau_{\ell}$ controls the smoothness of successive slope differences.  
We assign a half-normal prior, 
\(
\tau_{\ell} \sim \mathrm{Half\text{-}Normal}(0,\,0.5),
\)
which concentrates most of its mass below approximately 1 while allowing moderate flexibility in slope variation.

For the subject-level covariance matrix \( \boldsymbol{\Sigma} \), we use a 
standard deviation–correlation parameterization: marginal standard deviations 
receive \(\mathrm{Half\text{-}Cauchy}(0, 2.5)\) priors, and the correlation matrix is assigned 
an LKJ prior, ensuring positive definiteness and weak regularization.
For each risk factor \( \ell \), the cohort-level covariance matrix 
\( \boldsymbol{\Lambda}^{(\ell)} \) is specified analogously, with 
\(\mathrm{Half\text{-}Cauchy}(0, 2.5)\) priors on the marginal standard deviations and an LKJ 
prior on the correlation matrix.
For the error scale parameter \( \omega_{\ell}^{(p)} \), we specify a half-Cauchy prior with scale parameter 2.5, \( \omega_{\ell}^{(p)} \sim \text{Half-Cauchy}(0, 2.5) \), for \( \ell = 1, \ldots, L \) and \( p = 1, \ldots, P \), to avoid restrictive assumptions on variance components.
\subsection{\texorpdfstring{Nested $\hat{R}$}{Nested R-hat}} \label{supp:Nested_rhat}
In the nested structure, we implement $K=8$ superchains, each consisting of $M=16$ subchains initialized from the same starting values within each superchain. Each subchain includes 100 samples, of which 60 are designated as warmup iterations. To obtain the initial values we fit our model on 8 distinct 10\% samples of the dataset, each drawn with replacement. This sampling approach allows us to efficiently capture a diverse set of starting values that reflect the posterior distribution without requiring full-dataset runs, which would be computationally intensive given the large size of our data. Using smaller, representative samples enables the superchains to converge from informed starting points, enhancing the accuracy of the nested \( \widehat{R} \) diagnostic in assessing convergence across the full parameter space.

 Let \( \theta_{nmk} \) denote the \( n \)-th draw from the \( m \)-th chain in the \( k \)-th superchain, and \( \overline{\theta}_{\cdot \cdot k} \) represent the mean of the posterior draws in superchain \( k \). The nested \( \widehat{R} \) diagnostic calculates the between-superchain variance \( B_{\nu} \) and within-superchain variance \( W_{\nu} \) as follows
\[
B_{\nu} = \frac{1}{K - 1} \sum_{k=1}^{K} \left(\overline{\theta}_{\cdot \cdot k} - \overline{\theta}_{\cdots}\right)^2,
\]
where \( \overline{\theta}_{\cdots } \) is the overall mean across all superchains. The within-superchain variance \( W \) is computed as
\[
W_{\nu} = \frac{1}{K} \sum_{k=1}^{K} \left( B_k + W_k \right),
\]
where \( B_k \) and \( W_k \) represent the between-chain and within-chain variances within superchain \( k \), defined as
\[
B_k = \frac{1}{M - 1} \sum_{m=1}^{M} \left(\overline{\theta}_{\cdot mk} - \overline{\theta}_{\cdot \cdot k}\right)^2,
\]
\[
W_k = \frac{1}{M} \sum_{m=1}^{M} \frac{1}{N - 1} \sum_{n=1}^{N} \left(\theta_{nmk} - \overline{\theta}_{\cdot mk}\right)^2.
\]
The nested \( \widehat{R}_{\nu} \) statistic is then defined as
\[
\widehat{R}_{\nu} = \sqrt{\frac{W_{\nu} + B_{\nu}}{W_{\nu}}} = \sqrt{1 + \frac{B_{\nu}}{W_{\nu}}}.
\]
\section{Model Validation and Diagnostics} \label{supp:Model_Validation}
\subsection{Residual QQ Plots} \label{supp:Residual_QQ}
\begin{figure}[!ht]
    \centering
\subfloat[SBP]{{\includegraphics[width=0.49\textwidth,height=0.2\textheight]{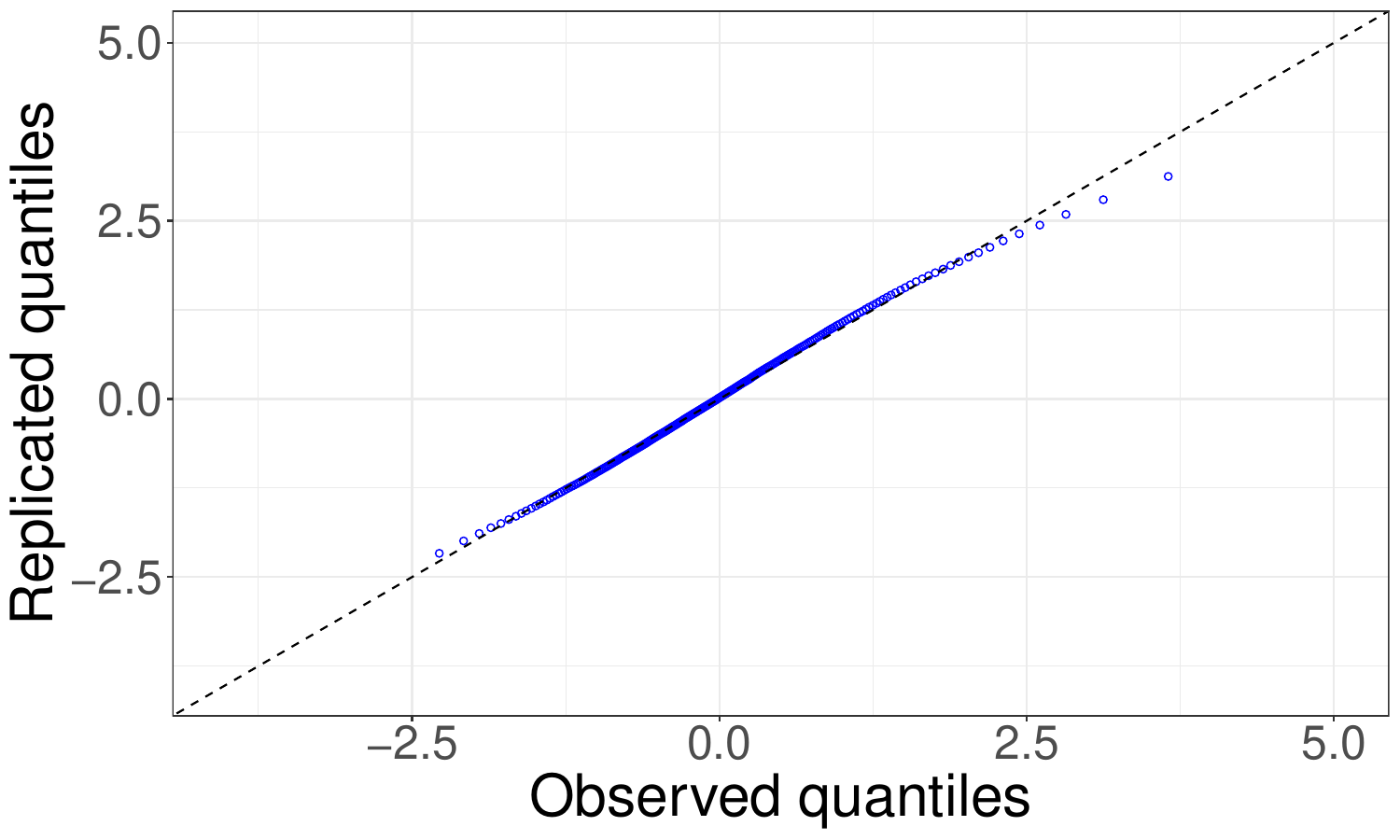}}}
\subfloat[DBP]{{\includegraphics[width=0.49\textwidth,height=0.2\textheight]{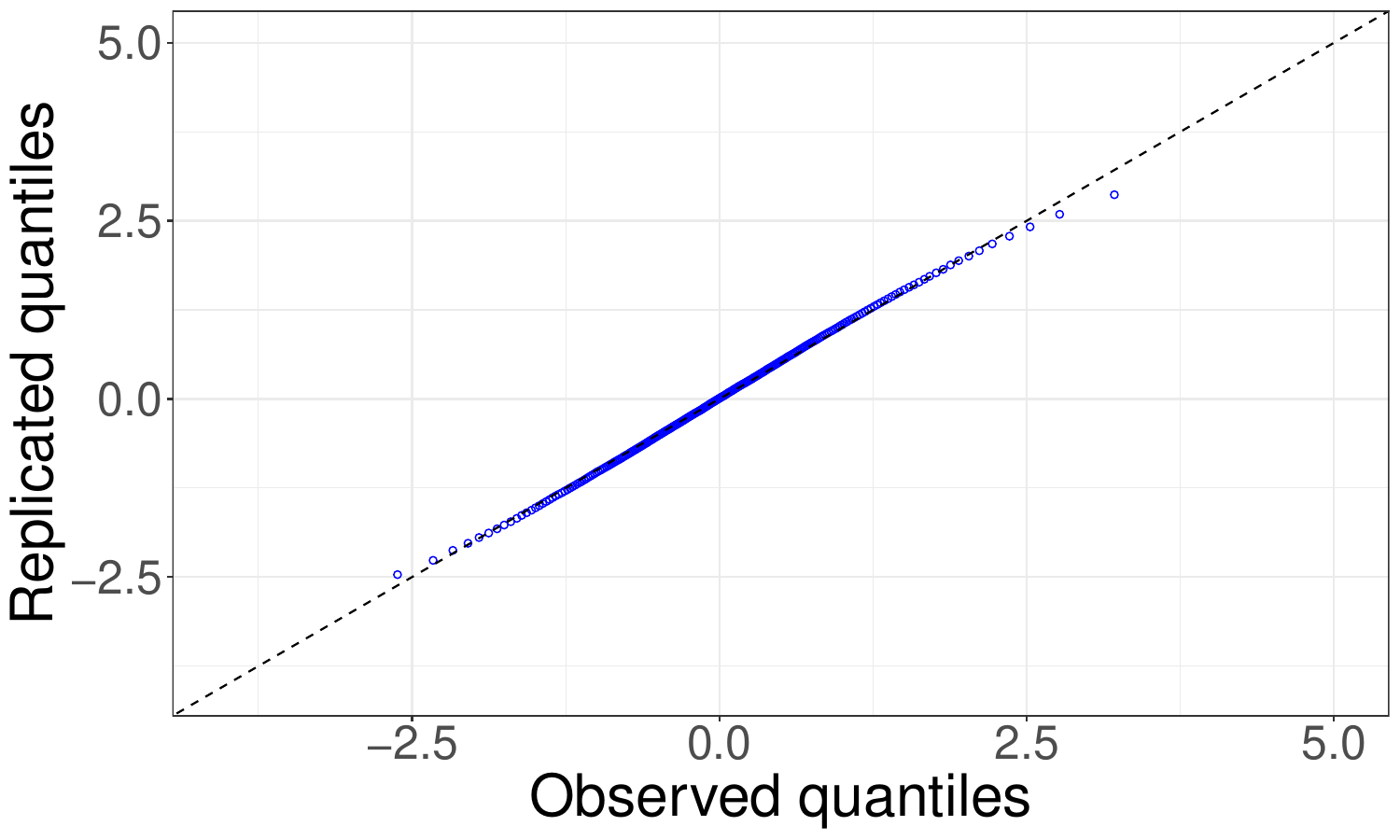}}} 

\subfloat[BMI]{{\includegraphics[width=0.49\textwidth,height=0.2\textheight]{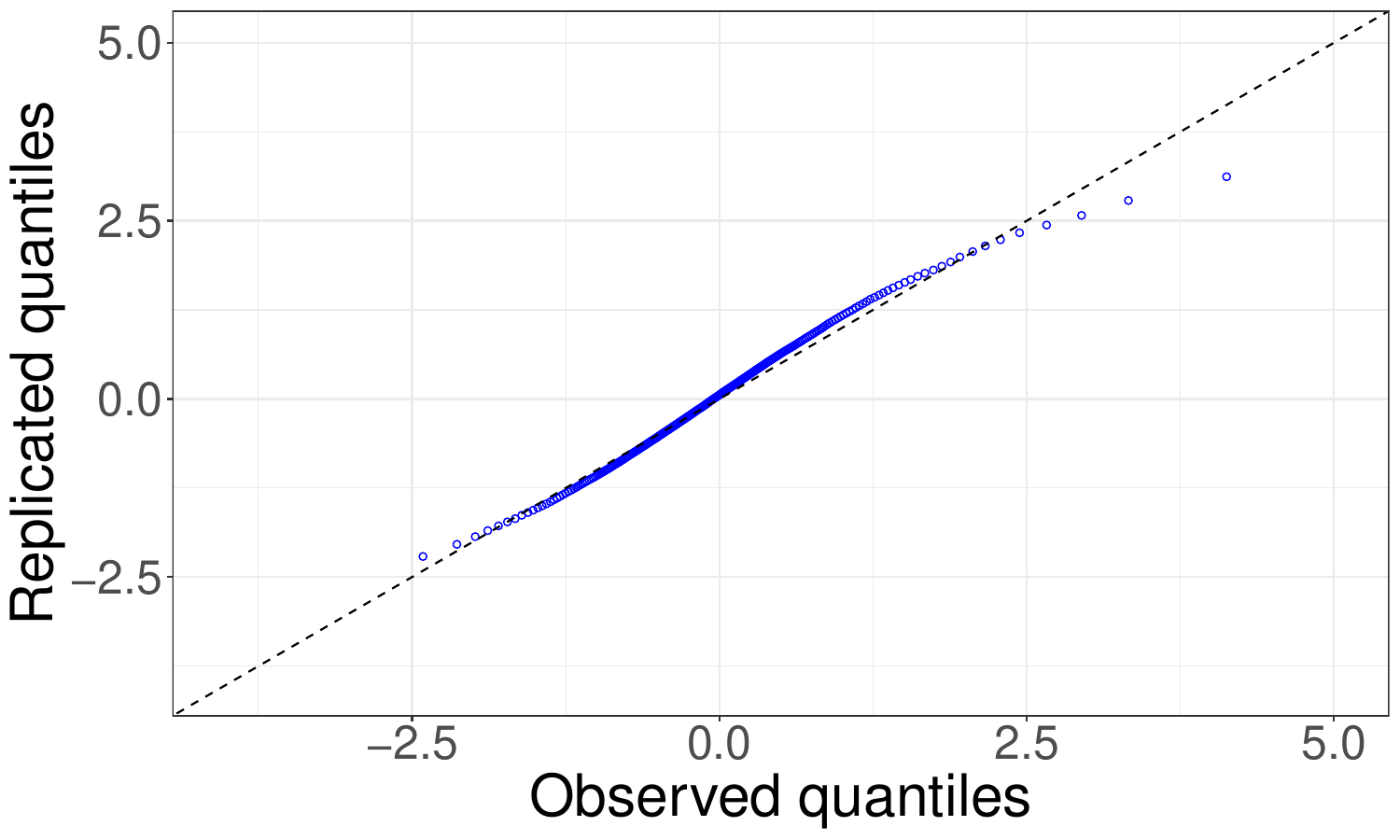}}}
\subfloat[TOTCHL]{{\includegraphics[width=0.49\textwidth,height=0.2\textheight]{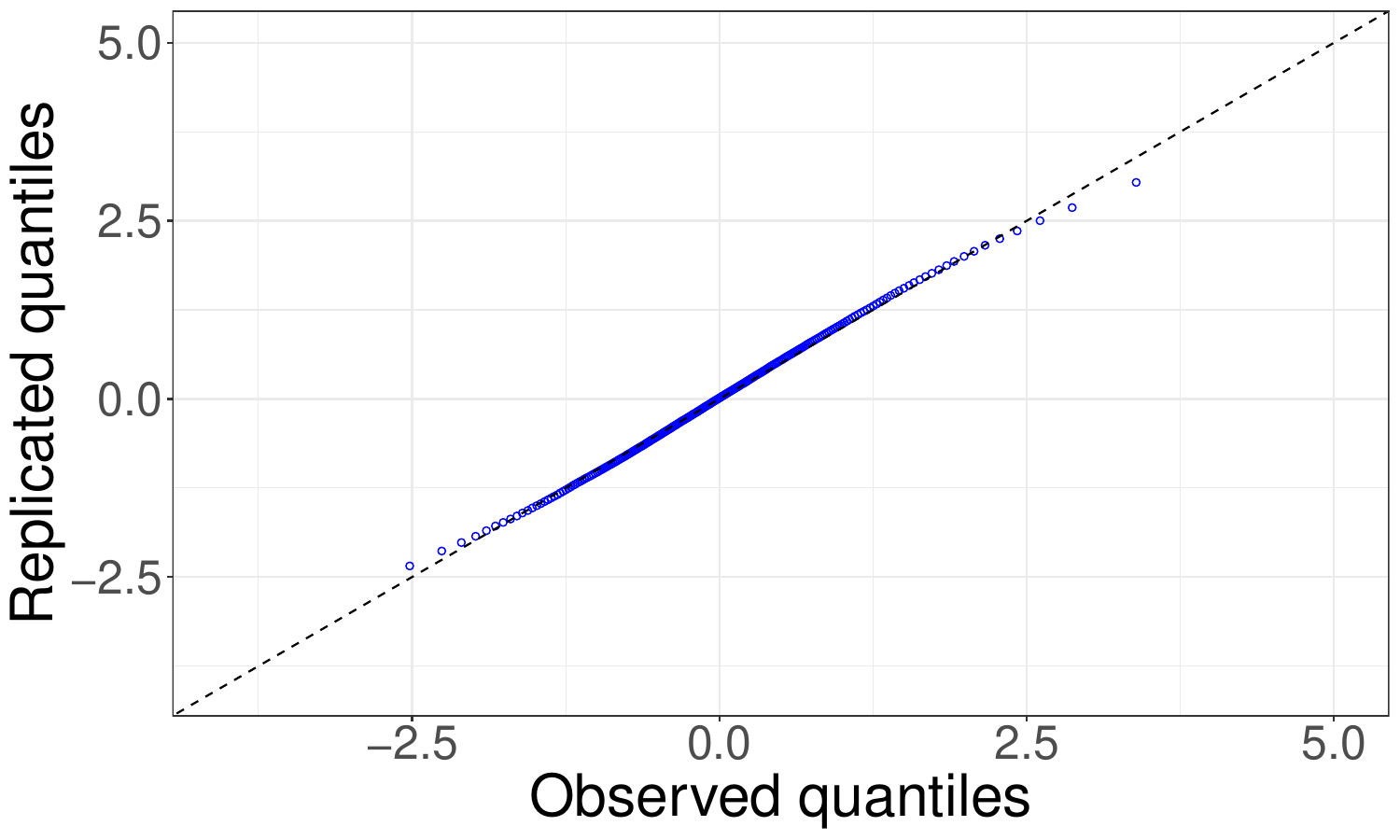}}} 

\subfloat[GLU]{{\includegraphics[width=0.49\textwidth,height=0.2\textheight]{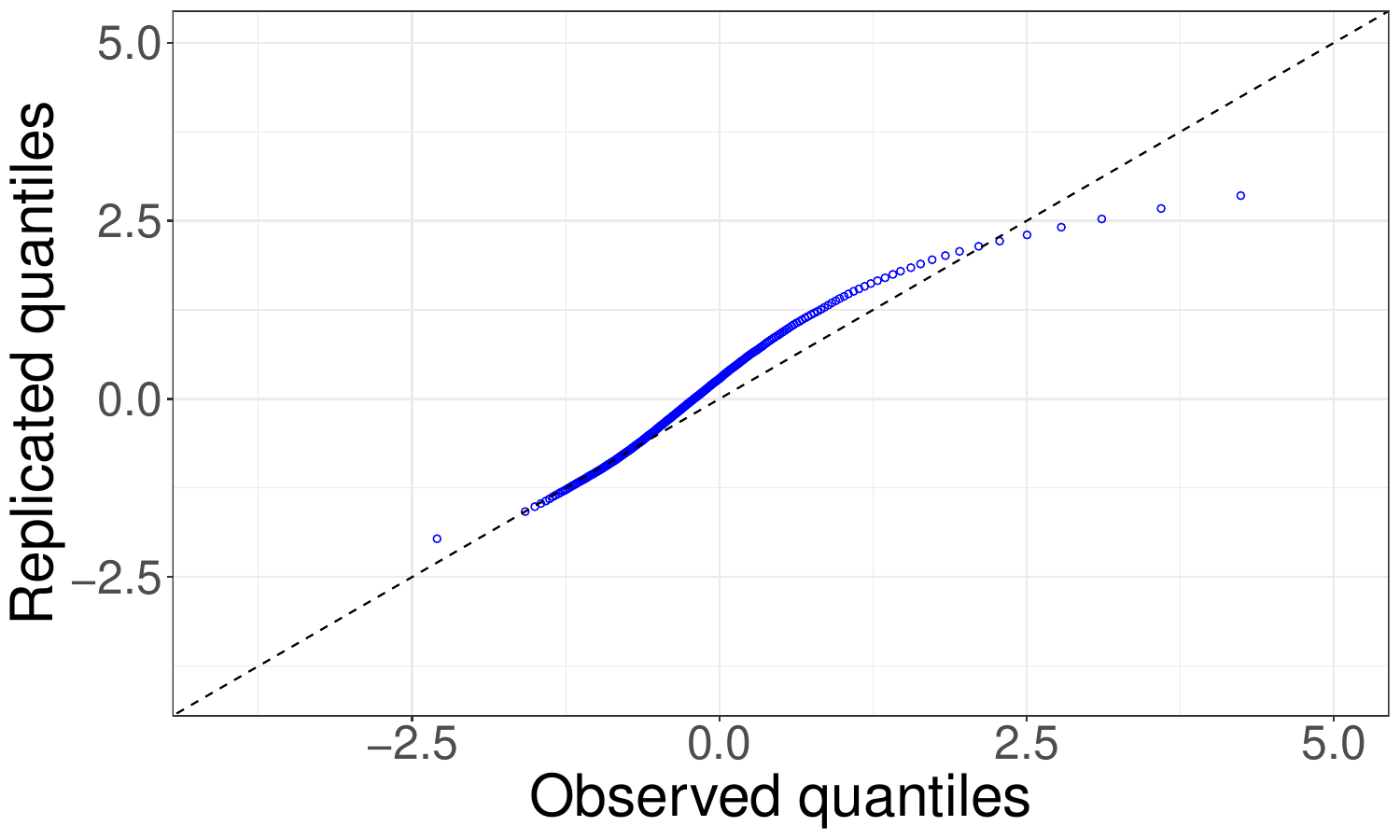}}}
\subfloat[HDL]{ {\includegraphics[width=0.49\textwidth,height=0.2\textheight]{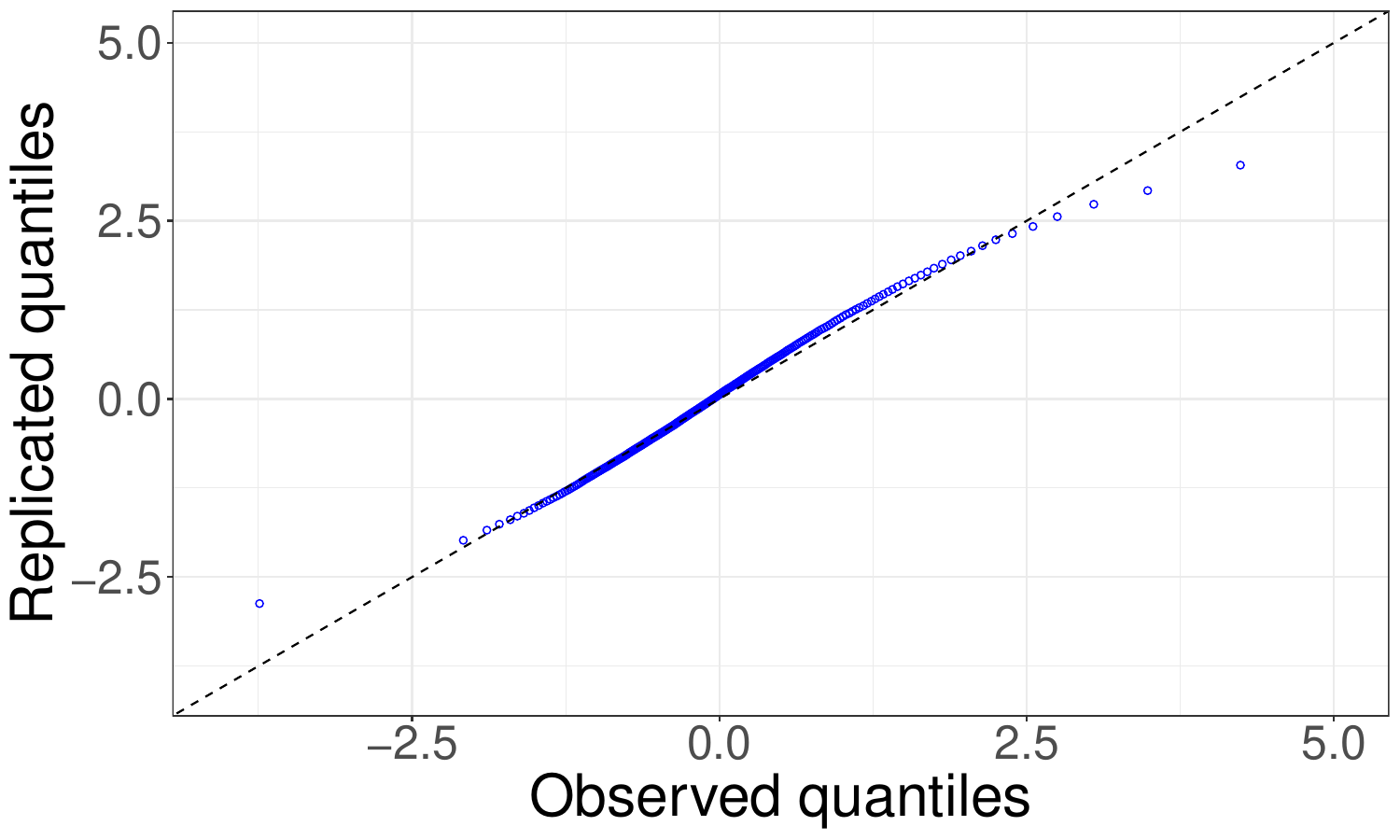}}} 

 \subfloat[TRIG]{{\includegraphics[width=0.49\textwidth,height=0.2\textheight]{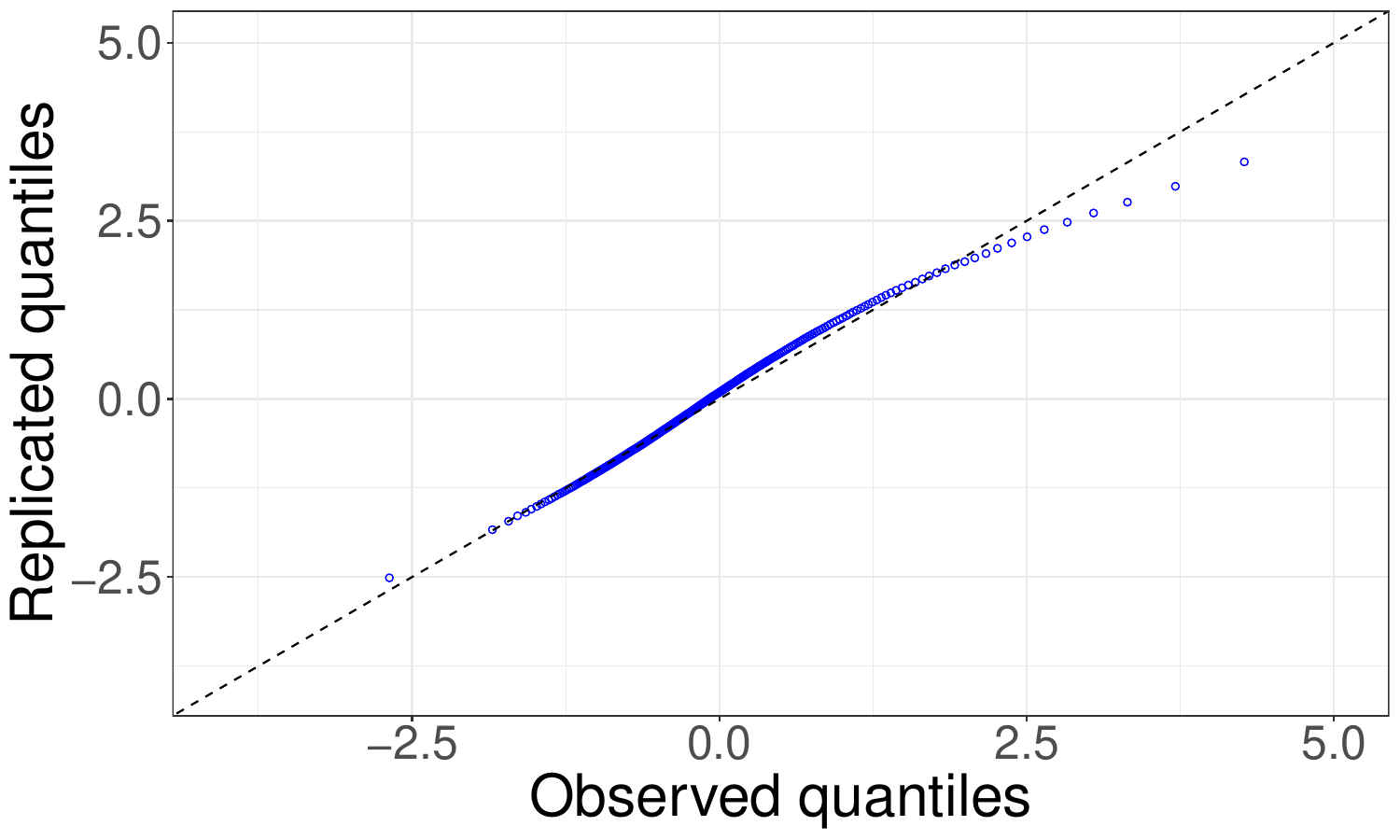}}}
    \caption{Standardized Residuals: Observed vs. Replicated (Men)}
\end{figure}

\begin{figure}[!ht]
    \centering
\subfloat[SBP]{{\includegraphics[width=0.49\textwidth,height=0.2\textheight]{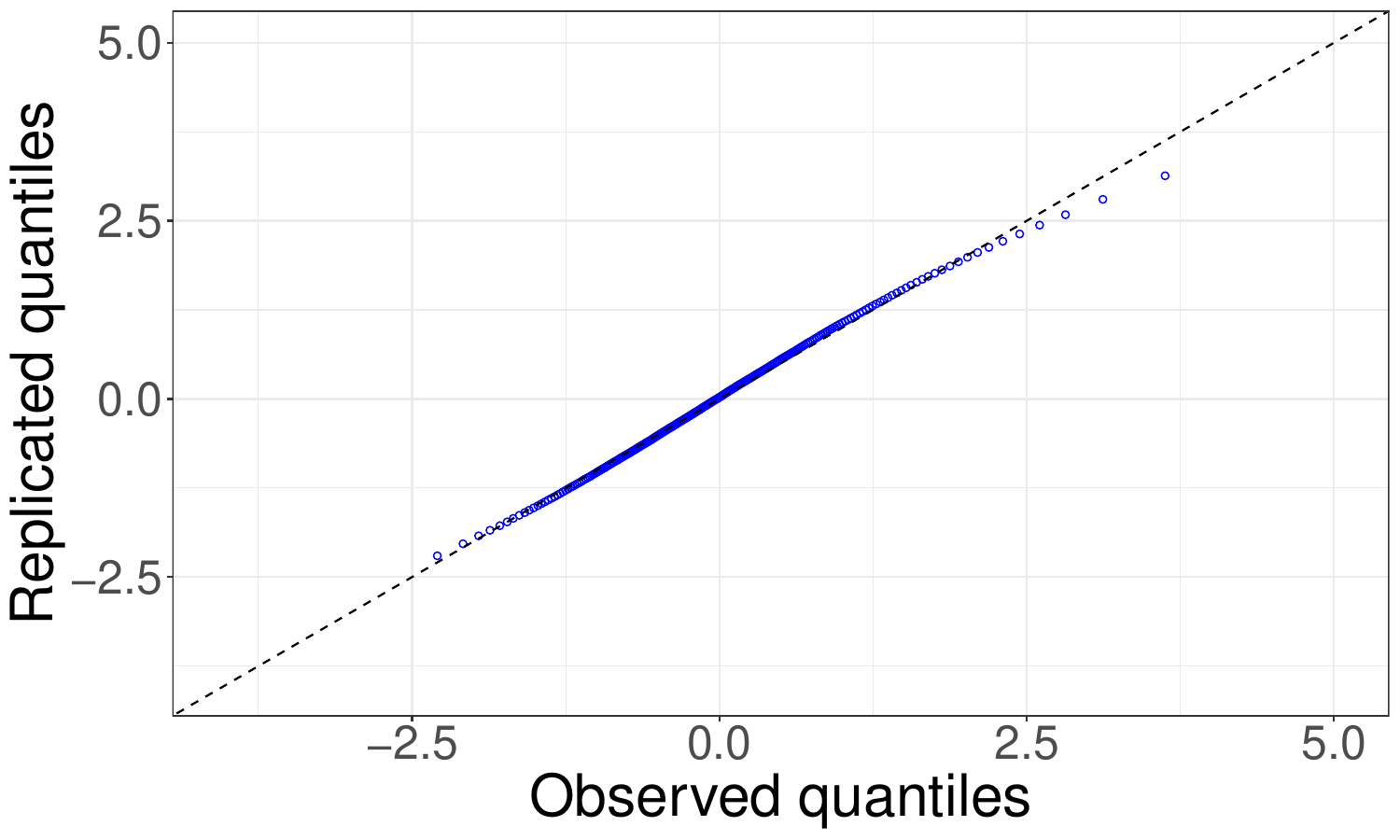}}}
\subfloat[DBP]{{\includegraphics[width=0.49\textwidth,height=0.2\textheight]{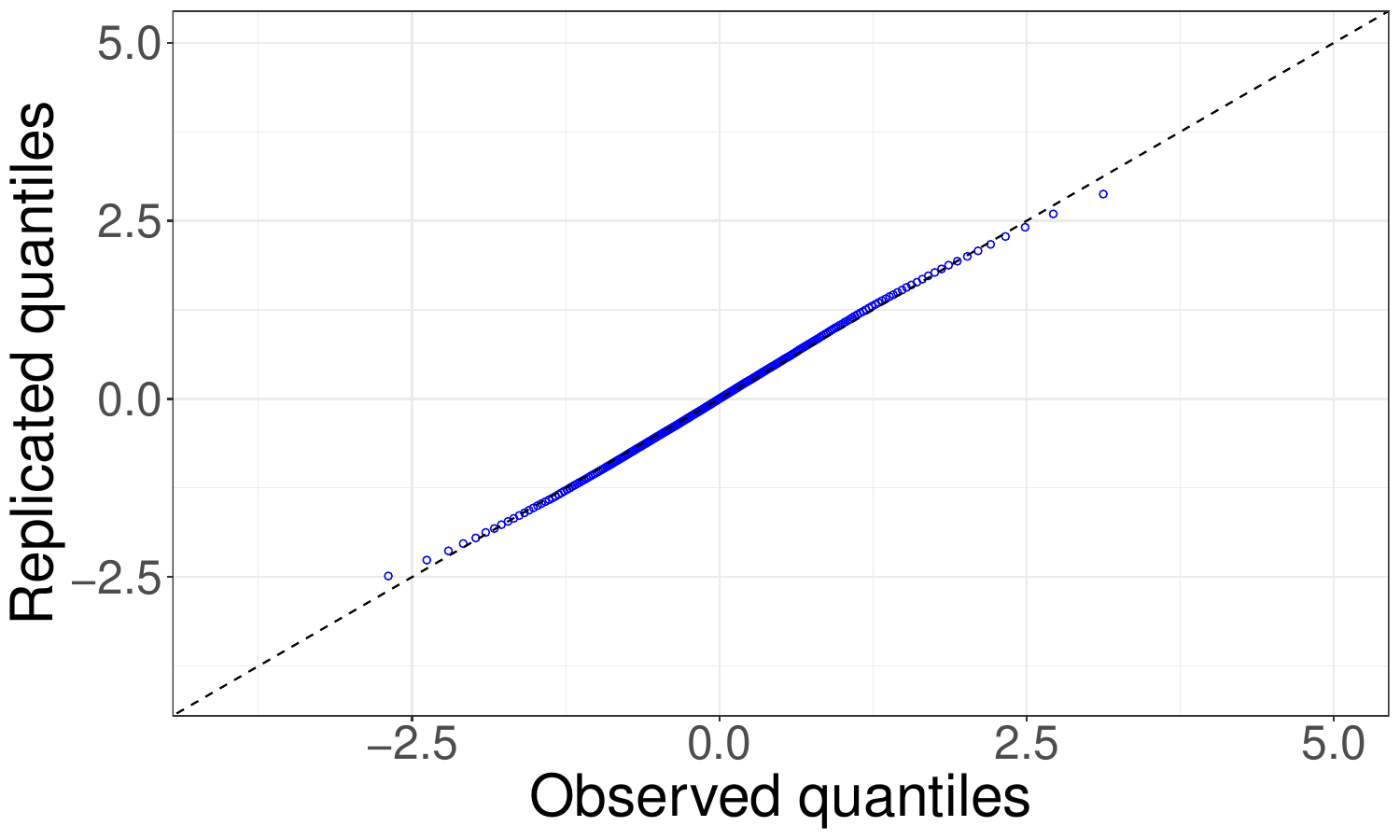}}}

\subfloat[BMI]{{\includegraphics[width=0.49\textwidth,height=0.2\textheight]{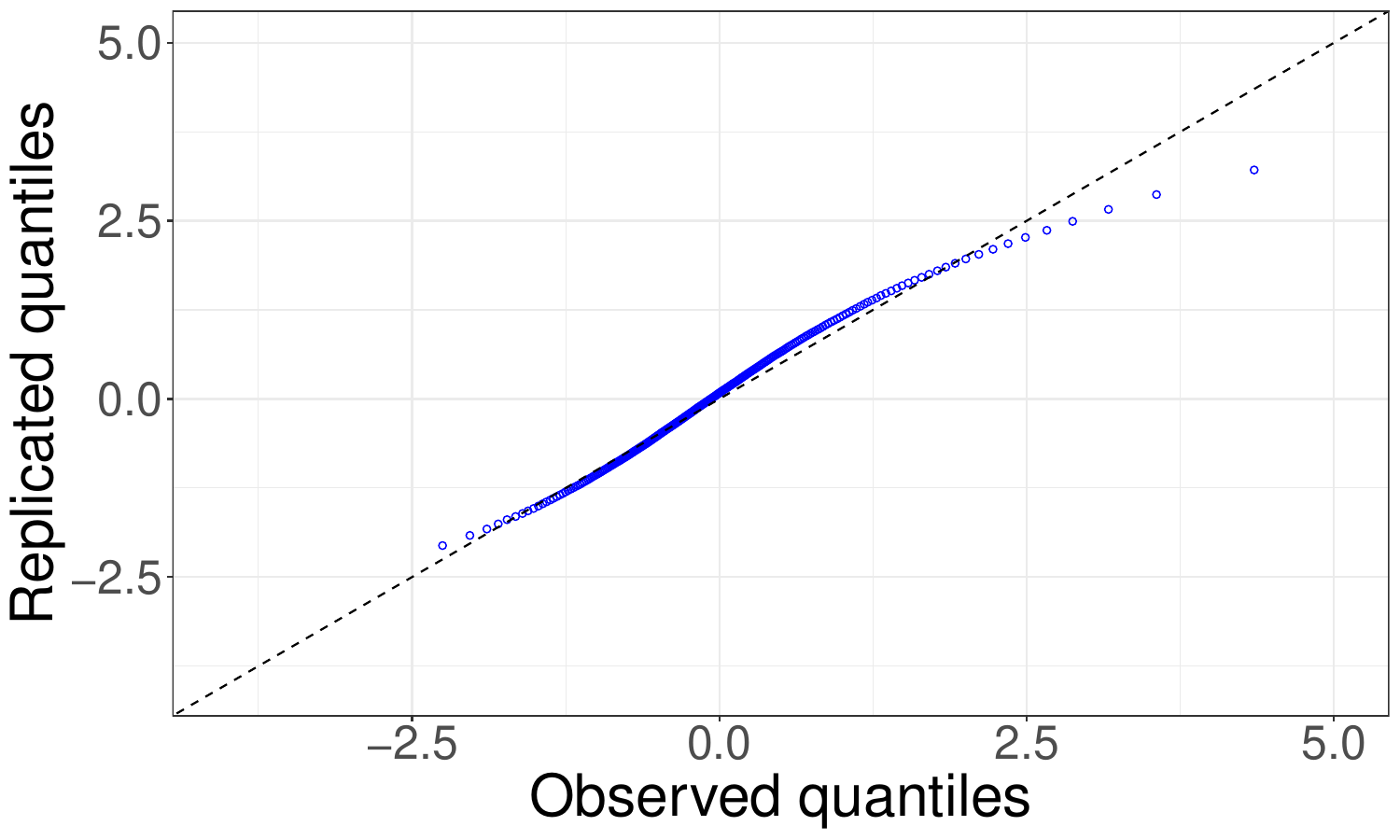}}}
\subfloat[TOTCHL]{{\includegraphics[width=0.49\textwidth,height=0.2\textheight]{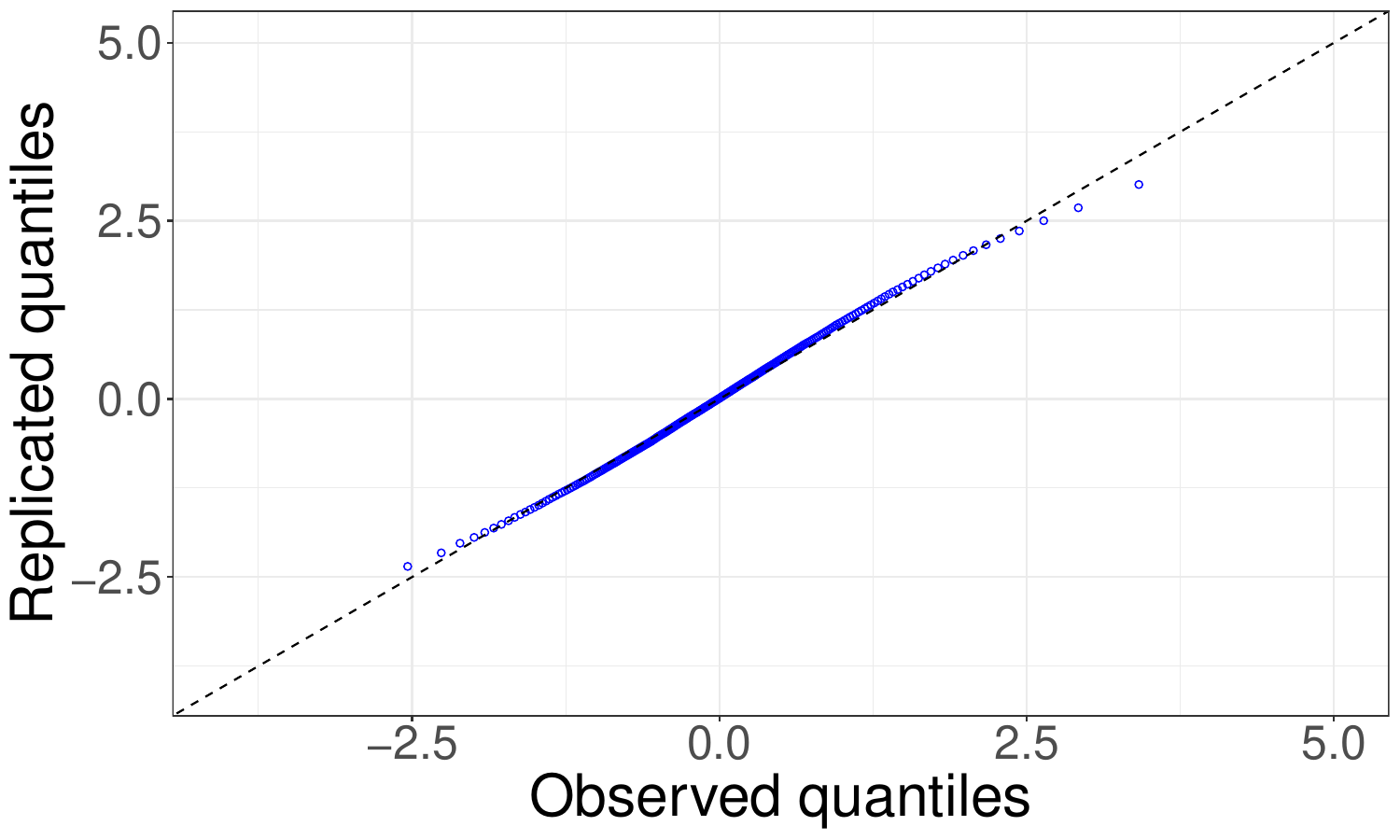}}}

\subfloat[GLU]{{\includegraphics[width=0.49\textwidth,height=0.2\textheight]{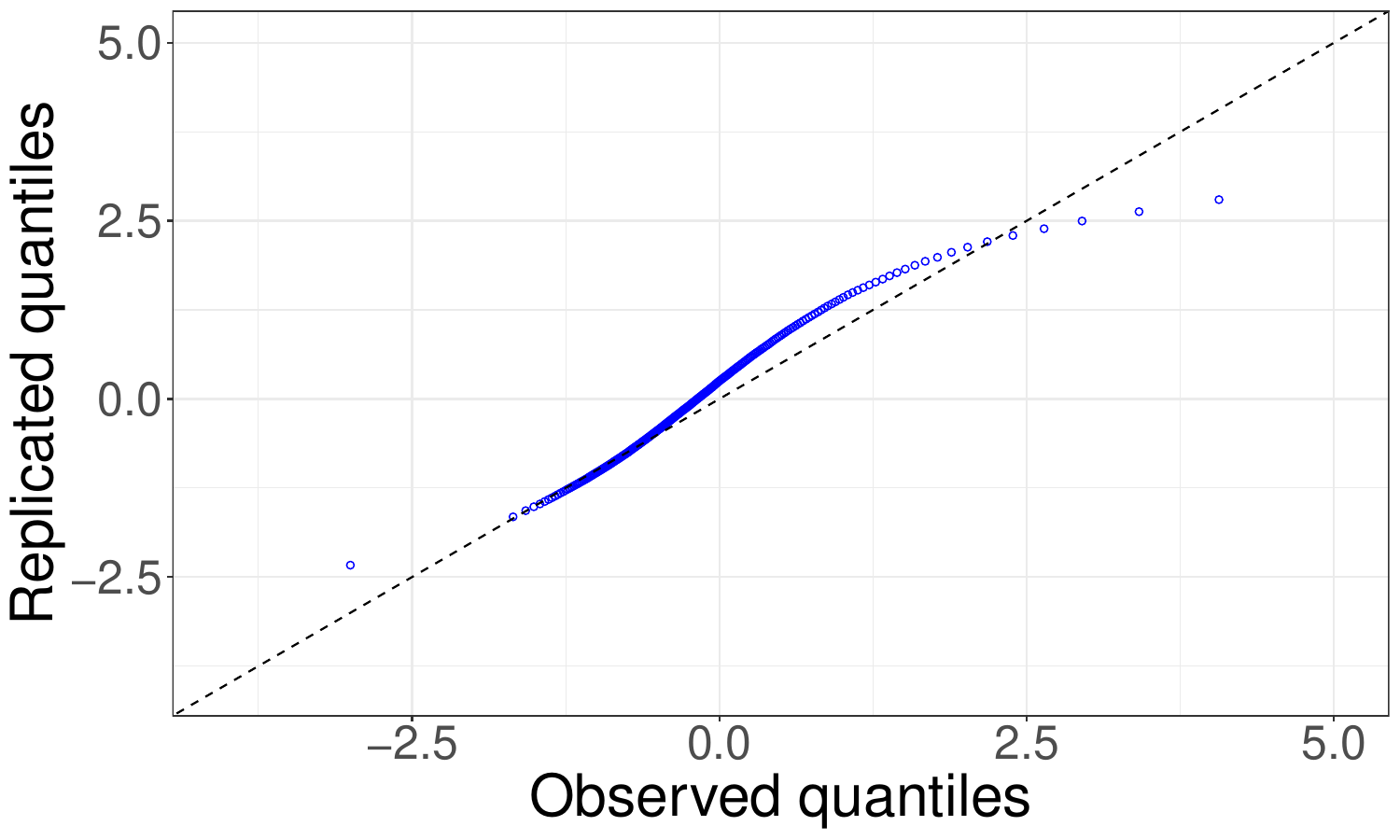}}}
\subfloat[HDL]{{\includegraphics[width=0.49\textwidth,height=0.2\textheight]{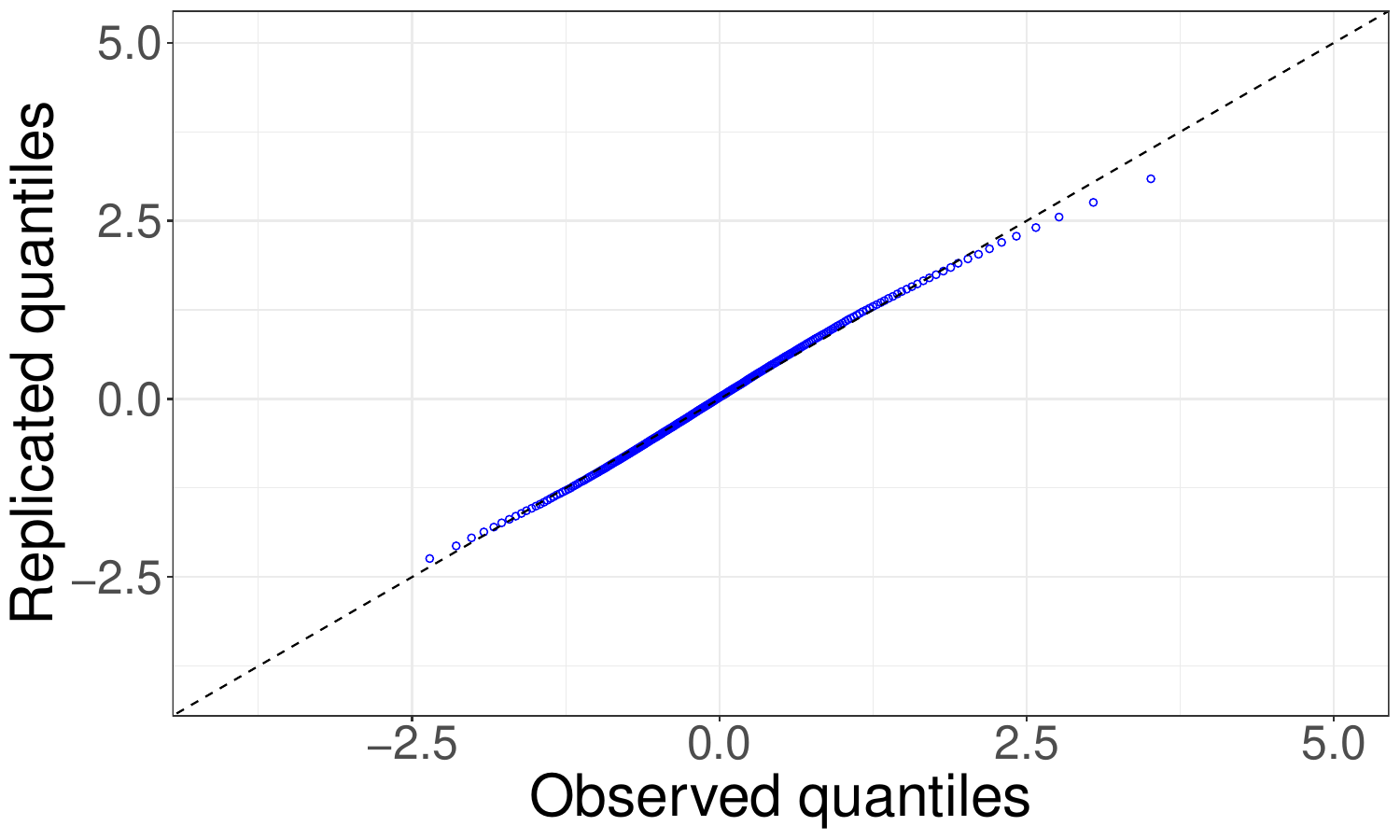}}}

 \subfloat[TRIG]{{\includegraphics[width=0.49\textwidth,height=0.2\textheight]{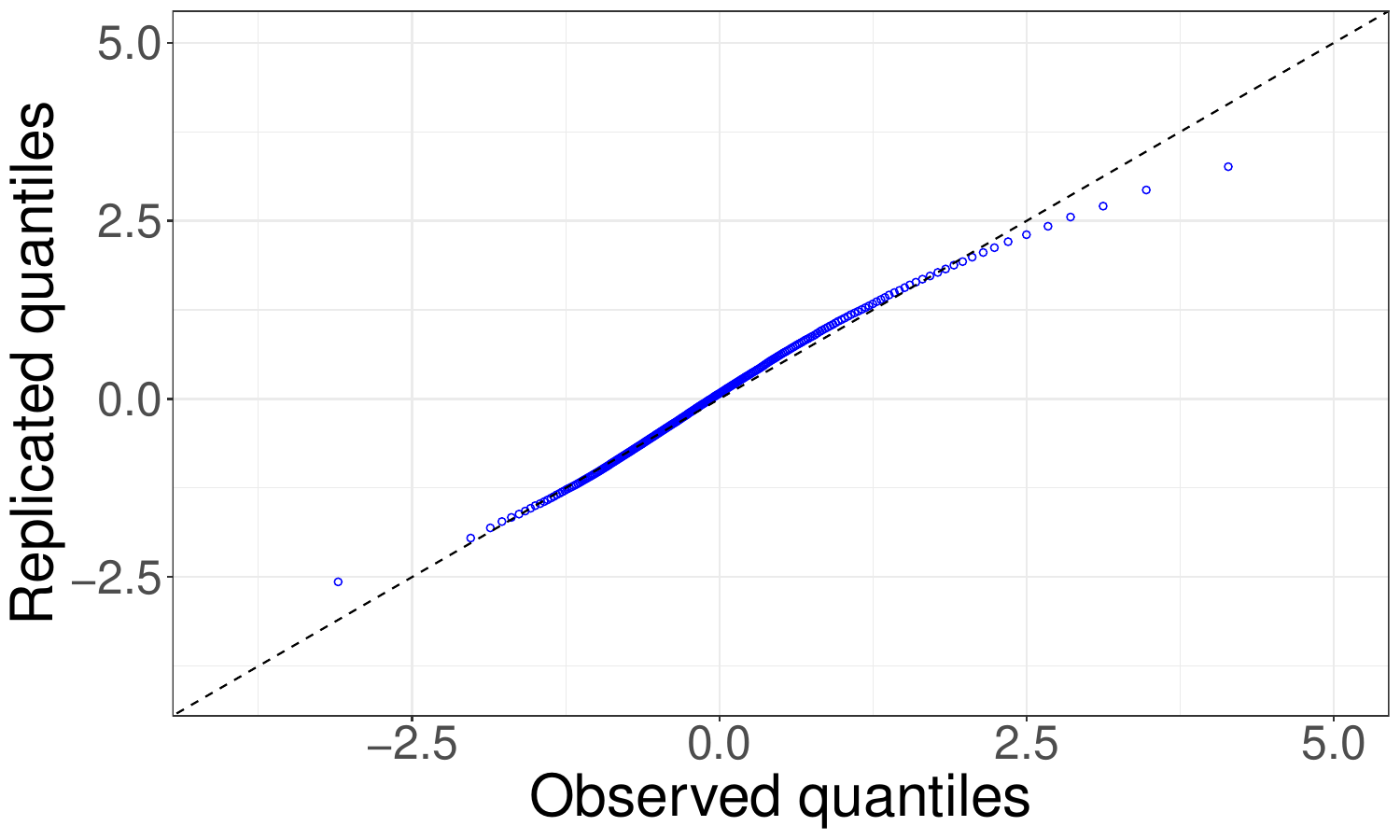}}}
    \caption{Standardized Residuals: Observed vs. Replicated (Women)}
\end{figure}

\newpage
\subsection{Predicted Mean Trajectories of CVD Risk Factors} \label{supp:Predicted_Mean}
\begin{figure}[!ht]
    \centering
\subfloat[SBP]{{\includegraphics[width=0.49\textwidth,height=0.2\textheight]{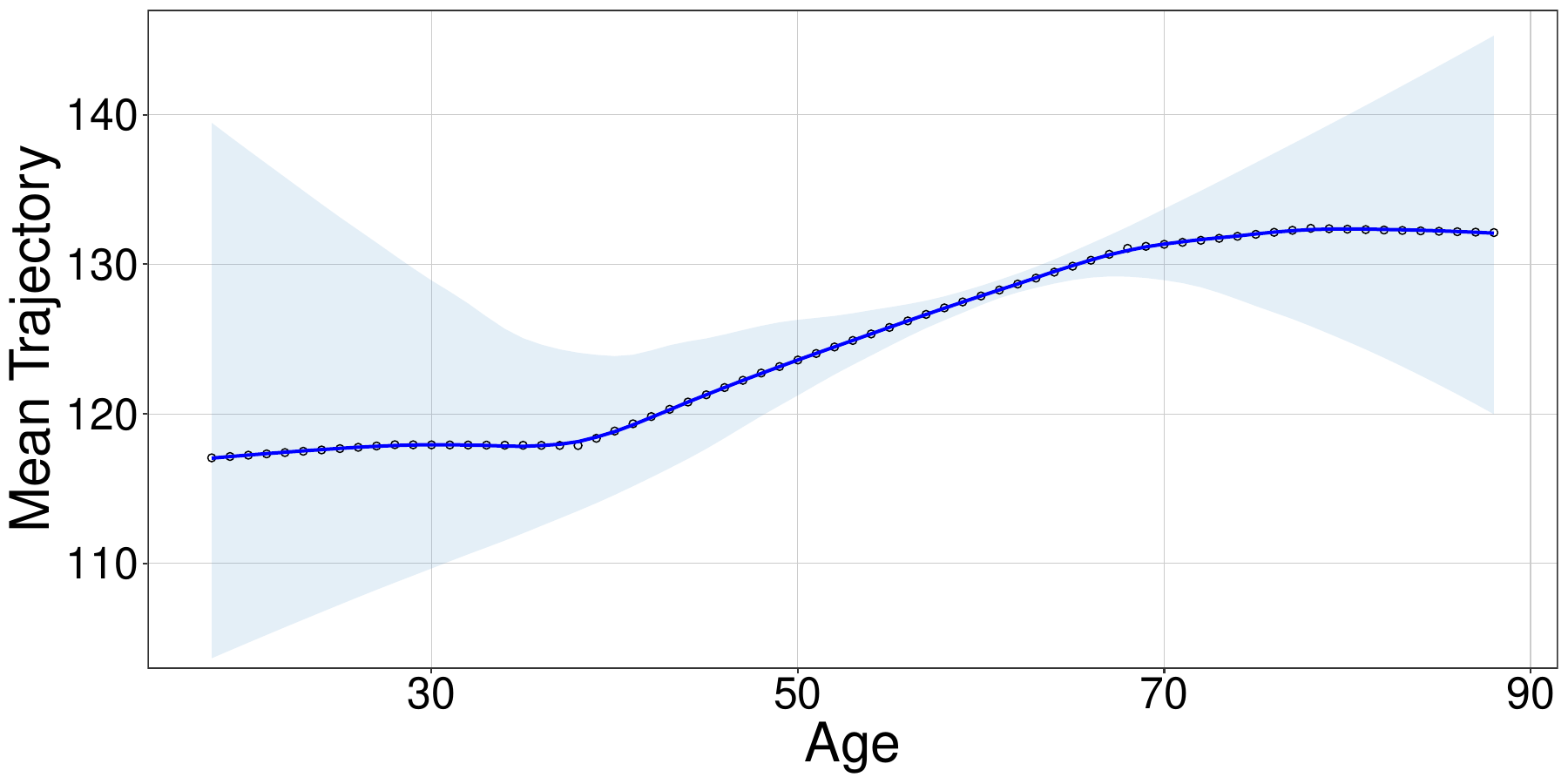}}}
\subfloat[DBP]{{\includegraphics[width=0.49\textwidth,height=0.2\textheight]{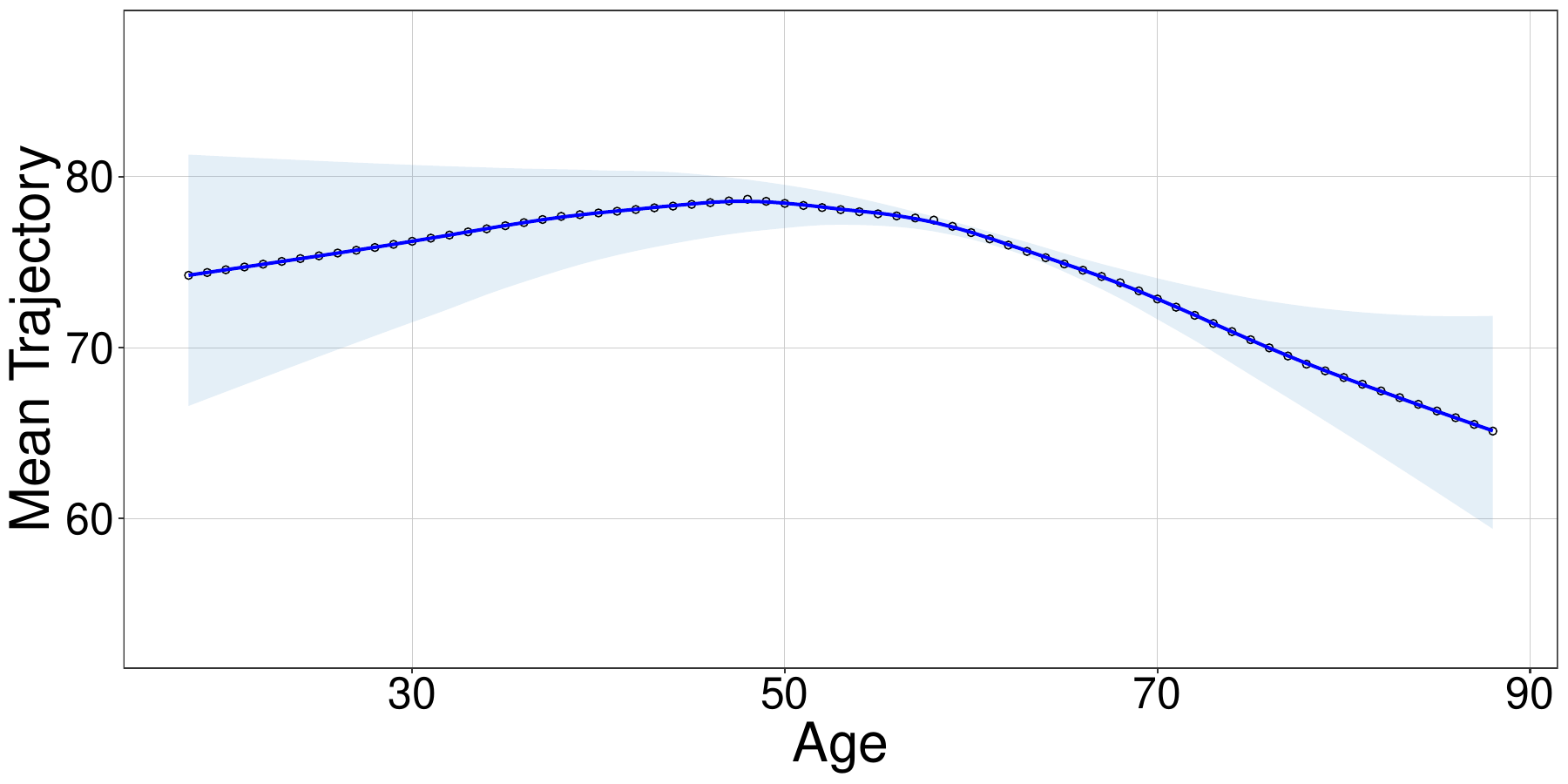}}} 

\subfloat[BMI]{{\includegraphics[width=0.49\textwidth,height=0.2\textheight]{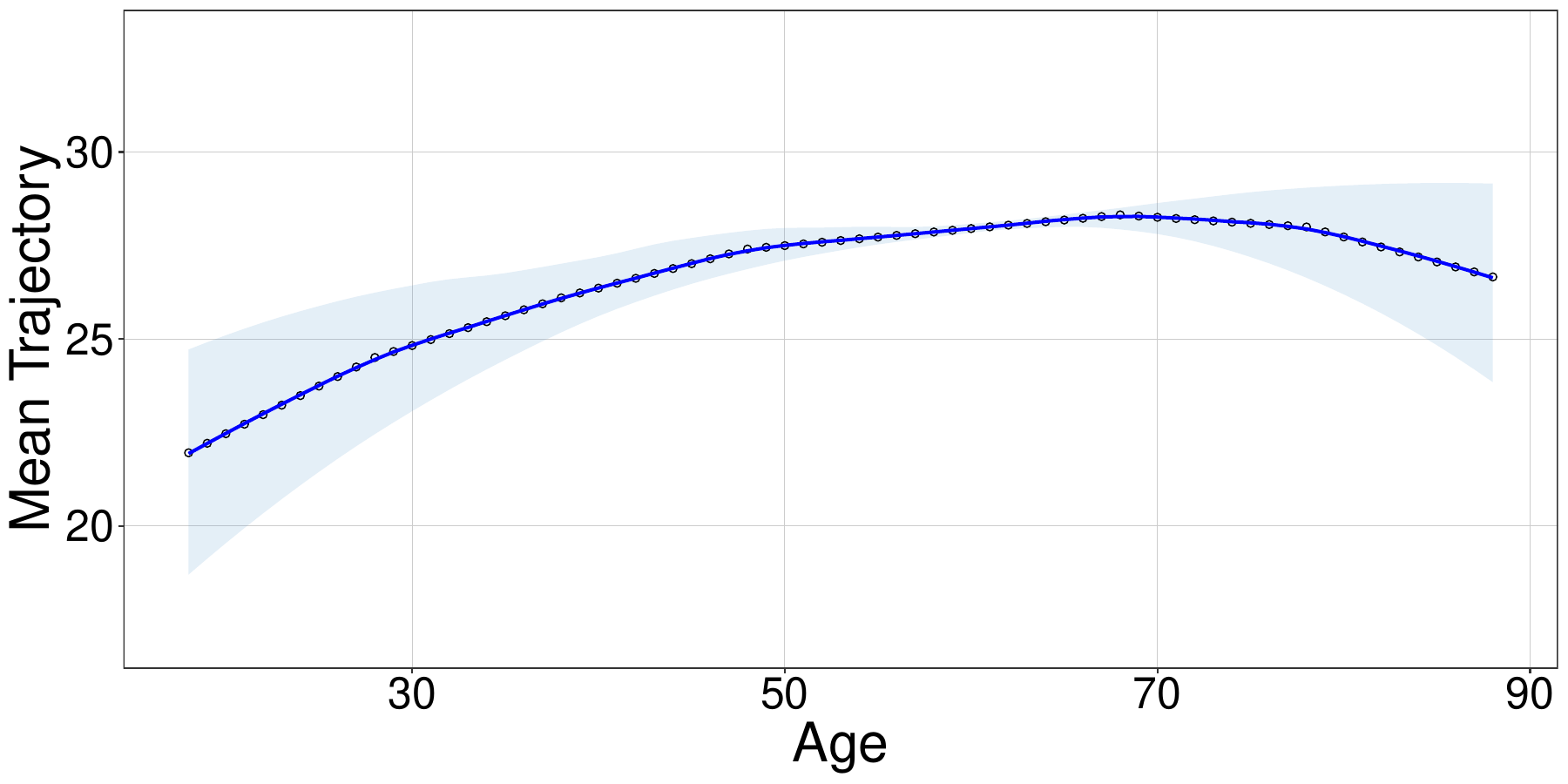}}}
\subfloat[TOTCHL]{{\includegraphics[width=0.49\textwidth,height=0.2\textheight]{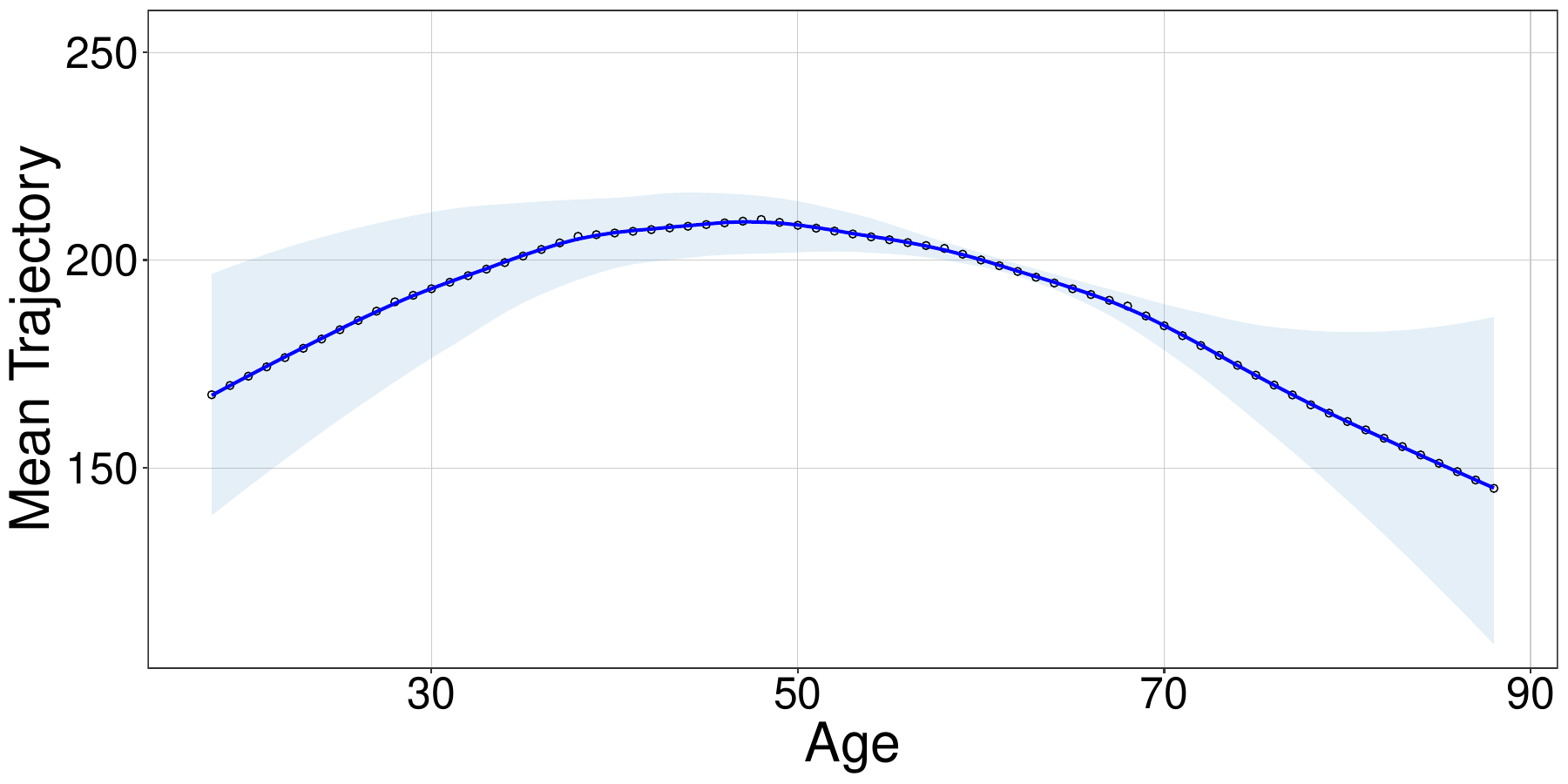}}} 

\subfloat[GLU]{{\includegraphics[width=0.49\textwidth,height=0.2\textheight]{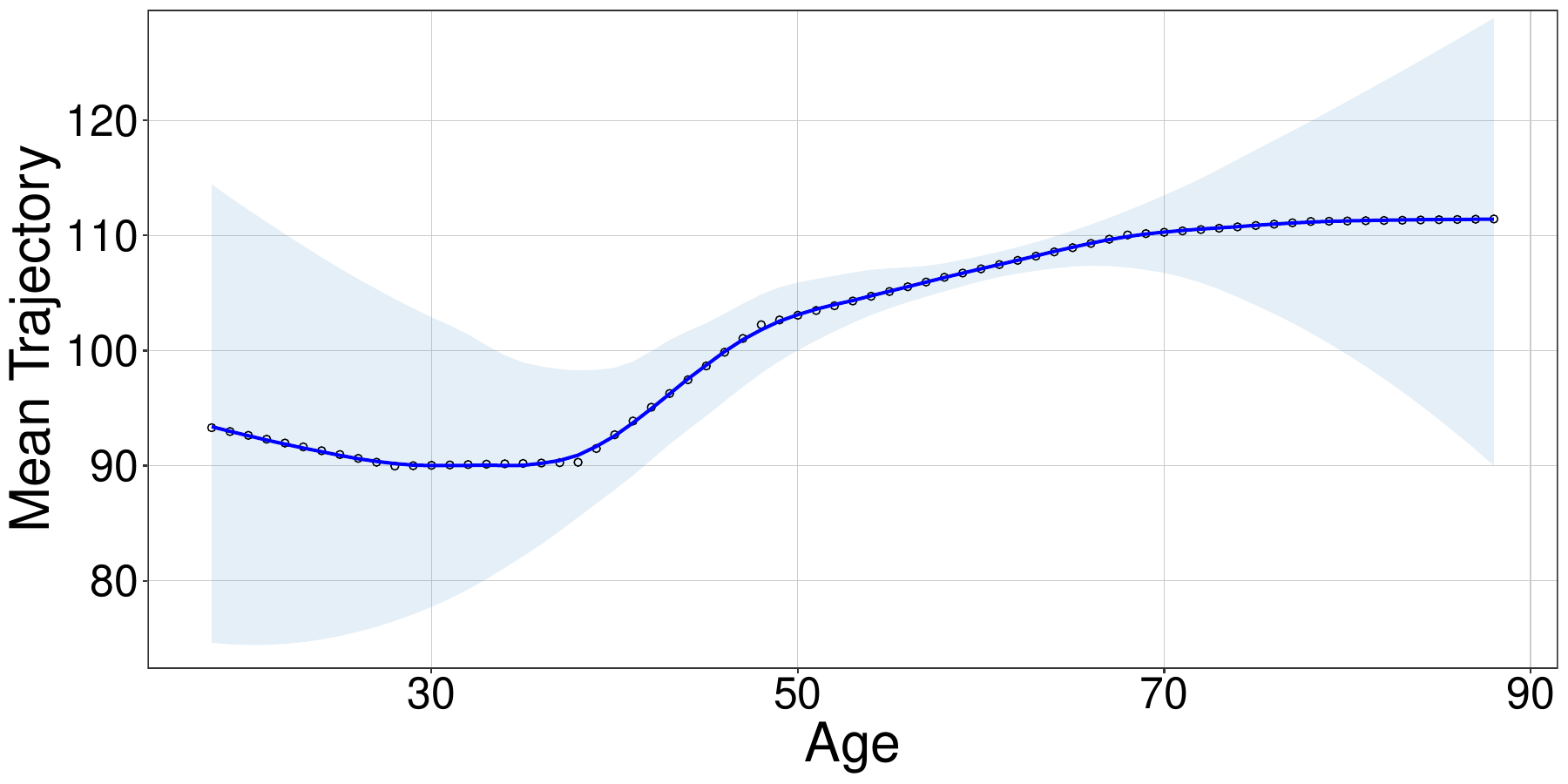}}}
\subfloat[HDL]{ {\includegraphics[width=0.49\textwidth,height=0.2\textheight]{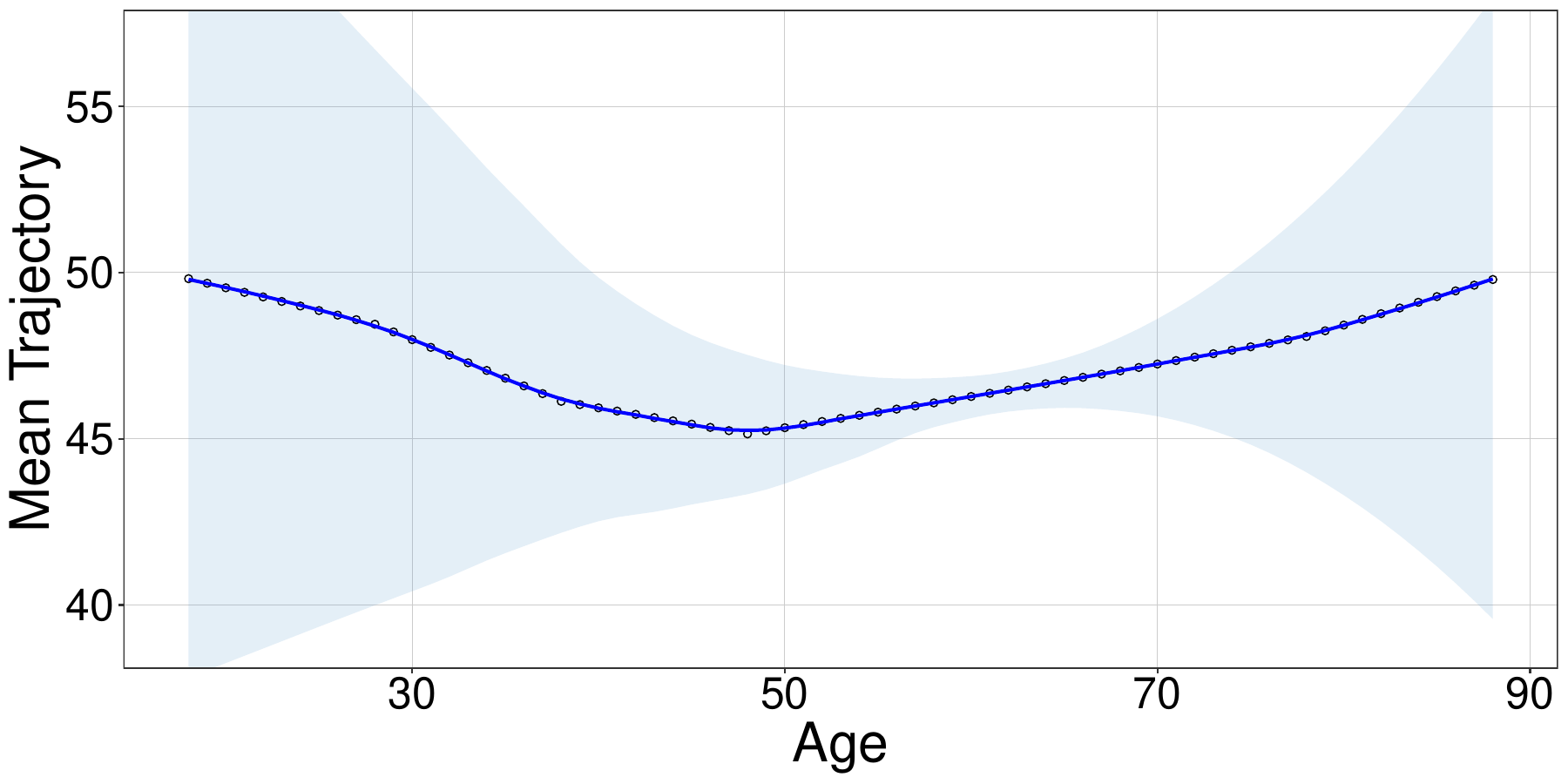}}} 

 \subfloat[TRIG]{{\includegraphics[width=0.49\textwidth,height=0.2\textheight]{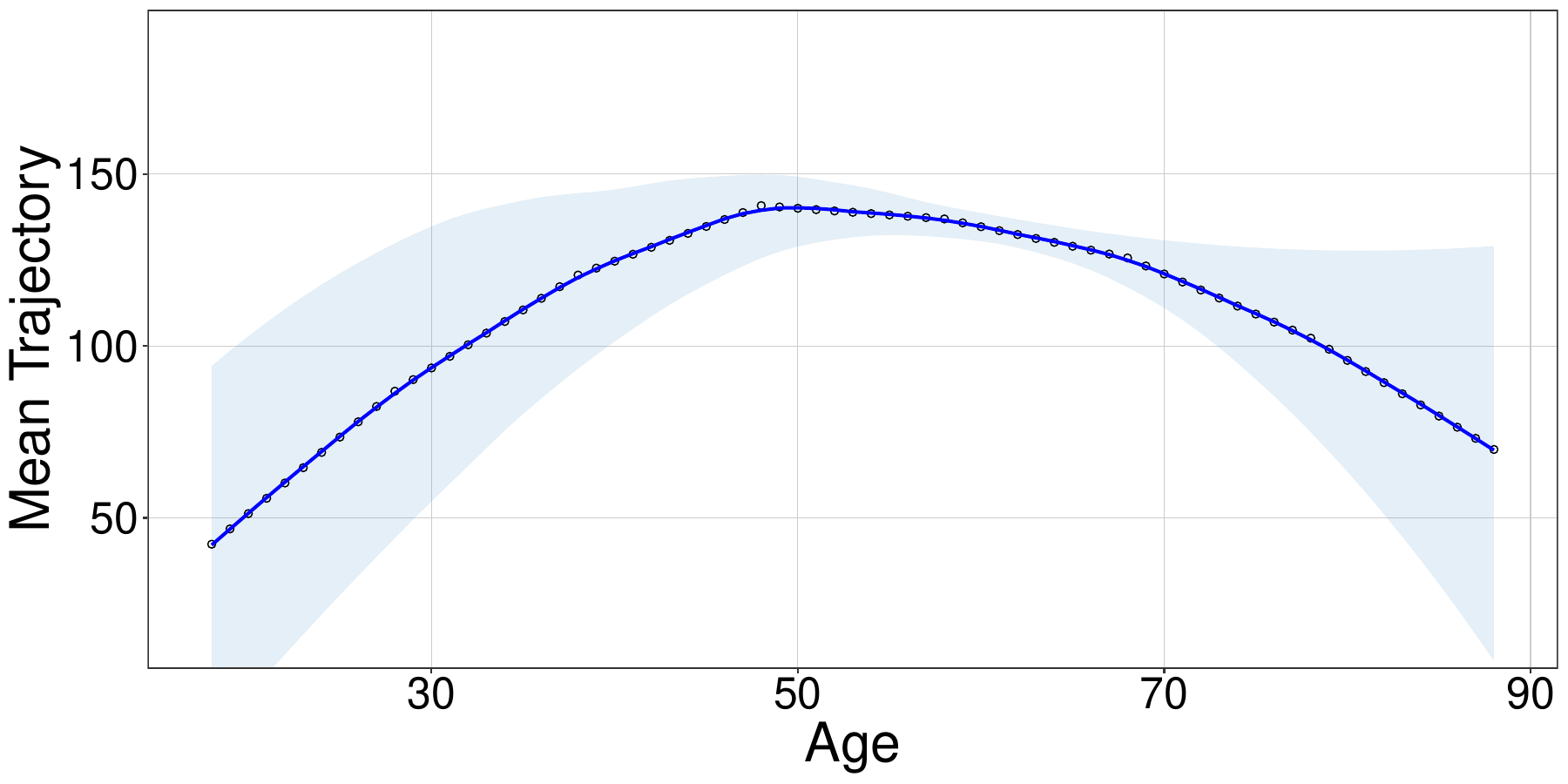}}}
    \caption{Predicted mean trajectories with 95\% credible intervals for seven CVD risk factors (Men)}
    \label{fig:mean_traj_men}
\end{figure}

\begin{figure}[!ht]
    \centering
\subfloat[SBP]{{\includegraphics[width=0.49\textwidth,height=0.2\textheight]{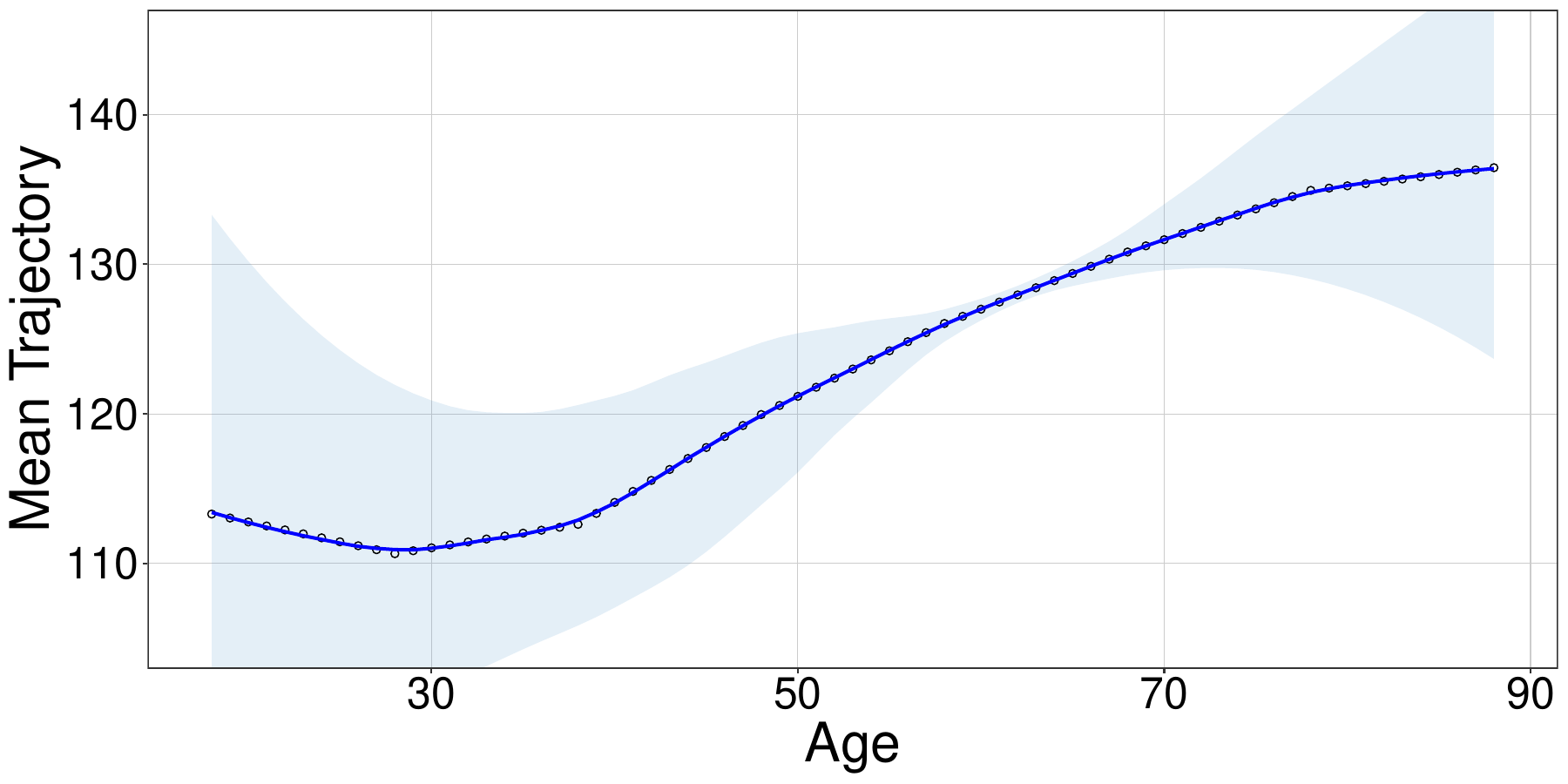}}}
\subfloat[DBP]{{\includegraphics[width=0.49\textwidth,height=0.2\textheight]{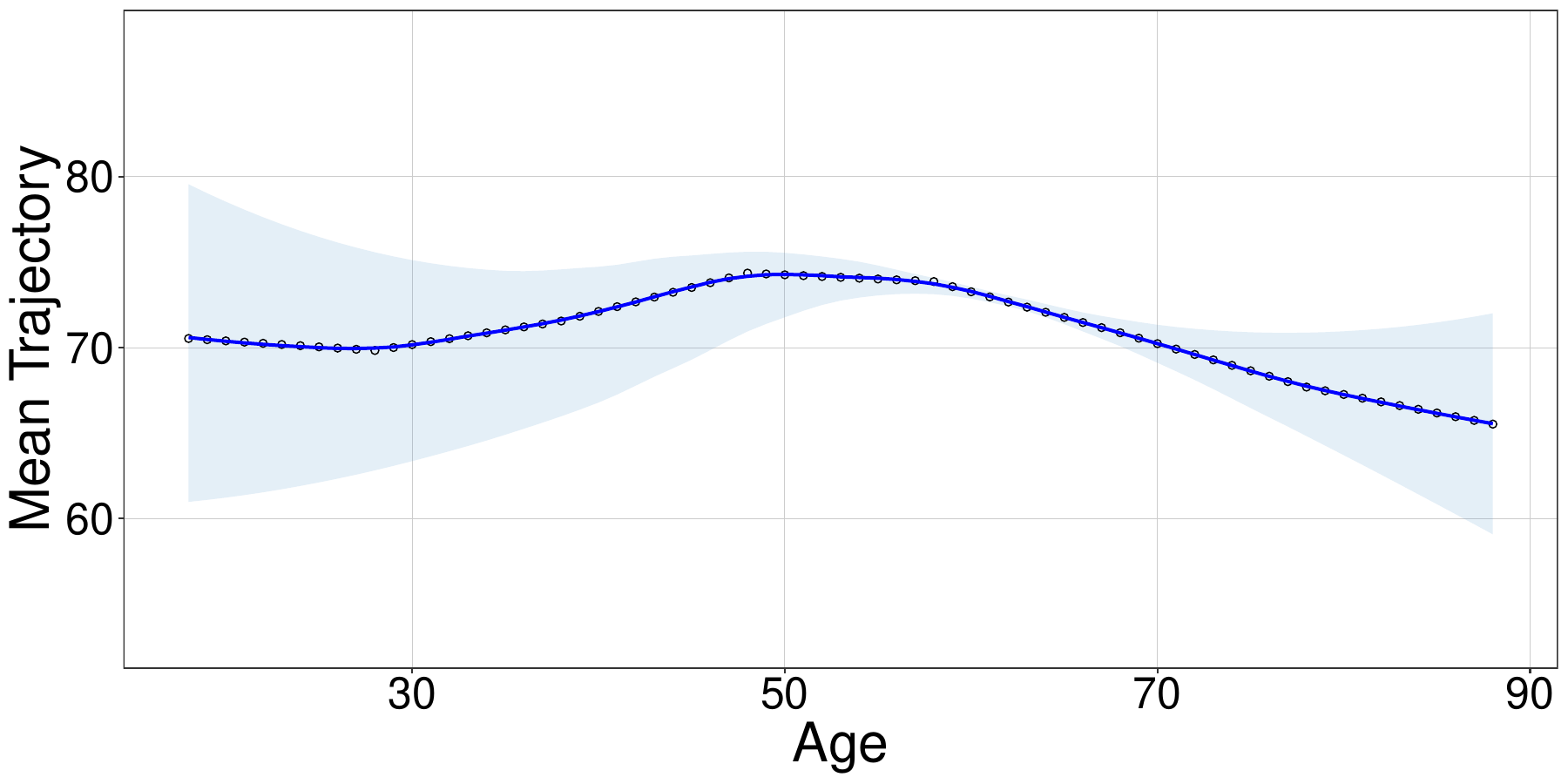}}}

\subfloat[BMI]{{\includegraphics[width=0.49\textwidth,height=0.2\textheight]{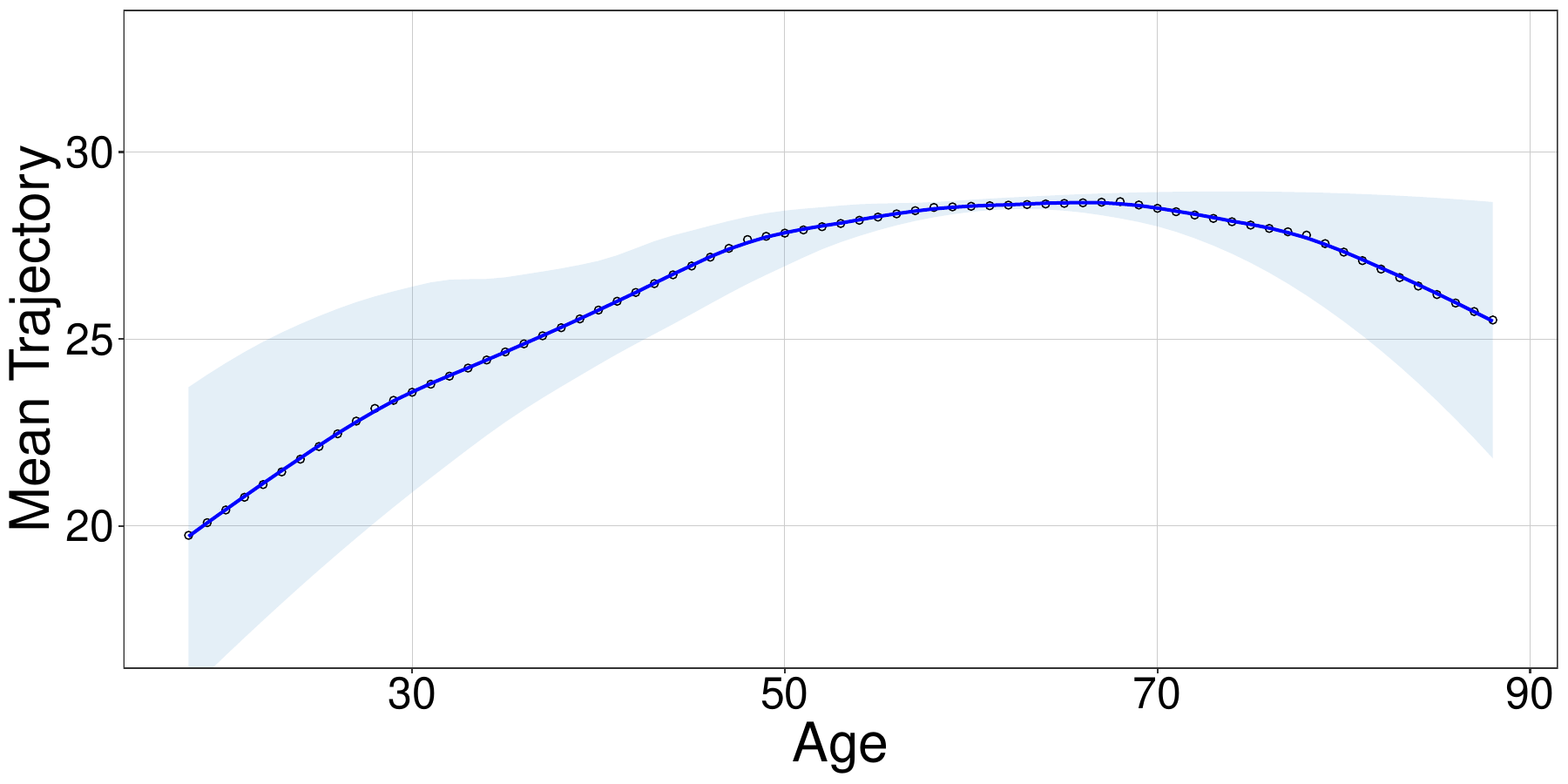}}}
\subfloat[TOTCHL]{{\includegraphics[width=0.49\textwidth,height=0.2\textheight]{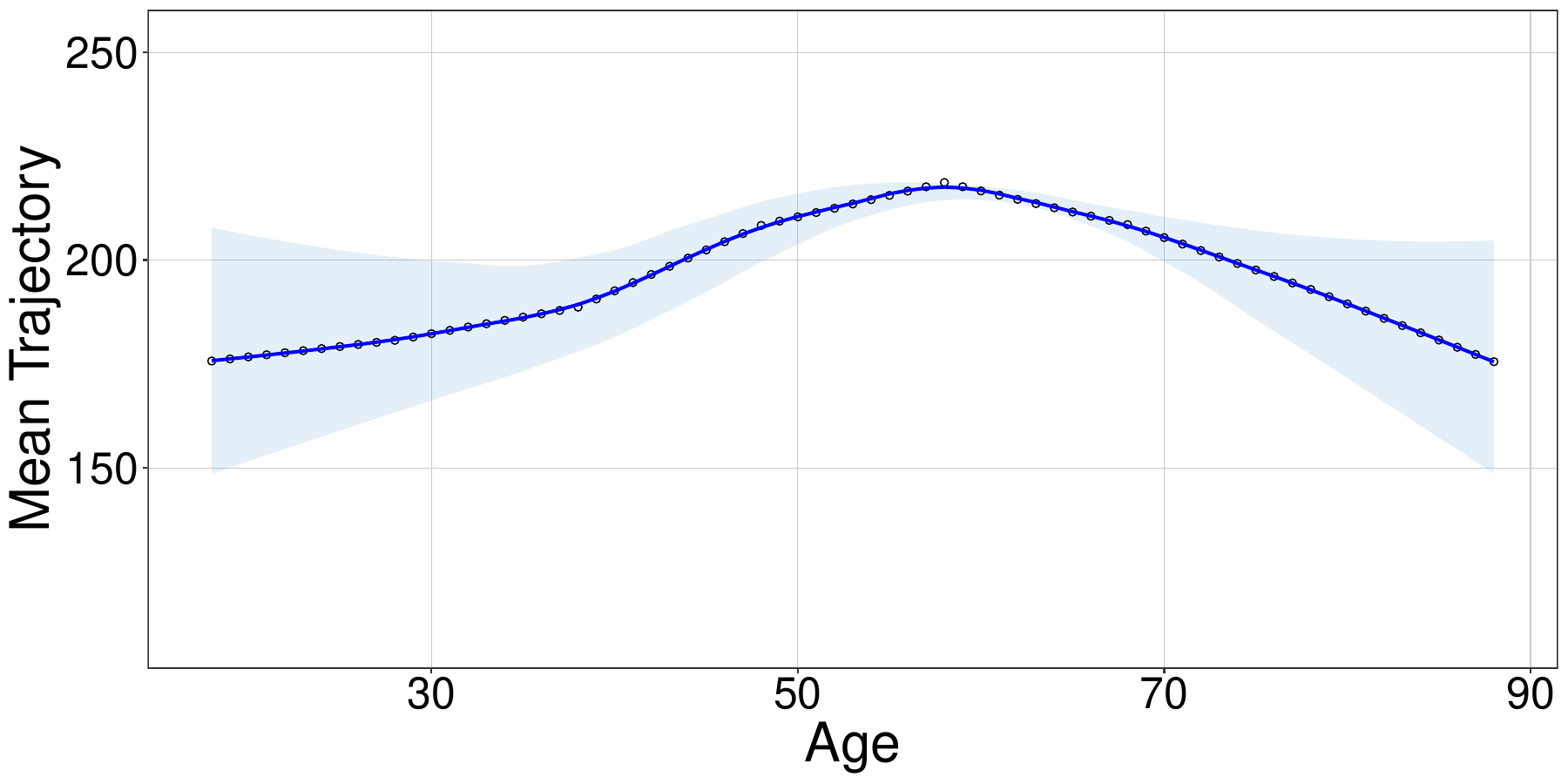}}}

\subfloat[GLU]{{\includegraphics[width=0.49\textwidth,height=0.2\textheight]{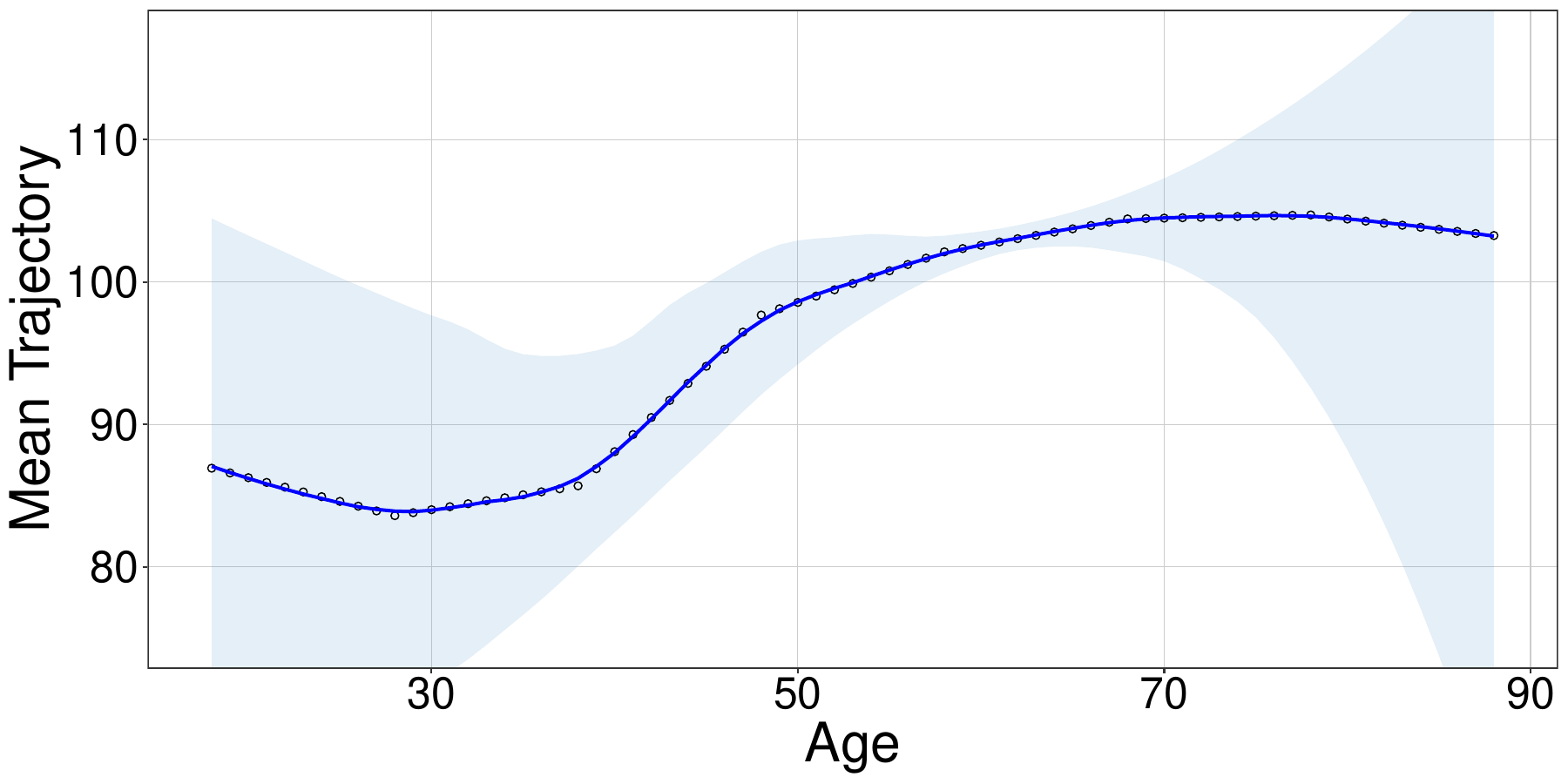}}}
\subfloat[HDL]{{\includegraphics[width=0.49\textwidth,height=0.2\textheight]{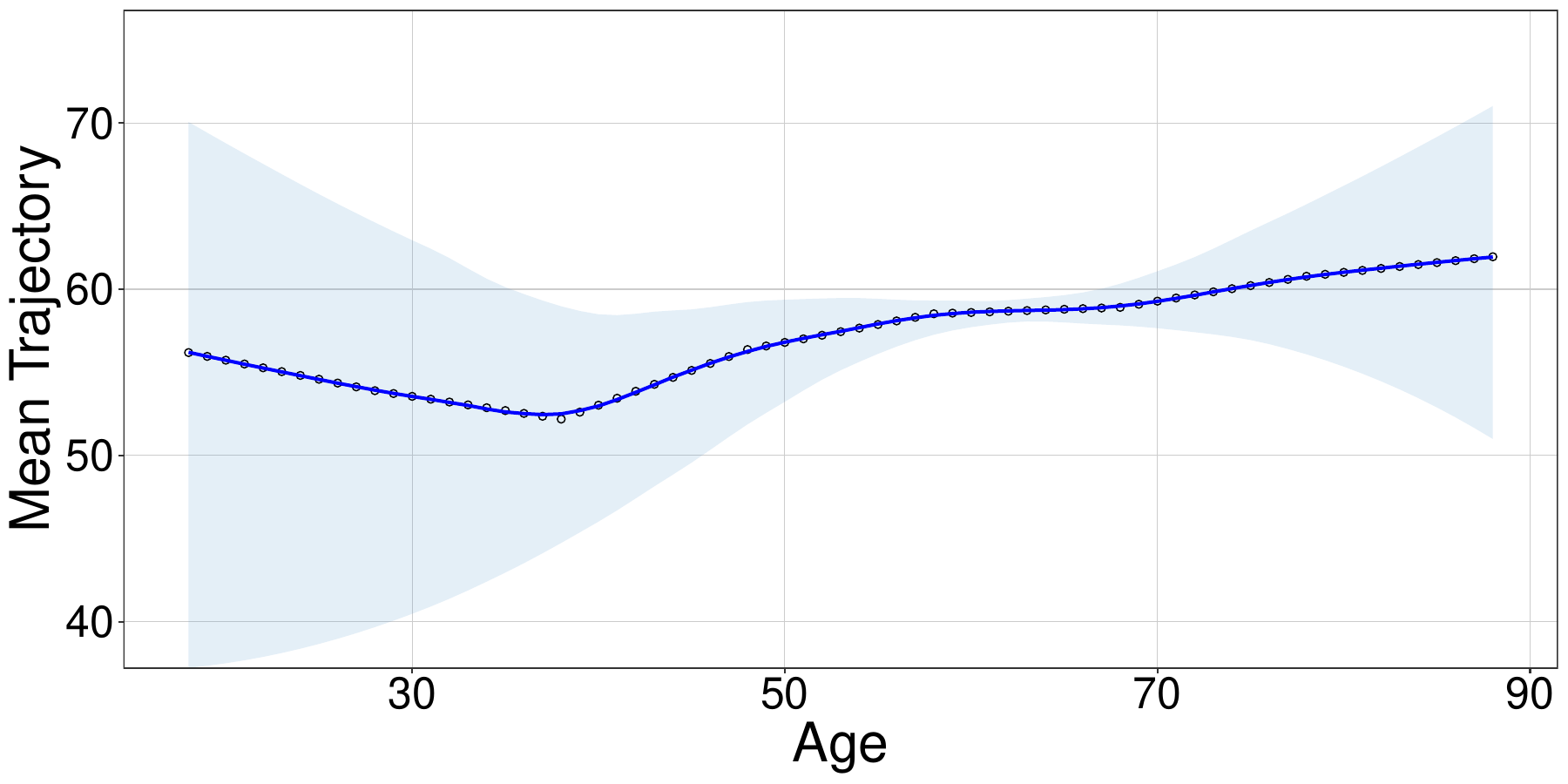}}}

 \subfloat[TRIG]{{\includegraphics[width=0.49\textwidth,height=0.2\textheight]{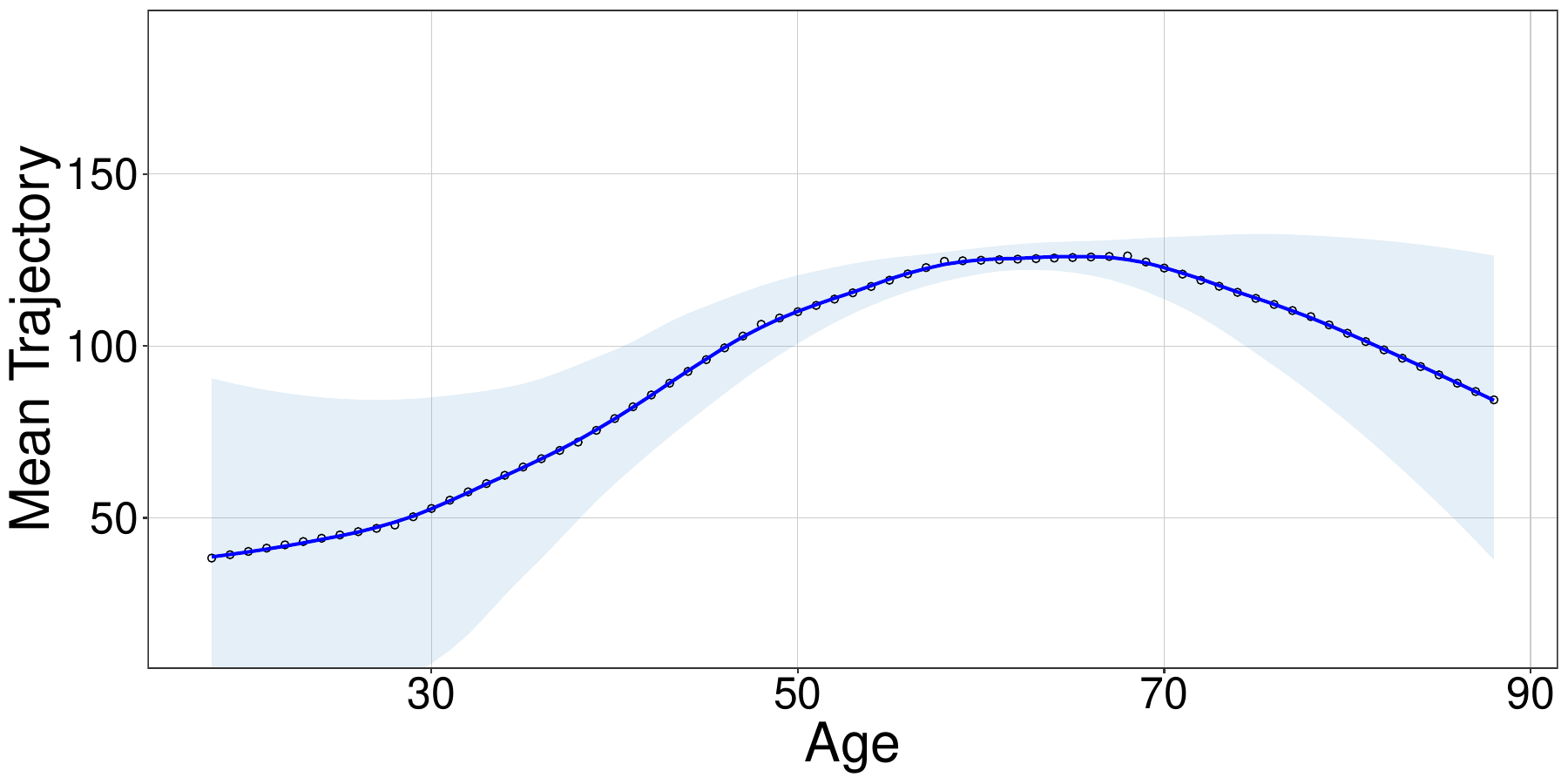}}}
    \caption{Predicted mean trajectories with 95\% credible intervals for seven CVD risk factors (Women)}
    \label{fig:mean_traj_women}
\end{figure}

We obtained predicted mean trajectories for each cardiovascular risk factor using posterior means from the proposed longitudinal model. 
For each participant, their posterior mean was calculated at ages 18–88~years conditional on observed baseline covariates (race, education, and birth year), and the resulting trajectories were averaged to yield population-level trends separately for men and women. 
The goal of this analysis was to assess whether our model captures realistic, biologically plausible age patterns and known sex-differences in cardiovascular risk factors when applied to the full data.
Figures~\ref{fig:mean_traj_men} and \ref{fig:mean_traj_women} display the estimated mean trajectories for the seven CVD risk factors, with 95\% credible intervals that reflect uncertainty and vary across age in accordance with the differing levels of information available across the age span. 

For men, the predicted trajectories follow expected physiological patterns across adulthood.
SBP rises steadily with age, showing a marked increase through midlife and a modest plateau in later life.
DBP increases until middle age and declines thereafter, consistent with arterial stiffening and widening pulse pressure at older ages.
BMI increases from early adulthood to around age~60 and gradually declines afterward, reflecting age-related changes in body composition.
TOTCHL and TRIG peak around midlife and decline in older ages, consistent with lipid metabolism changes and the widespread use of lipid-lowering medication.
GLU increases monotonically across the age span, a pattern that may partly reflect weight gain and age-related metabolic changes.
HDL exhibits a modest U-shaped pattern, decreasing through midlife and increasing again in older adulthood; its opposing pattern with TRIG is expected physiologically and often viewed as a favorable lipid profile.
Declines in several risk factors at older ages are also consistent with survival-related selection, particularly among individuals who died during follow-up.

For women, the trajectories show similar overall patterns but with characteristic sex differences.
Women exhibit lower SBP, DBP, and TRIG levels and higher HDL across ages compared with men.
The age-related rise in GLU is more gradual among women, while BMI and TOTCHL reach their peak slightly later in life.
HDL shows a more pronounced midlife dip among women, consistent with hormonal changes surrounding the menopausal transition, followed by an age-related rebound that may also be influenced by modest survival-related selection.
These trends are consistent with prior findings from large cohort studies and align closely with the empirically observed LRPP data \citep{franklin1997hemodynamic, wilkins2015data, lloyd2021coronary}, supporting the adequacy and interpretability of the mean model.

Overall, the predicted mean trajectories demonstrate that the model captures realistic, biologically plausible age patterns and sex-specific differences in cardiovascular risk factors, confirming both the flexibility and interpretability of the proposed modeling framework. The wider credible intervals for HDL, TRIG, and GLU at younger and older ages reflect the sparse cohort coverage and intermittent exam-level measurement of these risk factors across the LRPP, as detailed in Section \ref{supp:Missing_and} and Table \ref{Missing}.

\newpage
\subsection{Variance Ratios} \label{supp:Variance_Ratios}
\begin{figure}[!ht]
    \centering
    \subfloat[Men]{\includegraphics[width=13cm,height=9.5cm]{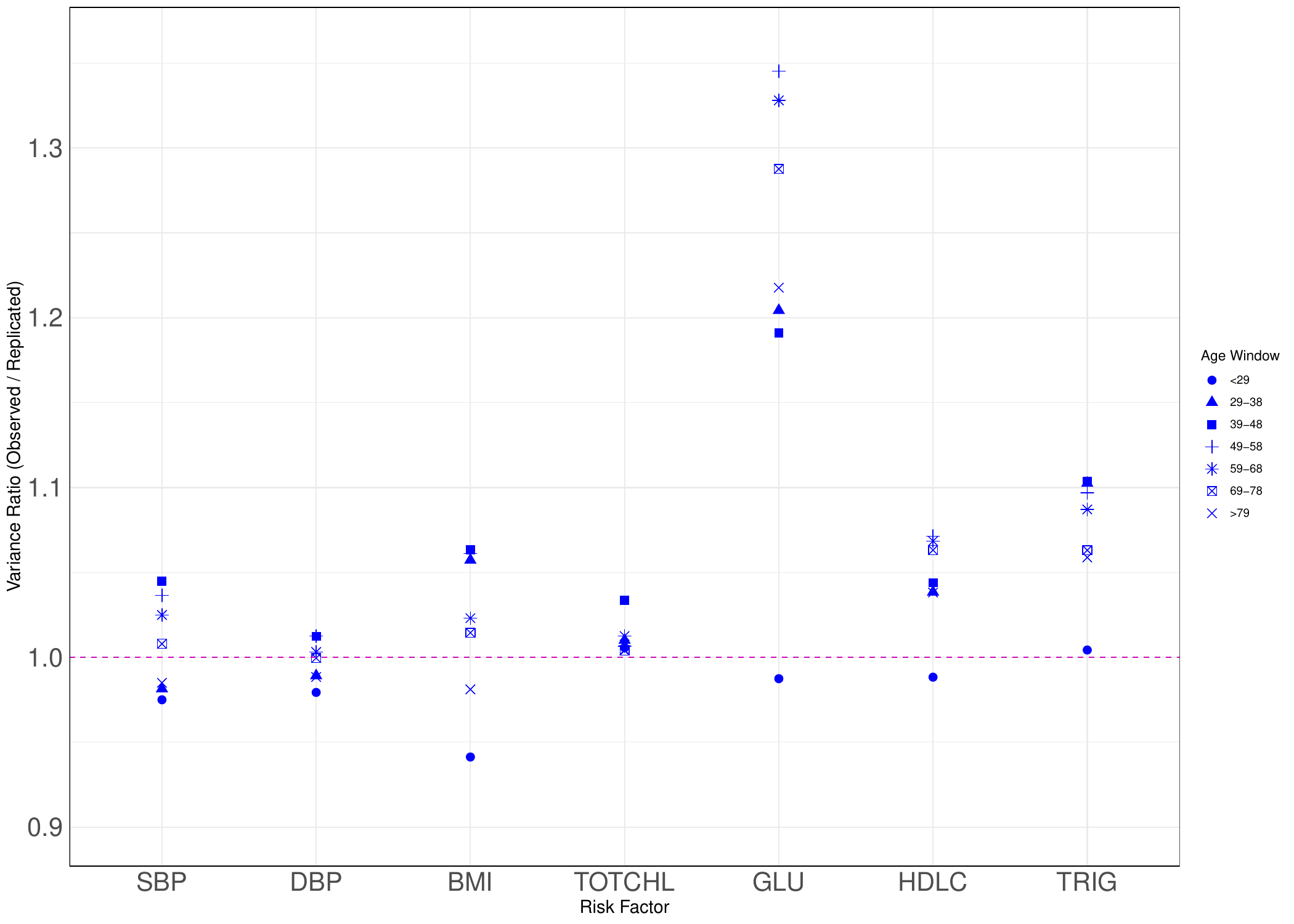}}

    \subfloat[Women]{\includegraphics[width=13cm,height=9.5cm]{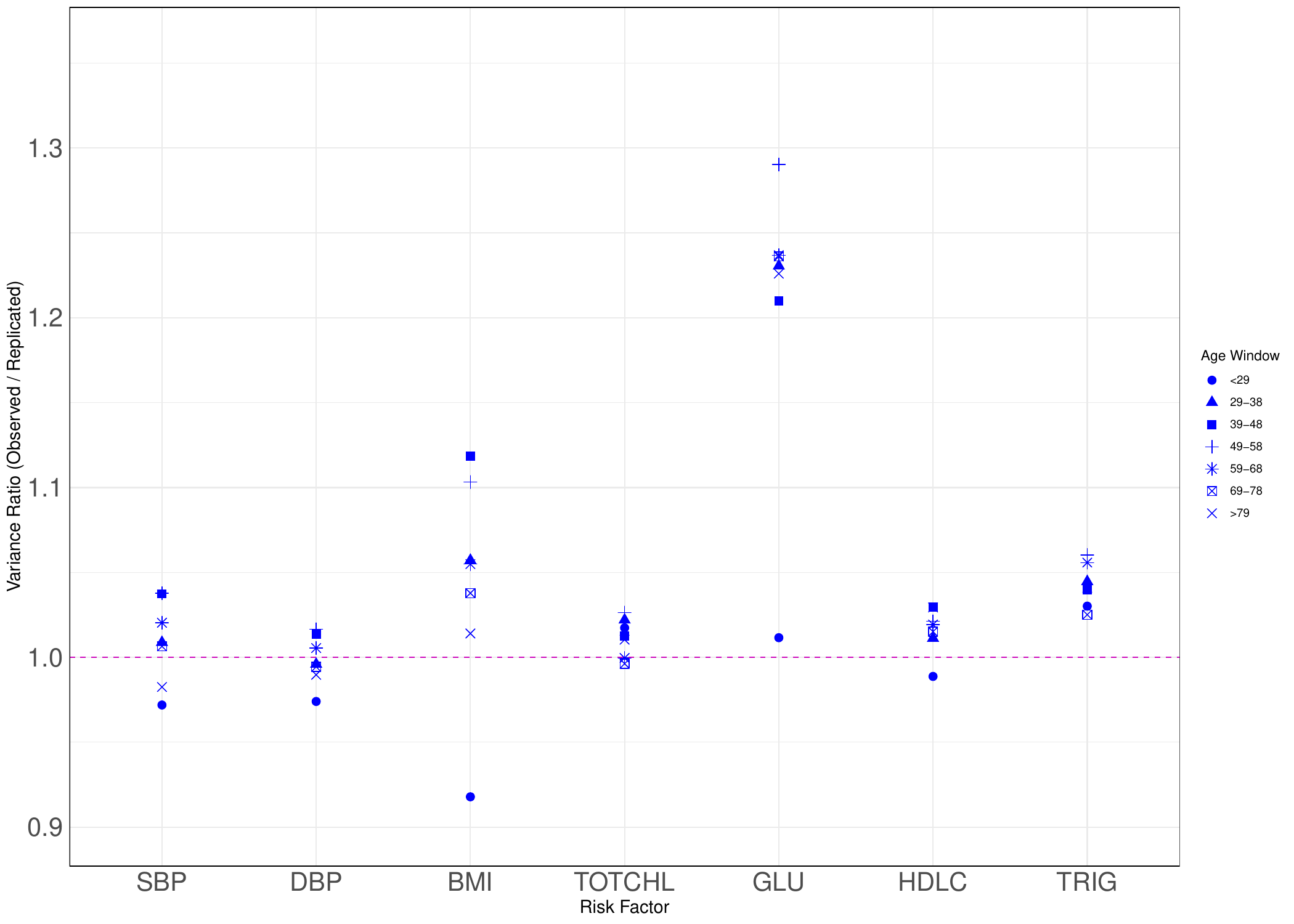}}

    \caption{Variance Ratio for (a) Men and (b) Women by Risk Factor and Age Window}
\end{figure}

\newpage
\subsection{Risk Factor Predictions Beyond Observed Ages in FOS} \label{supp:Risk_Factor}
\begin{figure}[!ht]
    \centering
    {\includegraphics[width=0.46\textwidth,height=0.22\textheight]{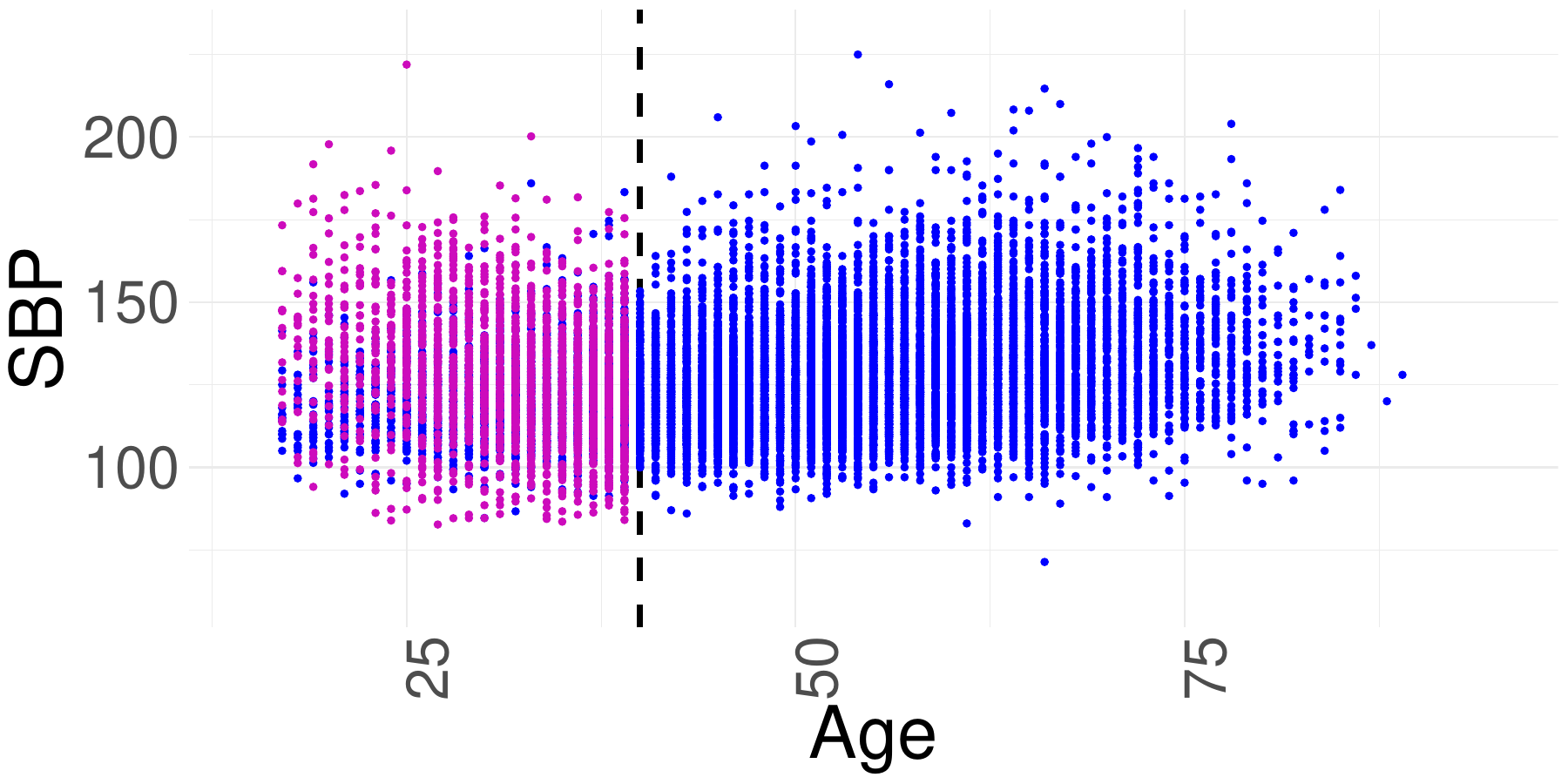}}
    {\includegraphics[width=0.46\textwidth,height=0.22\textheight]{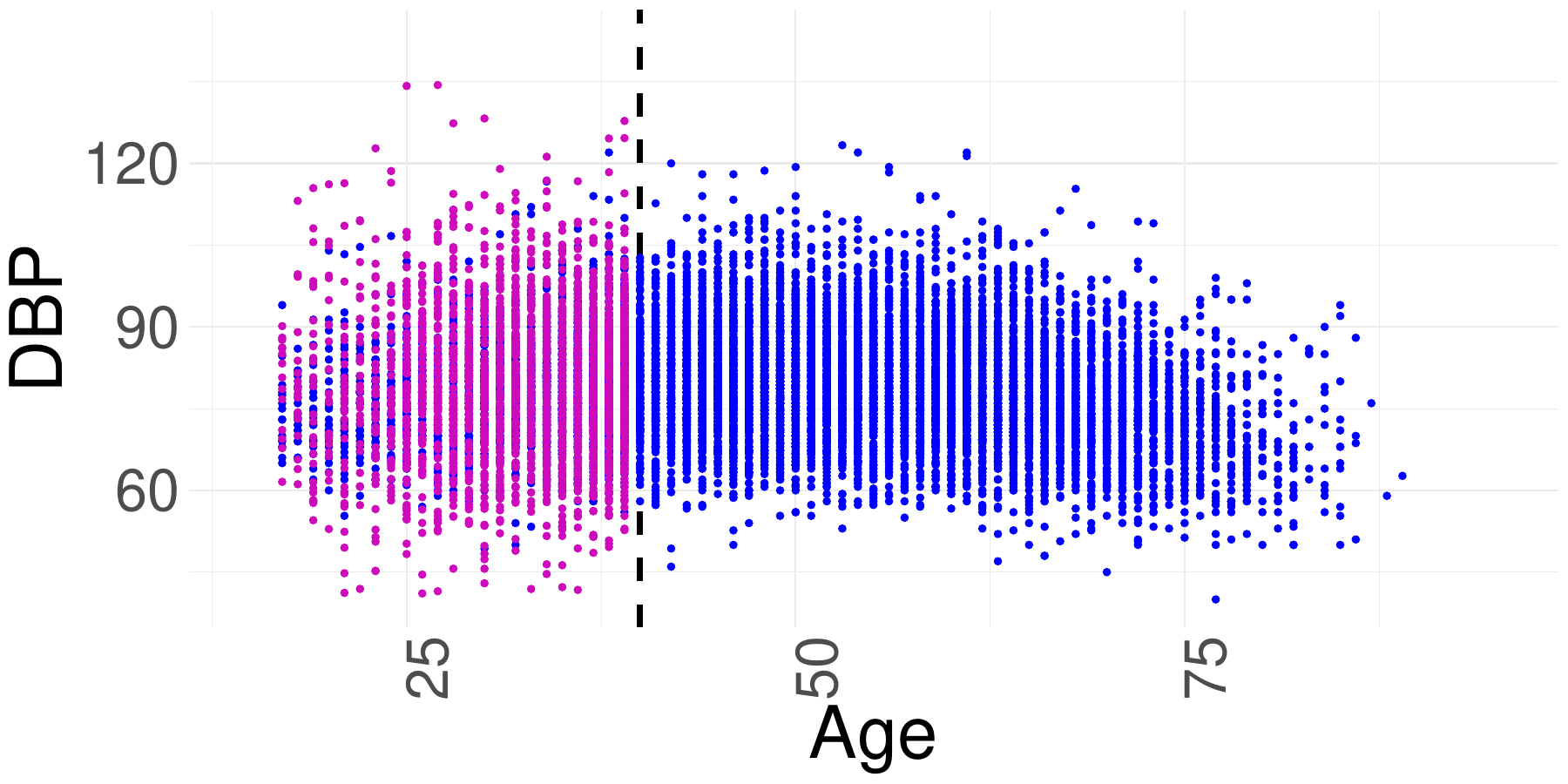}} 

{\includegraphics[width=0.46\textwidth,height=0.22\textheight]{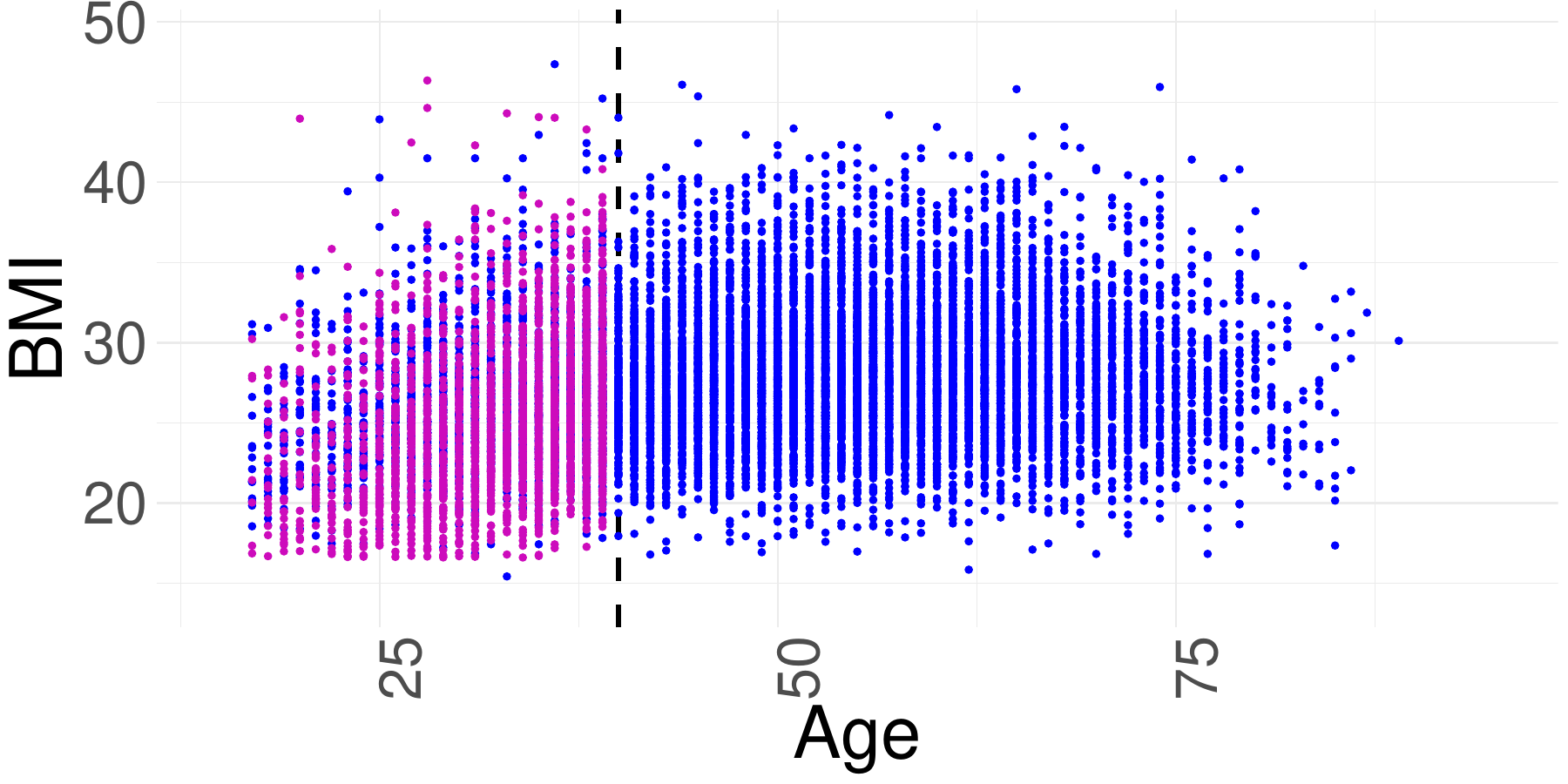}}
{\includegraphics[width=0.46\textwidth,height=0.22\textheight]{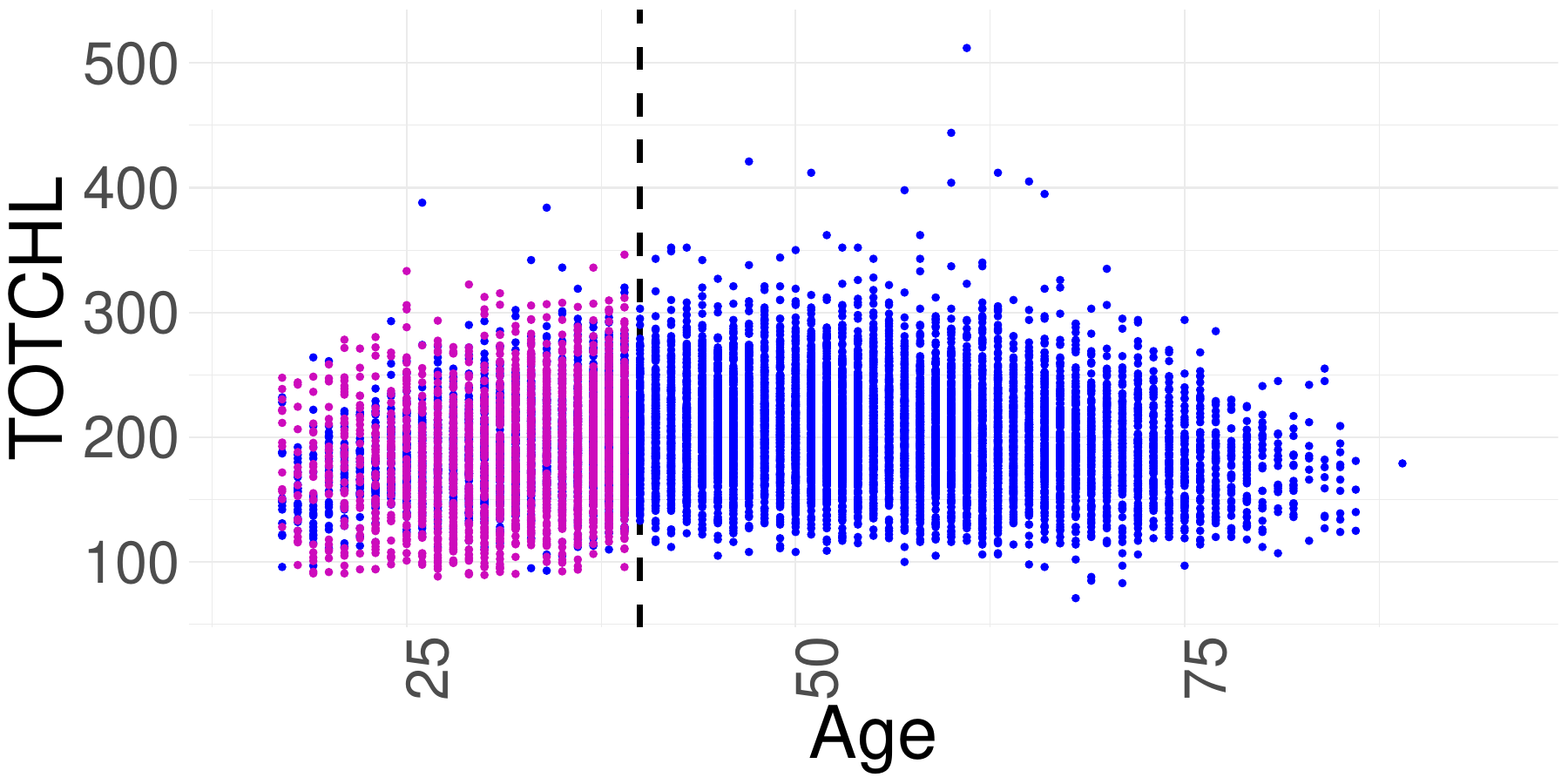}} 

    {\includegraphics[width=0.46\textwidth,height=0.22\textheight]{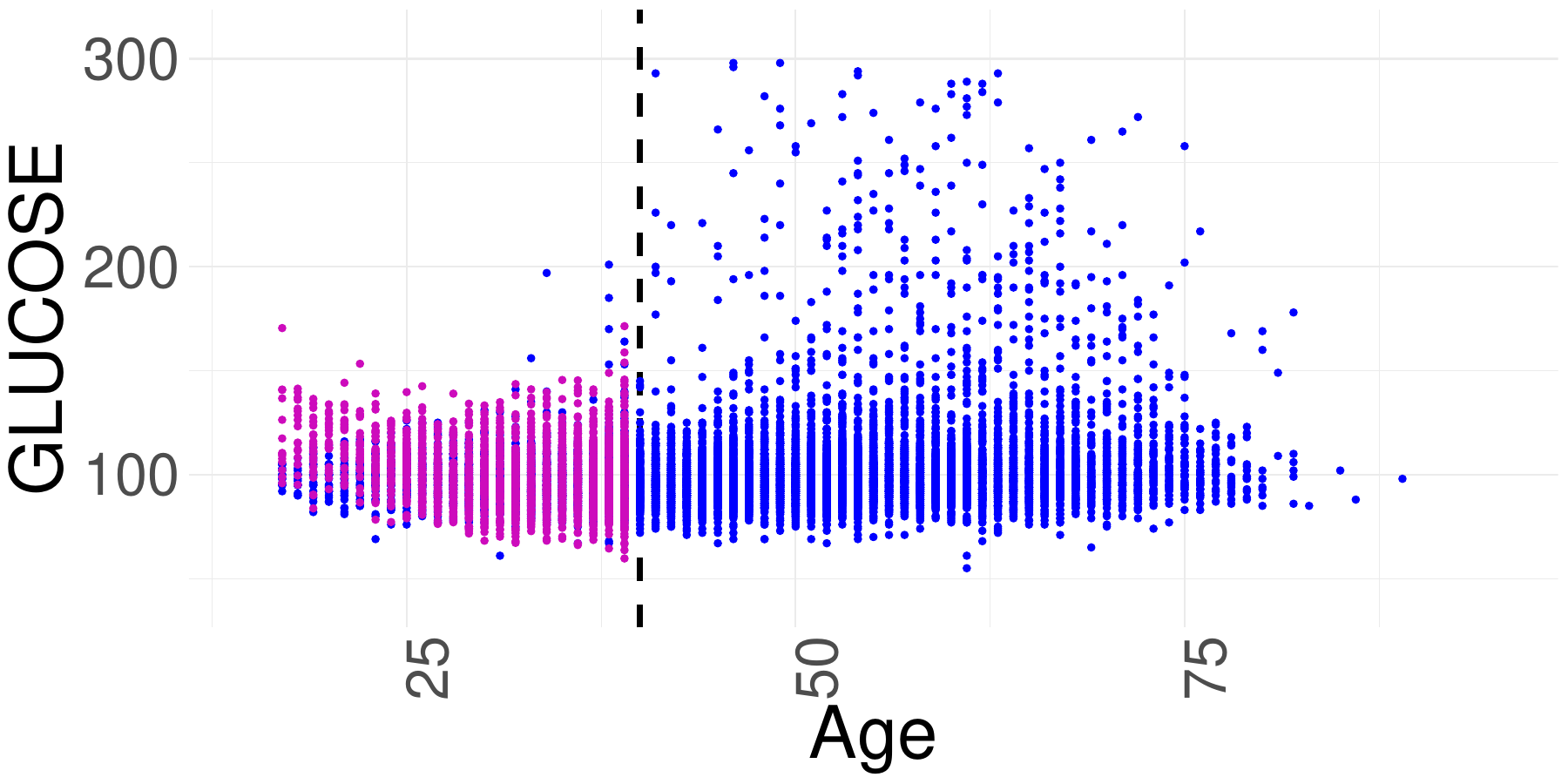}}
    {\includegraphics[width=0.46\textwidth,height=0.22\textheight]{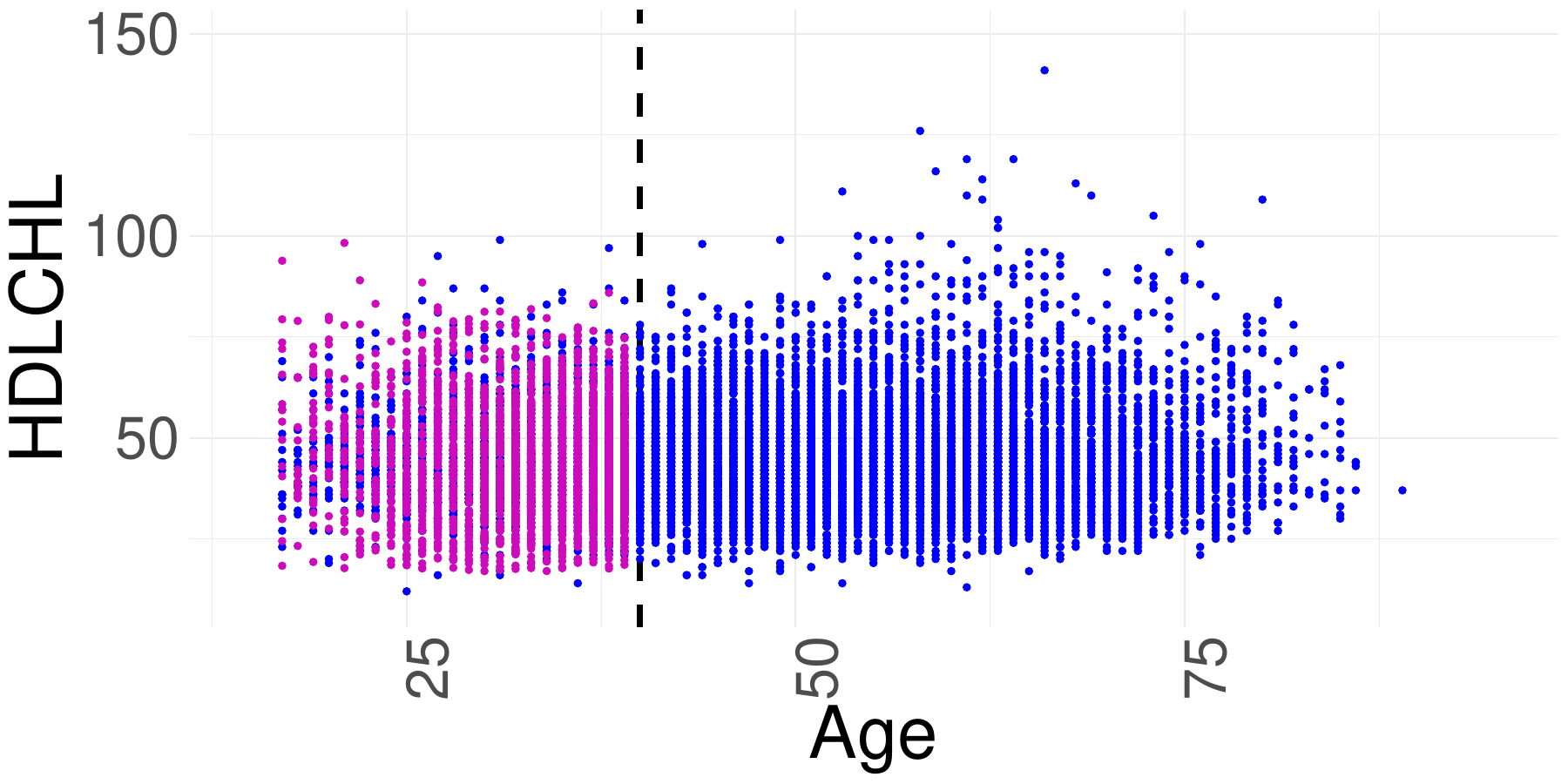}} 

    {\includegraphics[width=0.46\textwidth,height=0.22\textheight]{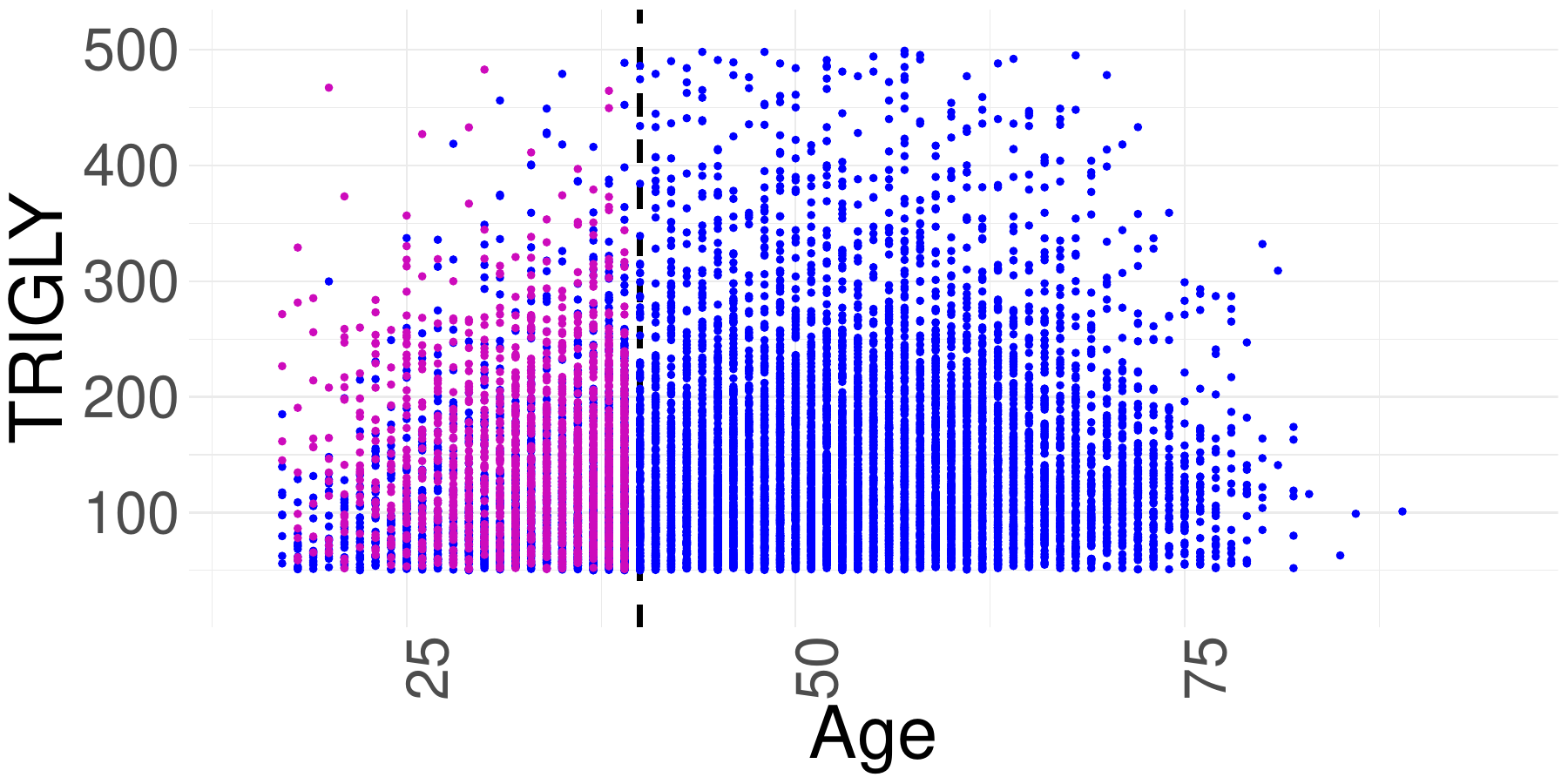}}
    \caption{Imputed (red) and observed (blue) values of risk factors against age for men in FOS.}
    \label{fig:men_L40}
\end{figure}

\begin{figure}[!ht]
    \centering
    {\includegraphics[width=0.46\textwidth,height=0.22\textheight]{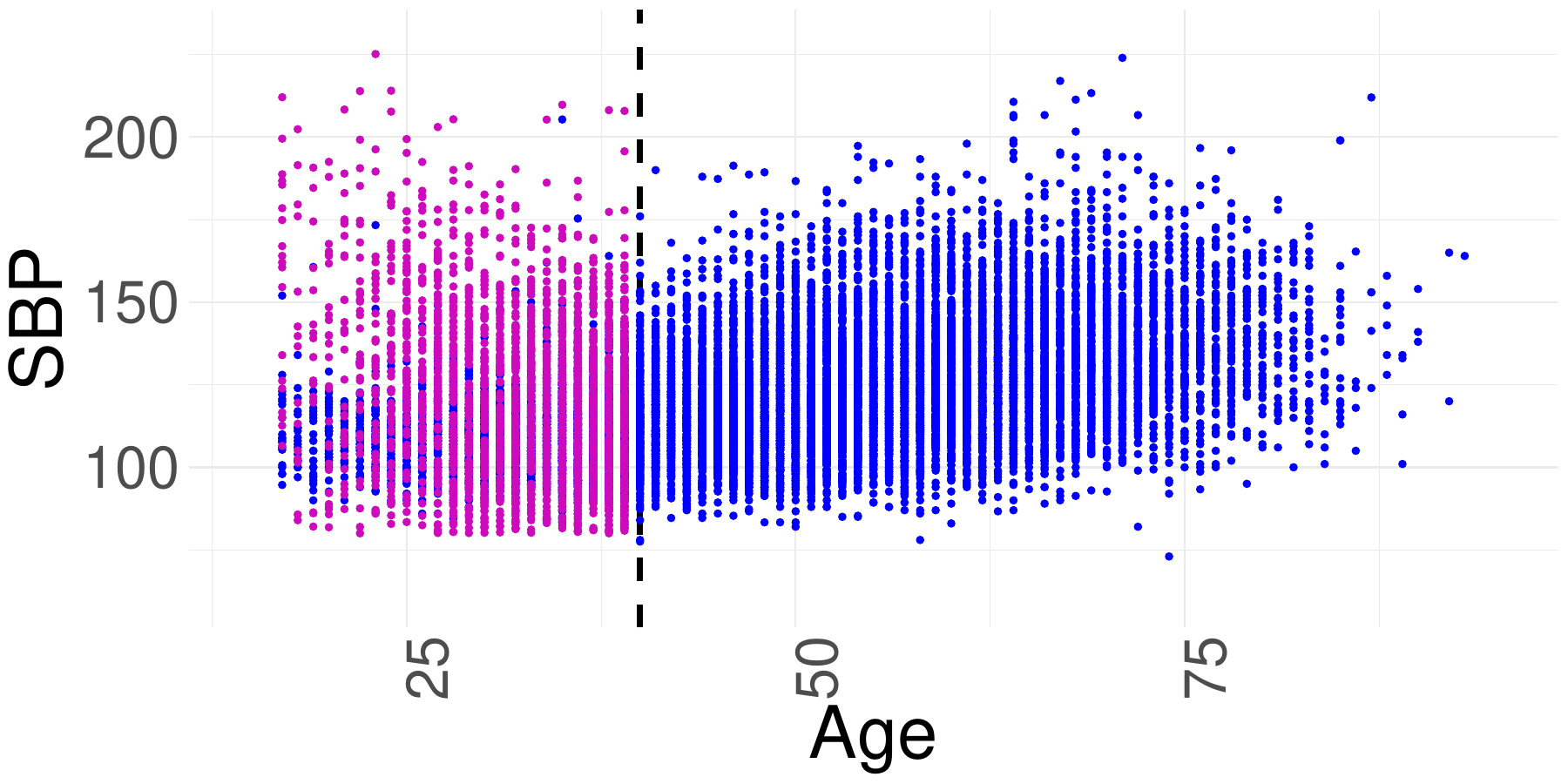}}
    {\includegraphics[width=0.46\textwidth,height=0.22\textheight]{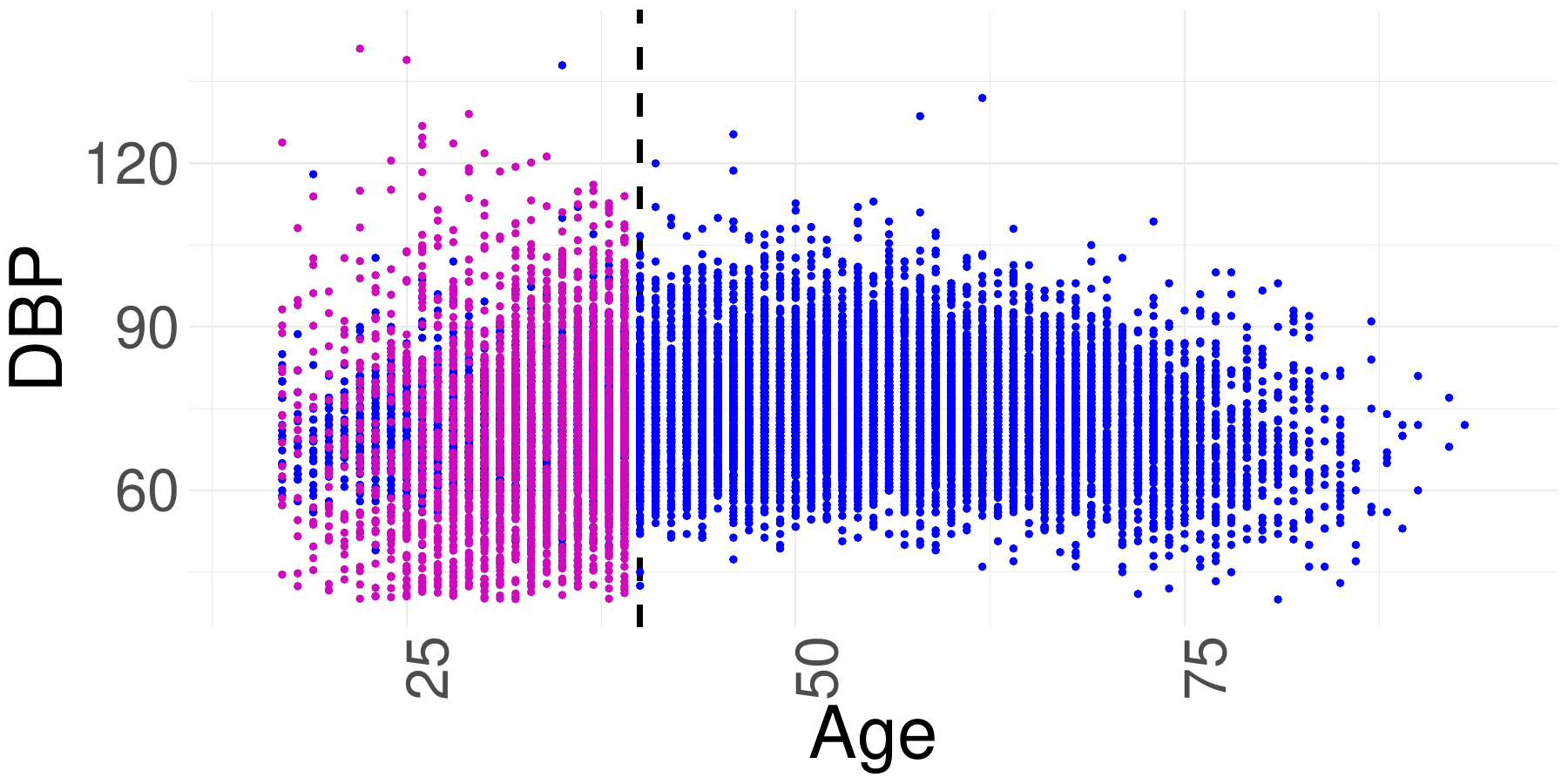}} 

{\includegraphics[width=0.46\textwidth,height=0.22\textheight]{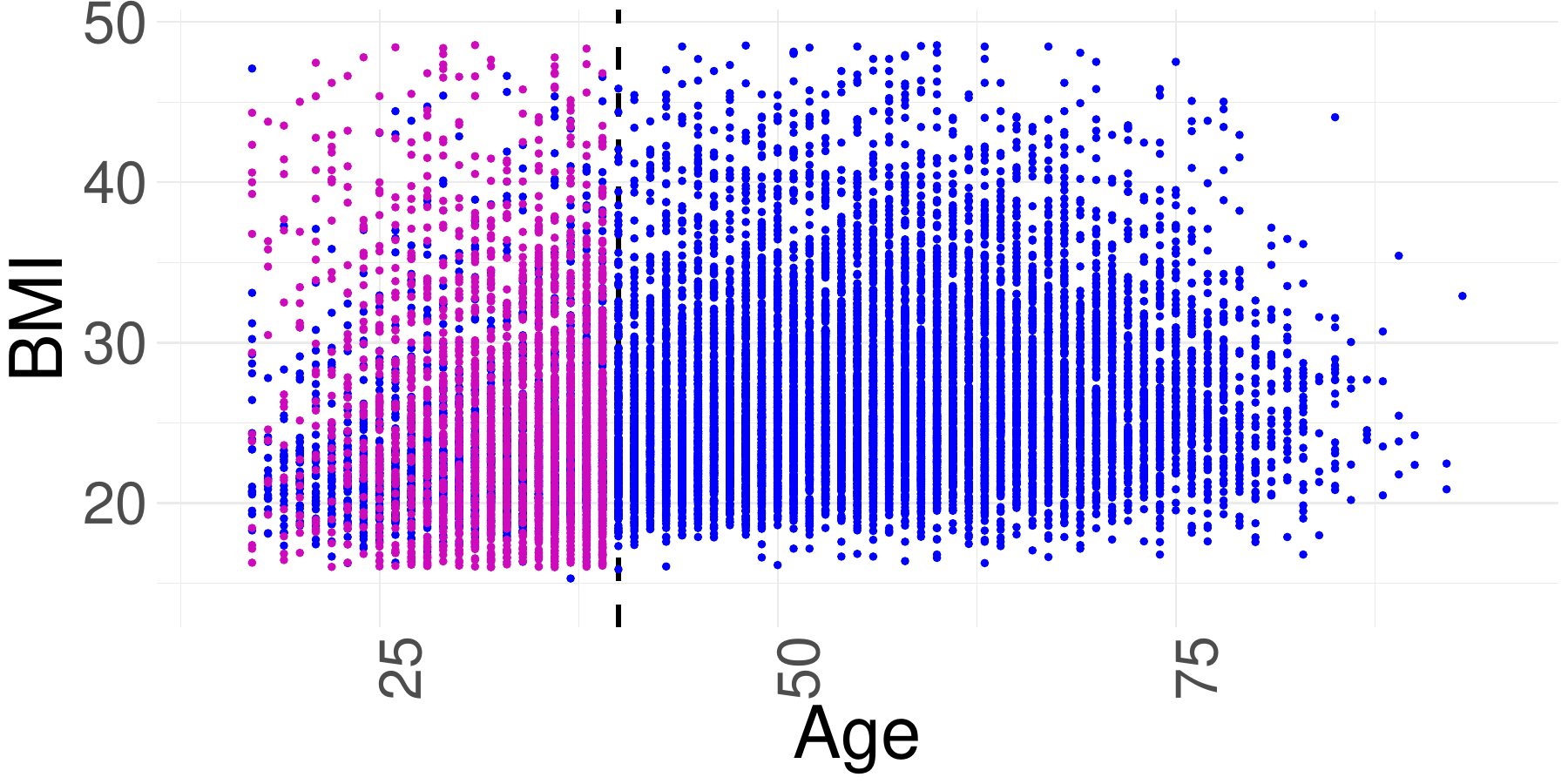}}
{\includegraphics[width=0.46\textwidth,height=0.22\textheight]{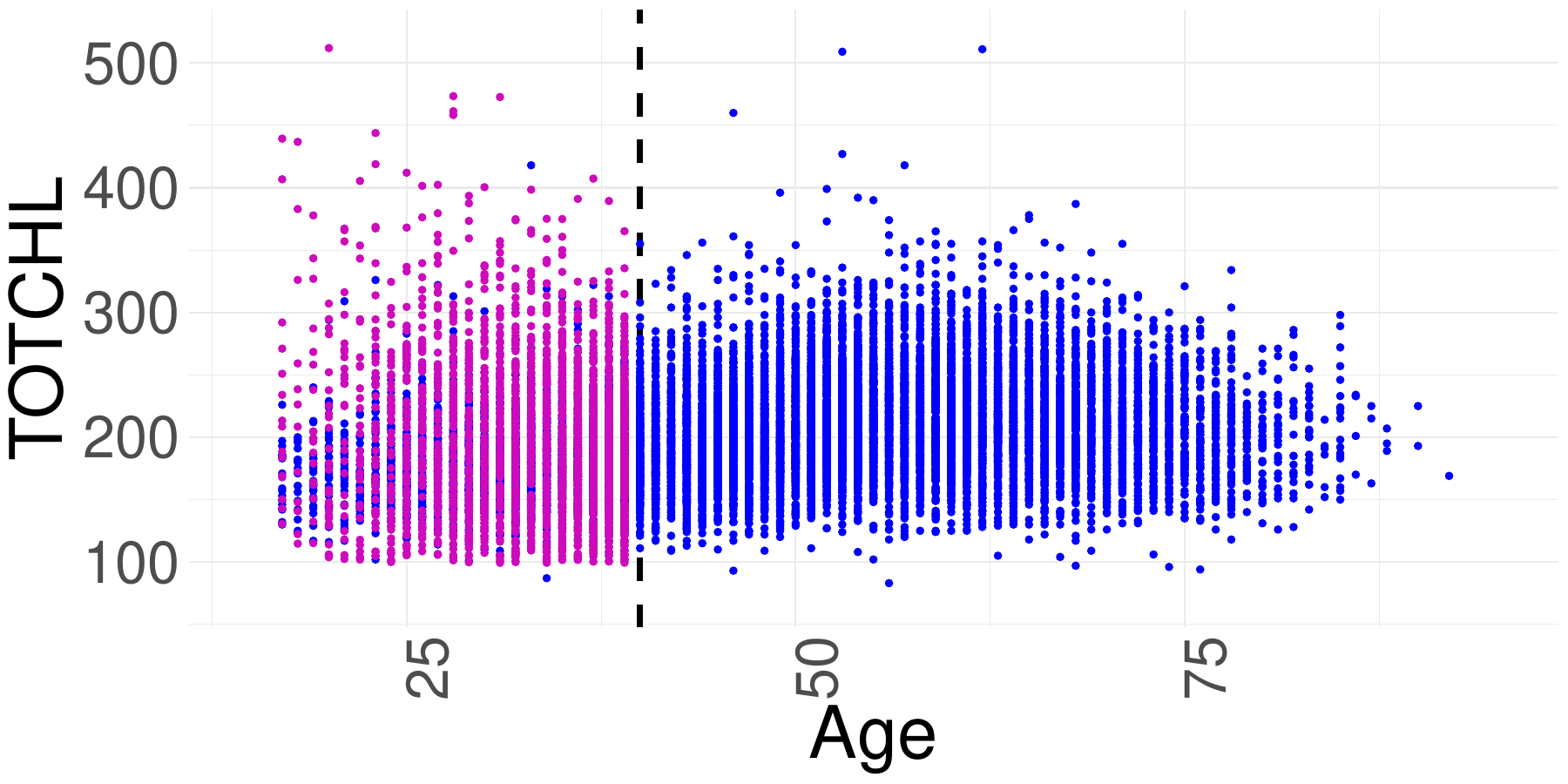}} 

    {\includegraphics[width=0.46\textwidth,height=0.22\textheight]{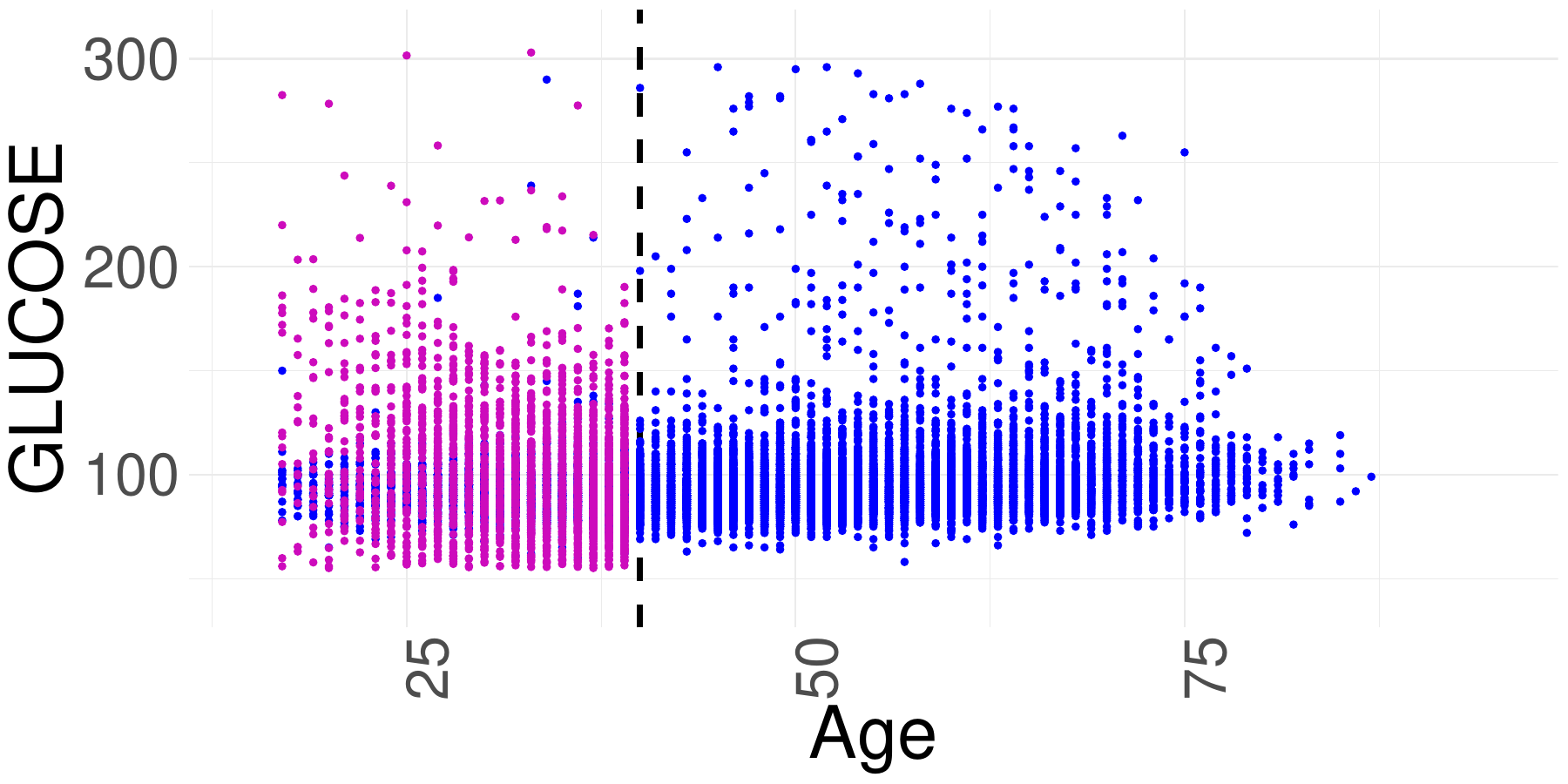}}
    {\includegraphics[width=0.46\textwidth,height=0.22\textheight]{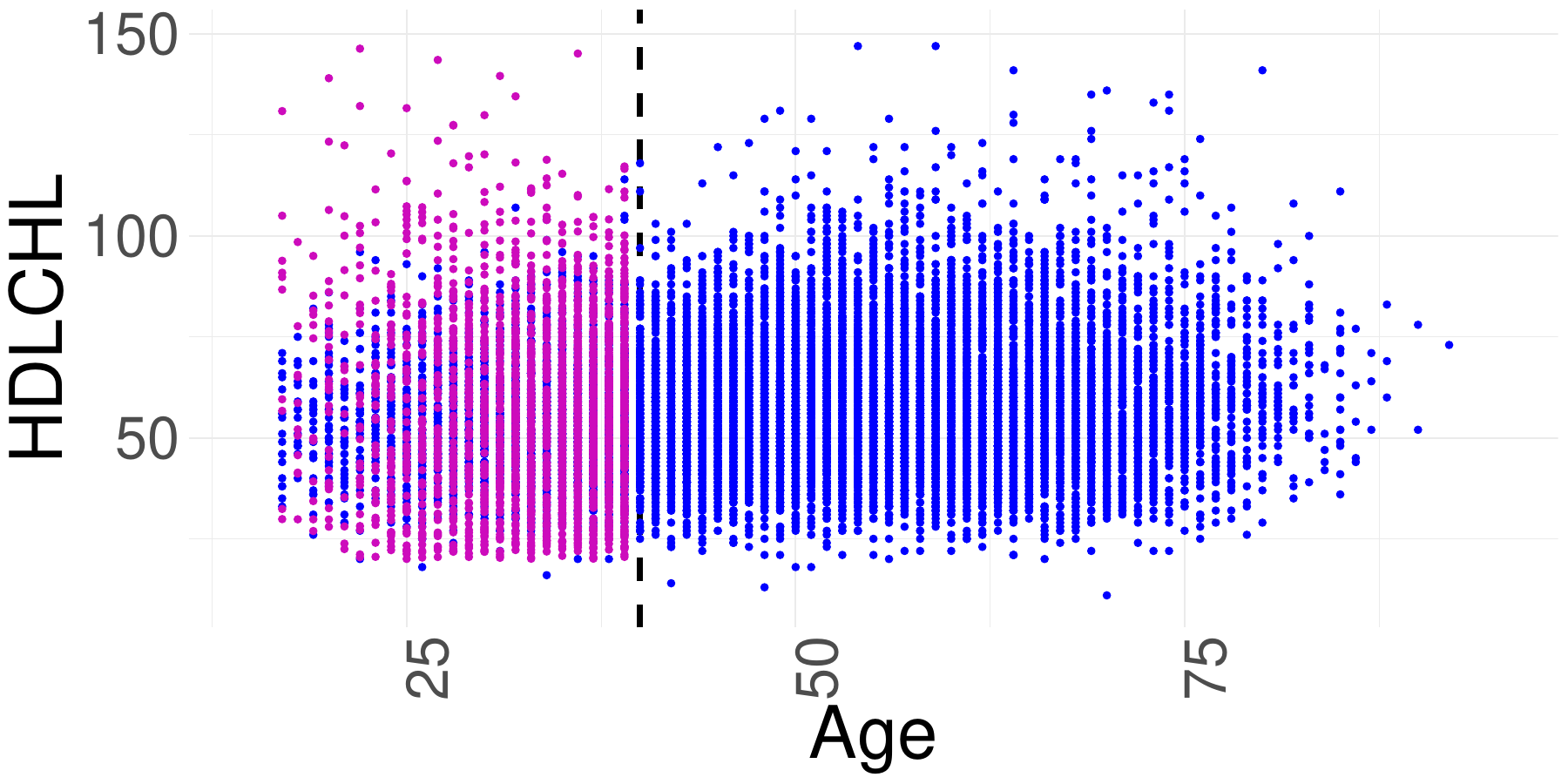}} 

    {\includegraphics[width=0.46\textwidth,height=0.22\textheight]{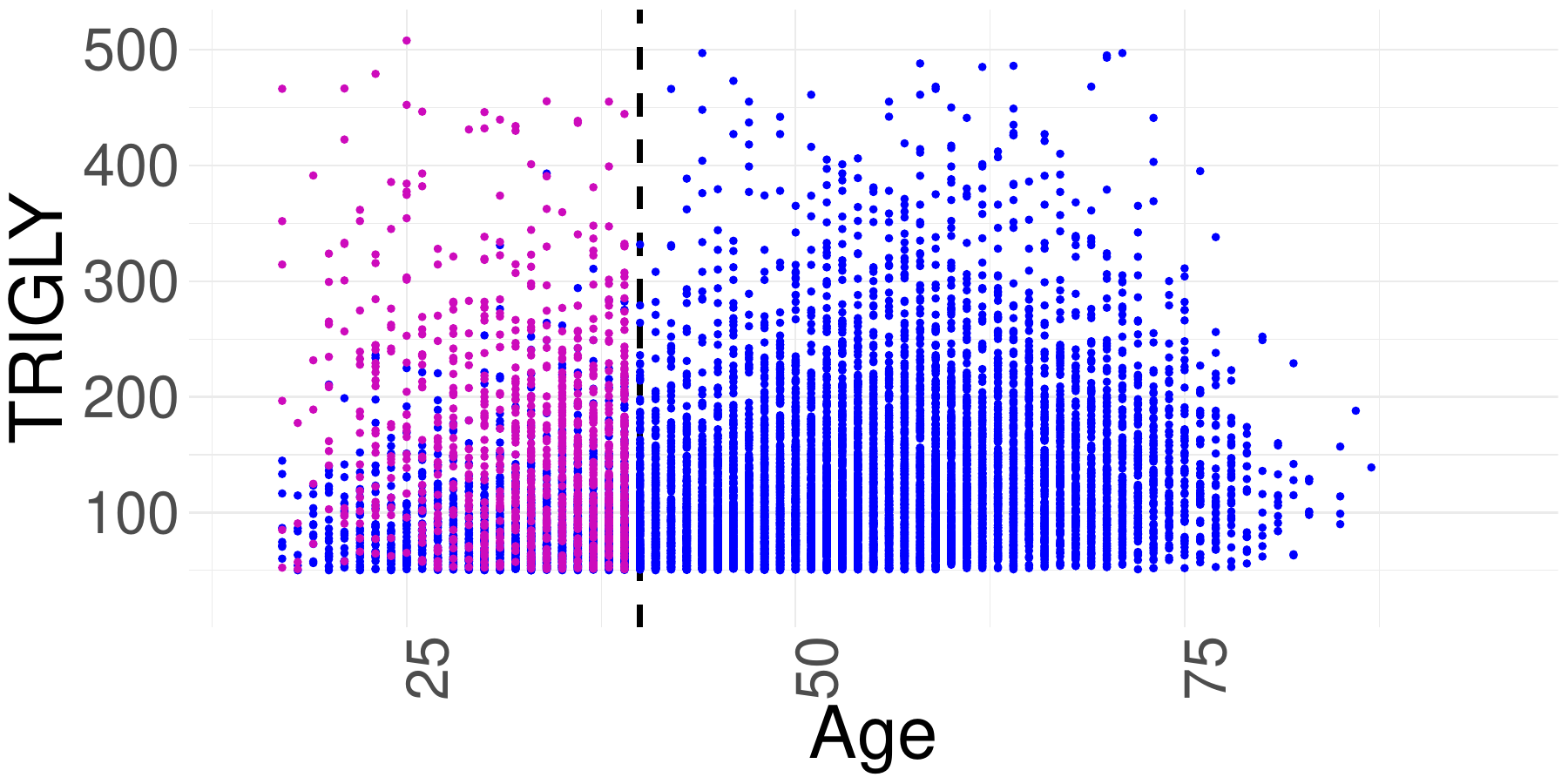}}
    \caption{Imputed (red) and observed (blue) values of risk factors against age for women in FOS.}
    \label{fig:women_L40}
\end{figure}

\subsection{Imputation Results for FOS Cohort} \label{supp:Imputation_Results}
For the FOS cohort, we applied Rubin’s rules to combine results across the imputed datasets. The mean AUC coefficient was 0.35 (SE = 0.15) for men and 0.77 (SE = 0.24) for women. In comparison, the real dataset yielded an AUC of 0.36 (SD = 0.12) for men and 0.68 (SD = 0.19) for women. Although imputation introduced some variability, the imputed estimates remained closely aligned with the observed data, demonstrating that the model effectively captured the underlying risk factor trajectories in the FOS cohort.

\section{Simulation Study} \label{supp:Simulation_Study}
To evaluate the performance of the proposed modeling framework, we conducted a simulation study based on the longitudinal model described in Section~3, simplified to a single risk factor and excluding baseline covariates. The goal was to examine the model’s ability to recover the cohort-level correlation structure, associated with \(\boldsymbol{\Lambda}\) under both full and partial age-range coverage across cohorts.

\subsection{Simulation Design} \label{supp:Simulation_Design}
We generated six synthetic cohorts (\(K = 6\)), each comprising 1{,}000 individuals observed at 13--14 visits. This number of repeated measurements ensured that each simulated subject contributed observations across a wide span of adulthood. Individual ages spanned 18--88 years, with visit intervals randomly drawn from 3--5 years to mimic realistic follow-up schedules. Baseline exam ages were restricted to at most 30 years to ensure sufficient early-age representation. This constraint guaranteed that all individuals contributed data from young adulthood through at least midlife, providing stable age trajectories before any age-based modifications were applied later in the study design.

\begin{figure}[!t]
    \centering
    \subfloat[Full coverage]{\includegraphics[width=7.5cm,height=5cm]{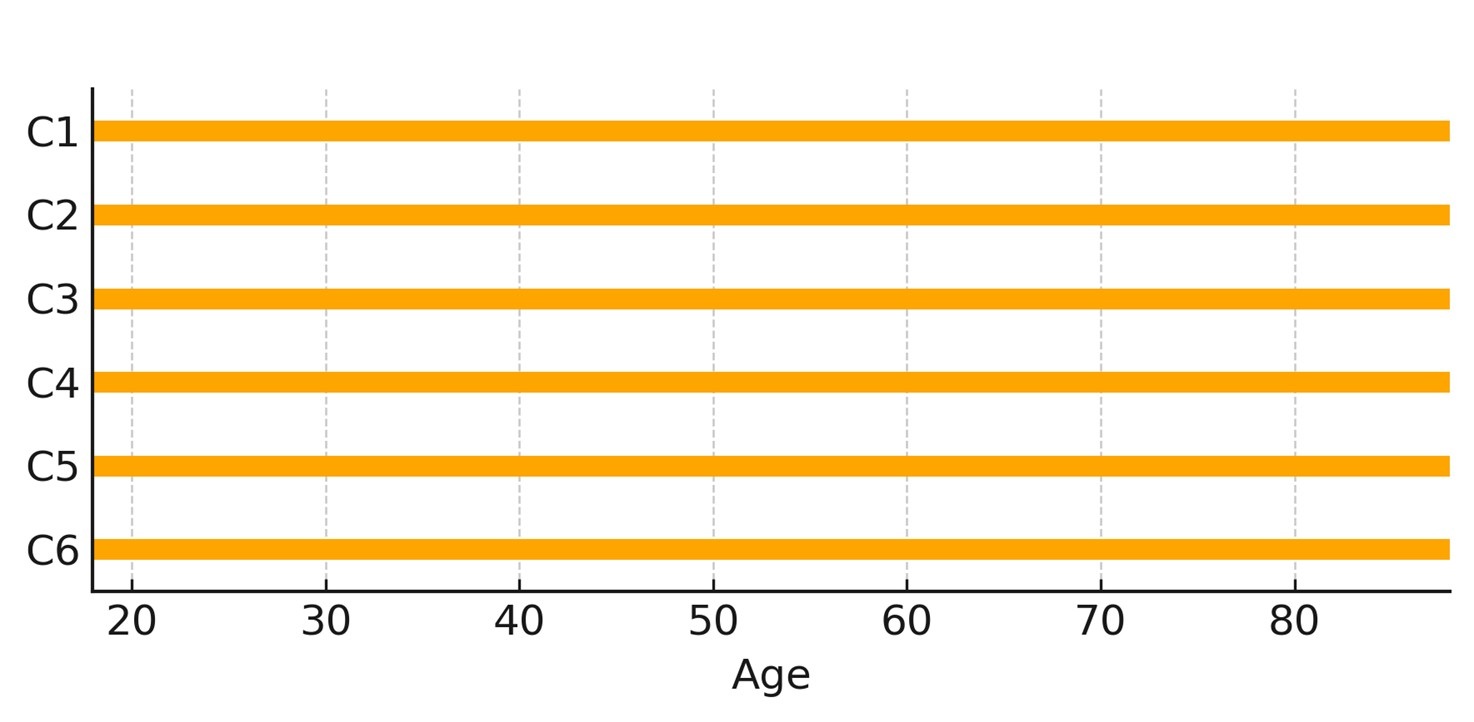}}
    \subfloat[Partial coverage]{\includegraphics[width=7.5cm,height=5cm]{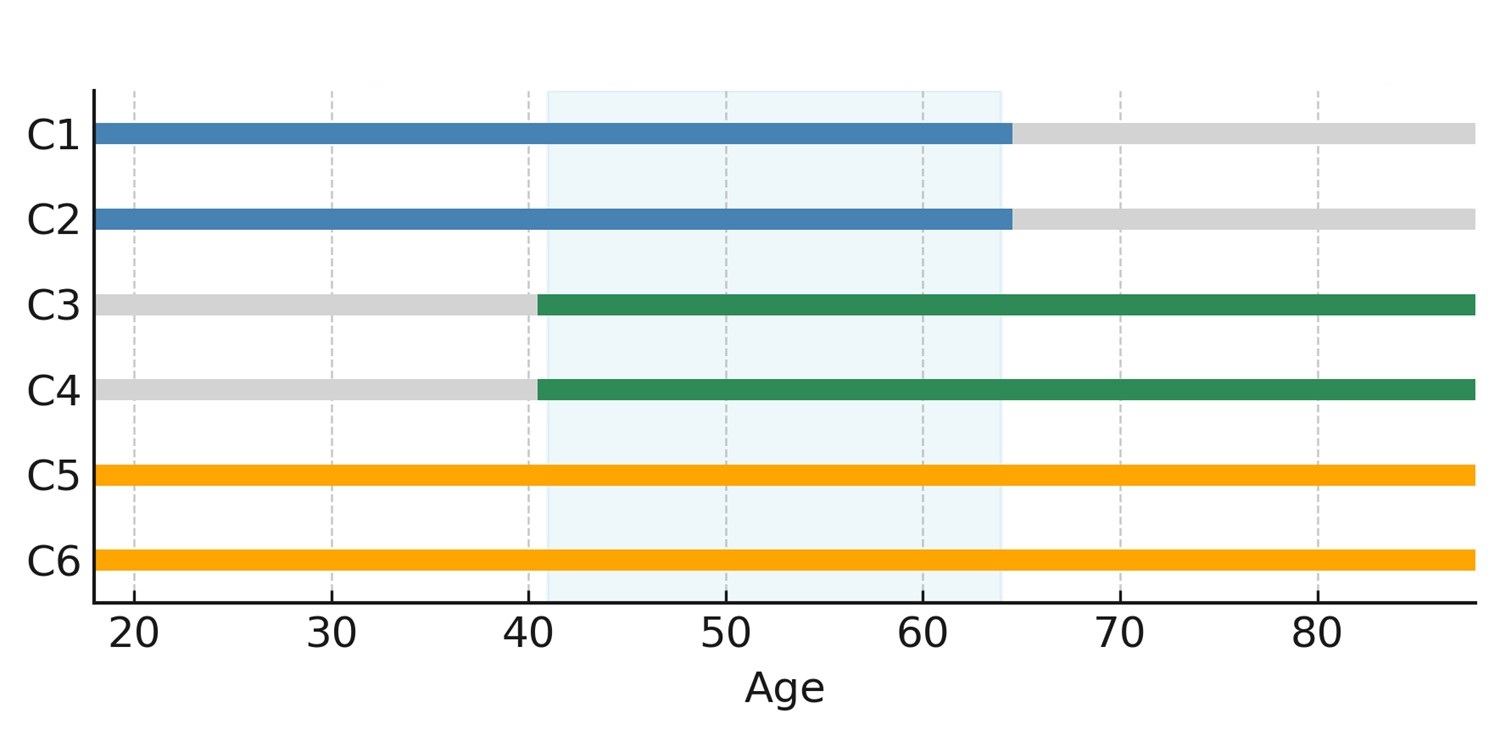}}

    \caption{
Illustration of simulated cohort overlap across the age range for six simulated cohorts (C1–C6). 
Panel~(a) the full‐coverage, 
and panel~(b) the partial‐coverage setting.}
\label{fig:sim_plots}
\end{figure}

Two simulation configurations were considered. 
In the \textit{full-coverage} scenario, all six simulated cohorts contributed data across the complete age range (18--88~years), representing a setting in which every cohort provides follow-up over the full life course (Figure~\ref{fig:sim_plots} (a)). 
In the \textit{partial-coverage} scenario, the same full-coverage datasets were modified by deleting observations outside each cohort’s designated age window: cohorts~1--2 provided data up to age~64, cohorts~3--4 contributed observations starting from age~41, and cohorts~5--6 covered the full range (Figure~\ref{fig:sim_plots} (b)). 
This configuration mimics the heterogeneity of age coverage across the LRPP cohorts, where CARDIA primarily represents younger participants, CHS captures older adults, and studies such as ARIC, MESA, FHS, and FOS span intermediate or full adult age ranges.
\subsection{Data-Generating Model} \label{supp:Data-Generating}
Data were simulated from the same piecewise linear spline model described in Section~3, 
restricted to a single risk factor and excluding baseline covariates. 
For subject \(i\) in cohort \(k\) at visit \(j\), the outcome was generated as
\[
y_{ik}(a_{ij}) = \boldsymbol{A}^{\top}(a_{ij})\boldsymbol{\beta}_{ik} + \varepsilon_{ik}(a_{ij}),
\]
where \(\boldsymbol{A}(a_{ij})\) is the spline basis vector consisting of an intercept, linear age, and truncated linear terms \((a - s_p)_+\) with knots at ages 18, 28, 38, 48, 58, 68, and 78. 
The coefficient vector \(\boldsymbol{\beta}_{ik}\) incorporates cohort- and subject-specific effects, as described Equation (4) of the manuscript.
Cohort- and subject-level covariance structures were specified to emulate realistic correlation patterns observed in the LRPP data. 
Residual errors \(\varepsilon_{ijk}\) were drawn from a skew-normal distribution with age-dependent scale parameters, allowing mild heteroskedasticity while maintaining realistic outcome ranges.

\subsection{Model Fitting and Evaluation} \label{supp:Model_Fitting}
A total of 200 simulated datasets were generated under the full-coverage configuration. 
Each dataset was first analyzed using the complete data (full overlap), after which selected age ranges were removed to create a corresponding partial-coverage version. Thus each simulated dataset has paired full and partial versions.
Both the full- and partial-coverage datasets were analyzed using the same model structure as described in Section~3, 
fitted via Bayesian inference. 
Convergence was assessed using nested \(\widehat{R} < 1.1\) diagnostics and trace plots. 

A primary focus was to assess how incomplete age coverage affected the accuracy of estimated cohort correlations. For each pairwise correlation, we computed bias, mean squared error (MSE), and nominal 95\% credible-interval (CrI) coverage under both simulation settings (Table~\ref{tab:Lambda_recovery}).

Comparing the two scenarios reveals the impact of incomplete age coverage on model performance. 
As expected, the full-coverage scenario, where all cohorts contribute observations across the entire adult age range, yielded relatively low bias and small MSE, with correlations showing mild attenuation toward zero. Coverage rates are generally above nominal, reflecting conservative interval estimates under the hierarchical model.
While performance was generally slightly better under full coverage, the differences between the two scenarios were modest, reflecting the robustness of the proposed modeling framework. In the partial-coverage scenario, where certain cohorts contribute only to younger or older age ranges, estimation remained accurate and well-calibrated, with only small changes in MSE and coverage across cohort pairs. These results demonstrate that the proposed model can recover the true correlation structure well even when overlap across cohorts is limited.
\begin{table}[!t]
\caption{Posterior summaries of cohort-level correlations under the two simulation settings, full and partial age coverage.}
\label{tab:Lambda_recovery}
\centering
\begin{tabular}{lcccccccc}
\hline
 & \multicolumn{1}{c}{True} &
 \multicolumn{3}{c}{Full coverage} &
 \multicolumn{3}{c}{Partial coverage} \\
\cline{3-5} \cline{6-8}
Cohort pair
 & Corr
 & Bias & MSE & Coverage
 & Bias & MSE & Coverage \\
\hline
C1--C2 & 0.55 & -0.049 & 0.013 & 0.995 & -0.038 & 0.013 & 0.995 \\
C1--C3 & 0.25 & -0.057 & 0.017 & 1.000 & -0.045 & 0.017 & 0.995 \\
C1--C4 & 0.25 & -0.049 & 0.015 & 0.995 & -0.054 & 0.017 & 0.995 \\
C1--C5 & 0.25 & -0.063 & 0.018 & 1.000 & -0.062 & 0.016 & 1.000 \\
C1--C6 & 0.25 & -0.042 & 0.014 & 1.000 & -0.048 & 0.015 & 0.995 \\
C2--C3 & 0.25 & -0.092 & 0.025 & 1.000 & -0.060 & 0.016 & 1.000 \\
C2--C4 & 0.25 & -0.065 & 0.019 & 1.000 & -0.063 & 0.020 & 1.000 \\
C2--C5 & 0.25 & -0.078 & 0.020 & 1.000 & -0.068 & 0.017 & 0.990 \\
C2--C6 & 0.25 & -0.056 & 0.015 & 1.000 & -0.060 & 0.017 & 0.995 \\
C3--C4 & 0.55 & -0.129 & 0.030 & 0.990 & -0.125 & 0.026 & 0.990 \\
C3--C5 & 0.25 & -0.084 & 0.018 & 1.000 & -0.084 & 0.022 & 0.990 \\
C3--C6 & 0.25 & -0.090 & 0.018 & 1.000 & -0.083 & 0.019 & 0.995 \\
C4--C5 & 0.25 & -0.082 & 0.018 & 1.000 & -0.091 & 0.022 & 0.990 \\
C4--C6 & 0.25 & -0.086 & 0.021 & 1.000 & -0.092 & 0.020 & 0.995 \\
C5--C6 & 0.55 & -0.168 & 0.040 & 0.990 & -0.163 & 0.039 & 0.995 \\
\hline
\end{tabular}
\end{table}
\end{document}